\documentclass[aps,pre,twocolumn,showpacs,showkeys,superscriptaddress,floatfix]{revtex4-1}
\pdfoutput=1

\usepackage[utf8]{inputenc}
\usepackage[english]{babel}
\usepackage{hyperref}

\usepackage{amsmath}
\usepackage{amssymb}
\usepackage{amscd}
\usepackage{mathtools}

\usepackage{color}
\usepackage{graphicx}
\usepackage{subfigure}

\usepackage{upgreek}
\usepackage{dcolumn}
\usepackage{natbib}

\newcommand{\rmd}{{\mathrm d}}

\newcommand{\dbar}{{\mathchar '26 \mkern-11mu {\mathrm d}}}

\begin{document} 

\title{Tension and chemical efficiency of Myosin-II motors}
\author{Pieter Baerts}
\author{Christian Maes}
\affiliation{KU Leuven, Instituut voor Theoretische Fysica, Celestijnenlaan 200D, B-3001 Leuven, Belgium}
\author{Jiří Pešek}
\author{Herman Ramon}
\affiliation{KU Leuven, BIOSYST-MeBioS, Kasteelpark Arenberg 30, B-3001 Leuven, Belgium}

\pacs{05.70.Ln, 87.16.A-, 87.16.Nn}

\begin{abstract}
Recent experiments demonstrate that molecular motors from the Myosin~II family serve as cross-links inducing active tension in the cytoskeletal network.  
Here we revise the Brownian ratchet model, 
previously studied in the context of active transport along polymer tracks, in setups resembling a motor in a polymer network, also taking into account the effect of electrostatic changes in the motor heads. 
We explore important mechanical quantities and show that such a model is also capable of mechanosensing. 
Finally, we introduce a novel efficiency based on excess heat production by the chemical cycle which is directly related to the active tension the motor exerts. 
The chemical efficiencies differ considerably for motors with a different number of heads, while their mechanical properties remain qualitatively similar. 
For motors with a small number of heads, the chemical efficiency is maximal when they are frustrated, a trait that is not found in larger motors. 
\end{abstract}

\maketitle 

\section{Introduction}
The cytoskeleton of a cell mainly consists of long semi-flexible actin polymers and myosin II molecular motors~\cite{mitchison1996actin}. 
These motors contain several heads that are able to bind to polymer filaments and walk along them~\cite{pollard1982structure}. 
The mechanism for walking relies both on a chemical cycle in the motor's head and on the asymmetry of a periodic electrostatic potential along such a polymer \cite{Reimann2002introduction}. 
The cytoskeleton serves multiple purposes: 
transporting large vesicles, providing the cell's stiffness, driving the separation of the cell during cytokinesis and governing the cell's motility, i.e. the ability of the cell to move and change its shape \cite{ross2008cargo,mitchison1996actin,rosenblatt2004myosin}.
The part of the cytoskeleton that plays a crucial role in the stiffness and motility of the cell is the actin-myosin cortex \cite{vicente2009non}, which will serve as the exemplary system in this paper.

Recently, it has been argued that the main role of myosin~II motors inside the cytoskeleton is to generate tension in the actin polymer network \cite{ma2012nonmuscle,chugh2017actin,monier2010actomyosin}.
Therefore, the main goal of this work is to capture the behavior of myosin~II motors in such a network 
and to introduce an efficiency of the motor that is not related to transport, but rather to tension in the motor.
We show that this efficiency is directly related to the net supply of energy provided by the chemical cycle 
and we investigate how this chemical efficiency depends on ATP concentrations and on the external force on the system.
However, before we do so, we revisit certain features of molecular motors in a simple Brownian ratchet model \cite{reimann2002brownian} with constant chemical rates and relate the force-velocity relation to the mean force generated by the ratchet potential and to the tension in the motor.

Cytoskeletal polymers exhibit a sequence of monomers that cause a periodic structure \cite{yogurtcu2012mechanochemical}. 
These monomers are polarized, so that when chained together in a polymer, they give rise to an asymmetric, ratchet-like, electrostatic potential along the length of the filament \cite{Nie2014,nie2014conformational}. 
A simplified version of such a potential, used in this work, is shown in Fig.~\ref{fig:energy}. 
The heads of a myosin~II motor also carry an electric charge \cite{barterls1993myosin} and  therefore feel the sawtooth potential of the filaments.
The chemical cycle that powers the molecular motor is driven by ATP molecules that can attach to a head of the motor, causing it to change the charge distribution of the head \cite{adelstein1980regulation}.
As a consequence, the head loosens its bond with the polymer filament. 
When bound to a motor head, the ATP molecule will hydrolyze to ADP and release a phosphate \cite{gajewski1986thermodynamics}. 
It this state, the head can attach to the polymer again, ADP will be released from the motor head and the chemical cycle can start over. 
Hence, the chemical state of a motor head in our model will determine how strongly it can be coupled to a cytoskeletal filament \cite{Nie2014,nie2014conformational}. 
Therefore, the head will experience a sawtooth potential that is stochastically flashing in time, changing the amplitude of the potential from high to low.

This mechanism provides a motivation for using ratchet models, 
where the motor heads interact with a periodic asymmetric potential that changes depending on the chemical state of the head. 
The basic cycle of the ratchet model works as follows:
Whenever a motor is in a state where no ATP is bound to heads, i.e., the state in which the motor is tightly bound to the filament, 
the head will most probably be very close to a minimum of the ratchet potential. 
On the other hand, while in the state where the ATP is bound to heads, it is easier for the motor to diffuse over a larger distance. 
Due to the asymmetry of the ratchet potential, the motor will be biased to explore one side of the ratchet potential rather than the other. 
Therefore, the motor will more likely jump to an adjacent period of the sawtooth in the preferred direction, indicated by the asymmetry of the ratchet potential \cite{reimann2002brownian}.
This will eventually lead to directed motion of the motor on long timescales \cite{hoffmann2016molecular,Reimann2002introduction,de2007symmetries}. 

Ratchet models have already been studied extensively in the context of molecular motors \cite{reimann2002brownian,astumian1994fluctuation,astumian1996mechanochemical,julicher1997modeling,Reimann2002introduction,julicher1997spontaneous,peskin1995correlation,huxley1969mechanism,huxley1971proposed}.
The models, used in these studies, are very similar to each other but often differ in some basic assumptions.
The specific model that we have studied is presented in section~\ref{sec:ratchet}. 
It has the important trait that the chemical rates do not depend on spatial degrees of freedom, 
the sawtooth potential has a non-zero amplitude in all states 
and motor heads can attach and detach anywhere along the polymer filament.
Furthermore, the rates in our model do not depend on external forces acting on the motor but solely on concentrations of chemicals involved in ATP hydrolysis. 
We assume that constant concentrations are maintained in the environment at all times.
While for discrete models force depended rates are deemed necessary to reproduce experimental results \cite{sung2017chapter}, 
in the scope of this study, constant rates will prove to suffice.
The type of motor that we study is non-muscle myosin~II, which has a small number of heads \cite{pollard1982structure}. 
These non-muscle myosin~II motors are crucial to the motility of most eukaryotic cells \cite{vicente2009non}.

In muscle cells and in experimental studies with synthetic myosin motors, 
the number of heads per motor is also much higher than in non-muscle cells \cite{brown2009cross-correlated}.
For those motors, mean-field calculations have led to analytic results for the force-velocity relation \cite{julicher1997modeling},
which play an important role in interpreting and understanding those in-vitro experiments.
For motors with fewer heads, mean-field methods are not suitable and numerical simulations prove to be useful.

In section~\ref{sec:model_verification} we describe the dynamics of this motor-polymer system in detail by means of numerical methods.
Specifically we study the mean velocity, force -- velocity relation and stall force of myosin~II motors in order to verify our proposed model. 

As the main role of the myosin~II motor is to generate tension in the actin polymer networks \cite{ma2012nonmuscle,chugh2017actin,monier2010actomyosin}, 
we investigate the tension in the motor in section~\ref{sec:tension} for multiple setups, 
mimicking the typical polymer environment, 
e.g., the coupling of the motor to an elastic environment.
Earlier studies have also shown that myosin~II motors are able to sense the stiffness of their environment and that they can adapt the active forces that they generate accordingly \cite{stam2015isoforms,Albert2014}.
In those studies, the chemical rates did depend on the load on the motor. 
We  show that our simple ratchet model with constant rates is not only capable of capturing this important biological trait, it naturally emerges from the dynamics of the system without any additional assumptions.

Another important aspect of molecular motors is their ability to convert chemical energy into mechanical work \cite{astumian1996mechanochemical}.
The chemical cycle of ATP hydrolysis consists of many consecutive reactions and after one full cycle a net amount of energy is provided to the motor head \cite{gajewski1986thermodynamics}.
However, often this cycle is represented by only two states \cite{julicher1997modeling,Reimann2002introduction}, e.g. ATP bound and ADP bound, 
in which case a stochastic process on these states always satisfies detailed balance and no chemical energy can be extracted. 
In other words, the energetic gain of one reaction is canceled by the cost of the following reaction.
Nevertheless, since the chemical process is coupled to a diffusion process, we do break detailed balance in our model, 
as the energy gain per transition varies while the rates remain constant.

In section~\ref{sec:energie}, we look at the energy balances in various setups of a myosin~II motor interacting with actin filaments.
We link the excess heat for the chemical cycle to the ratchet force and consequently to the active tension produced by the motor.
This consideration leads to a novel way of defining a chemical efficiency, which is presented in section~\ref{sec:efficiency}. 
This efficiency compares the net amount of energy that is extracted from the ATP cycle to the total energy that is supplied to the motor.
Simultaneously, we show that the chemical efficiency relates the chemical input, measured by the differences in potential energy in each jump, 
to its active tension, rather than to its displacement.
We investigate how this efficiency depends on the external load, number of heads of the motor and ATP concentration. 
Consequently, we find that motors with a low number of heads, contrary to those with a high number of heads, achieve maximal chemical efficiency when they are frustrated, i.e., close to maximal tension.
That contrasts with the observation that from the mechanical perspective there is no qualitative difference between them. 

\section{The Ratchet model}
\label{sec:ratchet}
\begin{figure}[t]
\centering
\includegraphics[width=0.45\textwidth,height=!]{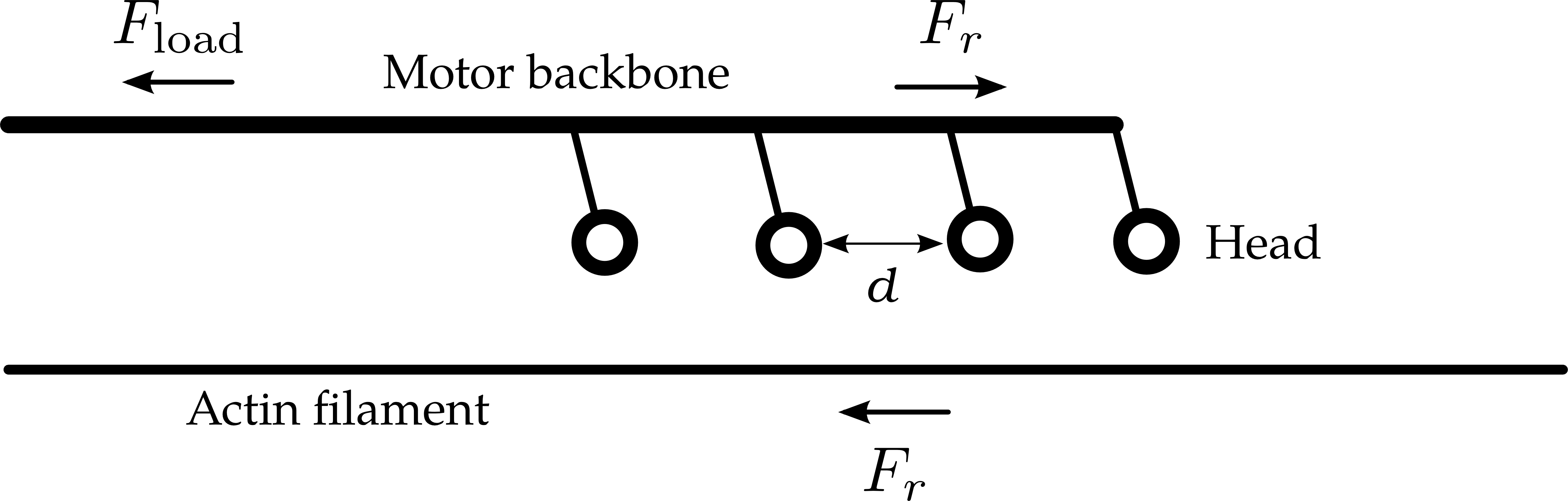}
\caption{
\label{fig:ratchet_setup}
Basic setup:
The backbone of the myosin motor is shown with four heads at distance $d$ from each other. 
The load $F_\text{load}$ is applied only on myosin and the actin filament is freely diffusing.
A force $F_\text{r}$ is generated by the ratchet interaction and causes movement in a preferred direction. 
} 
\end{figure}

To investigate the movement of myosin along the actin filament and the forces it generates, we consider the following one--dimensional model.
At first, we assume that the rigidity of the actin polymer and myosin backbone is high enough so that their elongations can be neglected.
Secondly, we assume that all myosin heads are attached directly to the myosin backbone at fixed equidistant positions; see Fig.~\ref{fig:ratchet_setup}.
These simplifications allow us to model both the motor and the filament as point particles undergoing overdamped diffusion in the inter-cellular environment \cite{vanKampen1981stochastic}.
Moreover, each myosin head has internal (chemical) states, e.g. telling
 whether or not there is an ATP bounded to the myosin head. 

\begin{figure}[t]
\centering
\includegraphics[width=0.45\textwidth,height=!]{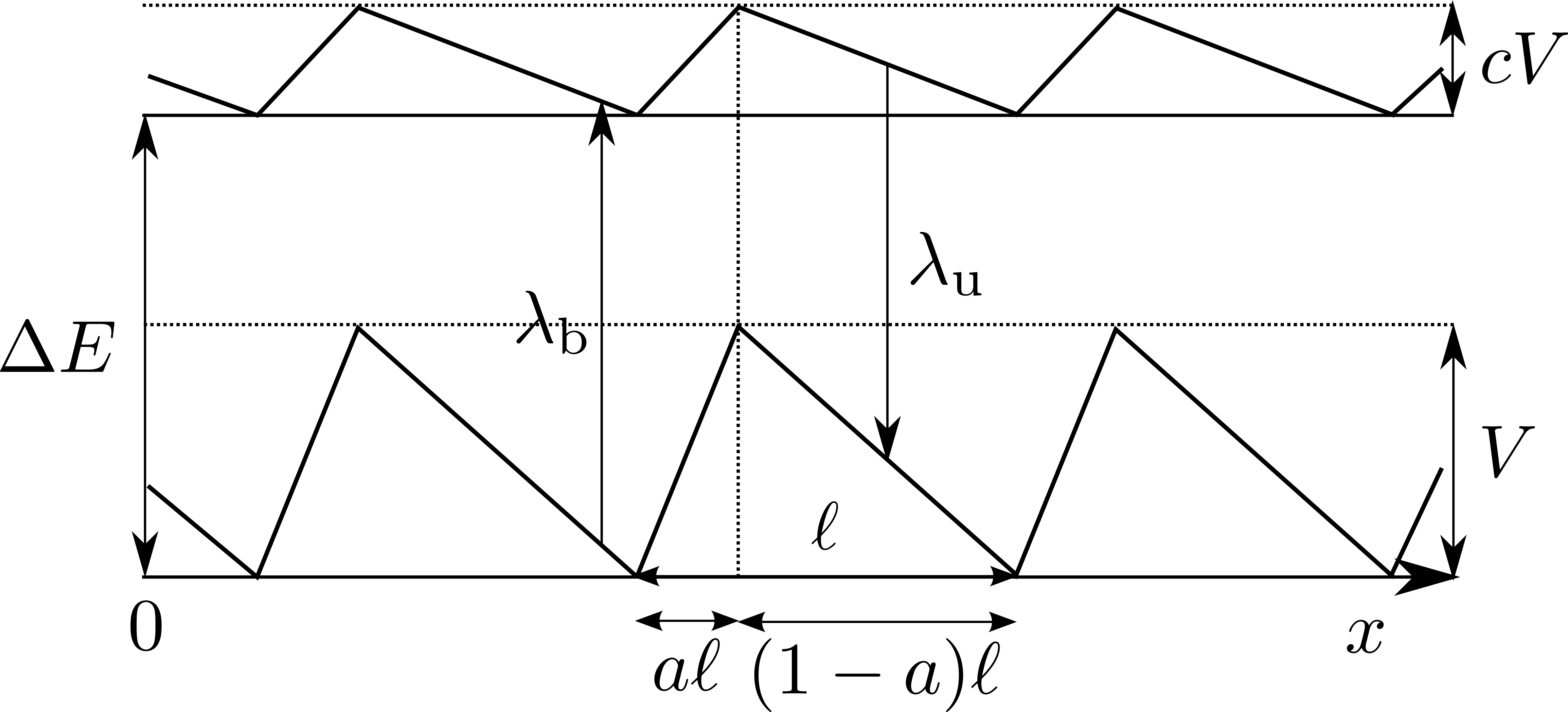}
\caption{
\label{fig:energy}
Ratchet model: the interaction between a myosin head and the actin polymer is given via a sawtooth potential with height $V$, period $\ell$ and skewness parameter $a$.
The transition between ATP \emph{bound} and \emph{unbound} state is governed by rates $\lambda_\text{b}$, $\lambda_\text{u}$ respectively, 
and is associated with the chemical energy $\Delta E$.
In the \emph{bound} state the ratchet potential is scaled down by factor $c$. 
}
\end{figure}

The dynamics of the positions is described by coupled overdamped Langevin equations 
\begin{align}
&\begin{multlined}[b]
\rmd x_\text{M} = 
- \frac{1}{\gamma_\text{M}} \left[ \nabla_\text{M} V_t(x_\text{M}(t) - x_\text{A}(t)) + F_\text{load} \right] \; \rmd t 
\\ 
+ \sqrt{ \frac{2 k_B T}{\gamma_\text{M}} } \; \rmd W_\text{M} ,
\end{multlined}
\label{eq:eom_M} \\
&\begin{multlined}[b][.42\textwidth]
\rmd x_\text{A} = 
- \frac{1}{\gamma_\text{A}} \nabla_\text{A} V_t(x_\text{M}(t) - x_\text{A}(t)) \; \rmd t 
\\
+ \sqrt{ \frac{2 k_B T}{\gamma_\text{A}} } \; \rmd W_\text{A} ,
\end{multlined}
\label{eq:eom_A}
\end{align}
where $x_\text{M}$ denotes the position of the motor's first head, 
$x_\text{A}$ denotes the position of an arbitrary reference point on the actin filament, 
$\gamma$ is the Stokes drag,
$T$ is the absolute temperature of the environment,
$W_\text{M}$ and $W_\text{A}$ denote independent Wiener processes,
$F_\text{load}$ is an external load on the motor, 
and the interaction potential $V_t$ represents the overall ratchet interaction, 
which is the sum over individual contributions from all motor heads,
\begin{equation}
V_t(x) = \sum\limits_{i=0}^{N-1} \zeta_t(i) \, V_\text{r} (x - i \, d ), 
\label{eq:ratchet_interaction}
\end{equation}
where $d$ is the distance between myosin heads, 
$N$ is the number of heads per motor,
$\zeta_t(i)$ represents the internal state, 
and $V_\text{r}$ is the periodical ratchet potential; see Fig.~\ref{fig:energy},
\begin{equation}
V_\text{r}(x) =  \begin{cases}
        V_\text{r}(x+\ell) & x < 0 \\[1ex] 
        \displaystyle \frac{ V x }{ a \ell } & 0 \leq x < a \ell \\[2ex]
        \displaystyle \frac{ V (\ell-x) }{ (1-a) \ell } & a \ell \leq x < \ell \\[2ex]
        V_\text{r}(x-\ell) & \ell \leq x  
   \end{cases} ,
   \label{eq:ratchet_potential}
\end{equation}
with total height $V$,
period $\ell$ 
and skewness parameter $a \in [0,1]$. 

The dynamics of the internal state of heads is governed by a continuous time Markov proces on two states with constant rates,
\begin{equation}
\zeta_t = 1 \overset{\lambda_\text{b}}{\underset{\lambda_\text{u}}{\rightleftarrows}} \zeta_t = c ,
\label{eq:transition}
\end{equation}
where $c\in\left[0,1\right]$ determines the amplitude of the ratchet potential when the motor is weakly interacting with the filament, 
which we use to account for changes in the free energy landscape \cite{Nie2014,nie2014conformational}.

This choice is justified by multiple considerations: 
First, in our model the internal state represents whether ATP (or one of its products) is \emph{bound} or \emph{unbound} to the myosin head,
rather than the binding of the myosin motor to the actin filament itself.
In other words, the internal state characterizes the electrochemical properties of the head, 
which further determine the magnitude of the interaction potential with the actin filament.

Secondly, we assume that whether the head itself is bound or not has no influence on the binding site of the ATP molecule.  In other words
 there is no effective dependence of the dynamics of the chemical cycle on the relative position of myosin and actin. 

Similarly, we assume that an external load does not influence the binding site of ATP on the motor head. 
It only acts on the motor backbone and opposes forces originating from the ratchet interaction. 

Finally, we simplify the full chemical network governing the internal states of myosin~II \cite{Bierbaum2011,Bierbaum2013} to only two states,  
which represent the long lasting states that significantly change the physical properties of the motor head.
Namely, we look at states where either ATP is bound or ADP  is bound to the myosin head, 
as escape rates from those states are small with respect to other transitions \cite{Bierbaum2011}. 
For these states, the estimated free energy landscapes \cite{Nie2014,nie2014conformational} very well reflect the changes in conformation and electric charge of the myosin head \cite{barterls1993myosin}.

Note that while the rates $\lambda_\text{b}$, $\lambda_\text{u}$ are independent of the load, 
they depend on the concentrations of the relevant chemical species involved in ATP hydrolysis. 
More specifically $\lambda_\text{b}$ is proportional to $[\text{ATP}]$ while $\lambda_\text{u}$ is inversely proportional to $[\text{ADP}]$ and $[\text{P}_\text{i}]$.

In order to obtain transition rates between these two states, we start from the three state model introduced by Albert et al. \cite{Albert2014},
which we further simplify by unifying the states corresponding to the hydrolized ATP (i.e. ADP+P) and ADP with P released, 
as this process occurs on a much shorter time-scale than other transitions.

While both the diffusion and the jump process are in equilibrium separately, 
the mechanism of the flashing ratchet brings the combined system out of equilibrium, \
leading to movement of the motor in a preferred direction, given by the asymmetry of the filament \cite{Reimann2002introduction,de2007symmetries}.

Fig.~\ref{fig:ratchet_setup} depicts the whole setup of the system
and Fig.~\ref{fig:energy} depicts the dynamics involved in flashing ratchet. 
Table \ref{tab:parameters} lists the values of the parameters used in this model. 
Some of them were taken from the literature while others were fitted, leading to realistic mean relative velocities~\cite{placcais2009spontaneous,saito1994movement} and stall force~\cite{kishino1988force,finer1994single} for myosin~II.
The friction coefficients have been calculated with Stokes' law~\cite{Broersma1960,Broersma1981}, using the known dimensions of the molecules~\cite{yogurtcu2012mechanochemical,pollard1982structure} and the viscosity of the intracellular medium~\cite{li2004diffusion}. 
\begin{table}[t]
\centering
\begin{ruledtabular}
\begin{tabular}{lcdcc}
Parameter & Symbol & \multicolumn{1}{c}{Value} & Units & Source\\
\hline
Thermal energy& $k_B T$ & 4.28 & pN nm & --- \\
Number of heads& $N$ & 4 & --- & \cite{pollard1982structure}\\
Head's spacing& $d$ & 15 & nm & \cite{pollard1982structure}\\
Step length& $\ell$ & 8 & nm & \cite{vilfan2003instabilities}\\
Binding rate& $\lambda_\text{u}$ & 40 & s$^{-1}$ & \cite{Albert2014} \\
Detaching rate& $\lambda_\text{b}$ & 80 & s$^{-1}$ & \cite{Albert2014} \\
Skewness& $a$ & 0.25 & --- & ---\\
Potential amplitude& $V$ & 40 & pN nm & ---\\
Potential scaling& $c$ & 0.3 & --- & \cite{Nie2014, nie2014conformational}\\
Chemical energy gap & $\Delta E$ & 30.4 & pN nm & \cite{gajewski1986thermodynamics}\\
Motor friction& $\gamma_\text{M}$ & 0.66 & pN $\upmu$s nm$^{-1}$ & \cite{Broersma1960,Broersma1981} \\
Actin friction& $\gamma_\text{A}$ & 0.97 & pN $\upmu$s nm$^{-1}$ & \cite{Broersma1960,Broersma1981} 
\end{tabular}
\end{ruledtabular}
\caption{
\label{tab:parameters}
Table of all parameter values and a reference to their source.
}
\end{table}

\subsection{Alternative setups}
\label{sec:other_setups}
The previously introduced setup, shown in Fig.~\ref{fig:ratchet_setup}, is most commonly studied \cite{reimann2002brownian,astumian1994fluctuation,finer1994single,julicher1997modeling,kishino1988force,peskin1995correlation,saito1994movement},
with the aim to characterize the transport induced by molecular motors.
However, recent experiments show that the main role of myosin~II motors is to cross-link the cytoskeleton \cite{ma2012nonmuscle}
and consequently produce an active tension inside the network \cite{chugh2017actin,monier2010actomyosin}.
As we want to investigate also the validity of our model under these conditions, 
we present a couple of natural extensions to this setup, more faithfully capturing the behavior of myosin~II motors inside these networks.
The original setup will be further referred to as ``load on the motor''.

\subsubsection{Load on the polymer}
\label{sec:load_on_polymer}
\begin{figure}[t]
\centering
\includegraphics[width=0.45\textwidth,height=!]{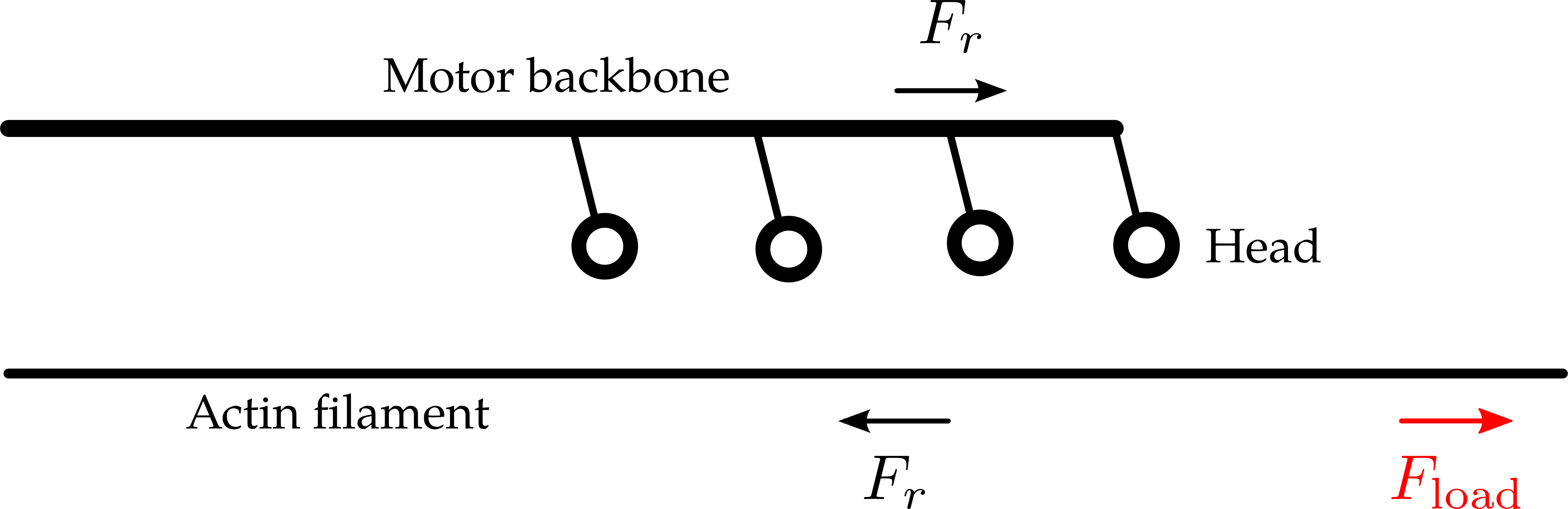}
\caption{
\label{fig:load_on_polymer setup}
In this illustration, a variation of the basic setup is shown.  
Now, the load $F_\text{load}$ is applied on actin polymer instead of the myosin motor. 
Also the orientation of the ratchet potential along the polymer is reversed.
} 
\end{figure}

The first alternative is very similar to the scenario of ``load on the motor'', 
but now with the external load applied to the polymer instead of the molecular motor. 
This corresponds to a situation where the motor is attached to the polymer, which itself is under the tension due to the action of other motors and complex interactions in a larger network.
As we  show later, in this setup the tension on the motor vanishes for large loads, which will be relevant for our discussion of the chemical efficiency. 
The basic scheme of this setup is depicted in Fig.~\ref{fig:load_on_polymer setup}.

As was advertised, the equations of motion are very similar to those of the ``load on the motor'', i.e. \eqref{eq:eom_M} and \eqref{eq:eom_A}.
\begin{align}
&\begin{multlined}[b][.42\textwidth]
\rmd x_\text{M} = 
- \frac{1}{\gamma_\text{M}} \nabla_\text{M} V_t(x_\text{M}(t) - x_\text{A}(t)) \; \rmd t 
\\ 
+ \sqrt{ \frac{2 k_B T}{\gamma_\text{M}} } \; \rmd W_\text{M} ,
\end{multlined}
\label{eq:eom_M_lp} \\
&\begin{multlined}[b][.42\textwidth]
\rmd x_\text{A} = 
\frac{1}{\gamma_\text{A}} \left[ F_\text{load} - \nabla_\text{A} V_t(x_\text{M}(t) - x_\text{A}(t)) \right] \; \rmd t 
\\
+ \sqrt{ \frac{2 k_B T}{\gamma_\text{A}} } \; \rmd W_\text{A} .
\end{multlined}
\label{eq:eom_A_lp}
\end{align}
In these equations, the argument of the ratchet potential is preserved, 
while the load $F_\text{load}$ is applied on the polymer and its direction is reversed with respect to the ``load on the motor'' setup. 
This ensures that the mean relative velocity of the motor with respect to the polymer is directly comparable in both setups. 
The equations of motion of this setup can be mapped into the ``load on the motor'' setup by interchanging the label of myosin and actin 
along with inverting the x-axis, 
i.e. $\gamma_\text{M}\leftrightarrow\gamma_\text{A}$ and $x_\text{M}\leftrightarrow - x_\text{A}$.

\subsubsection{Tug of war}
\label{sec:tug_of_war}
\begin{figure}[t]
\centering
\includegraphics[width=0.45\textwidth,height=!]{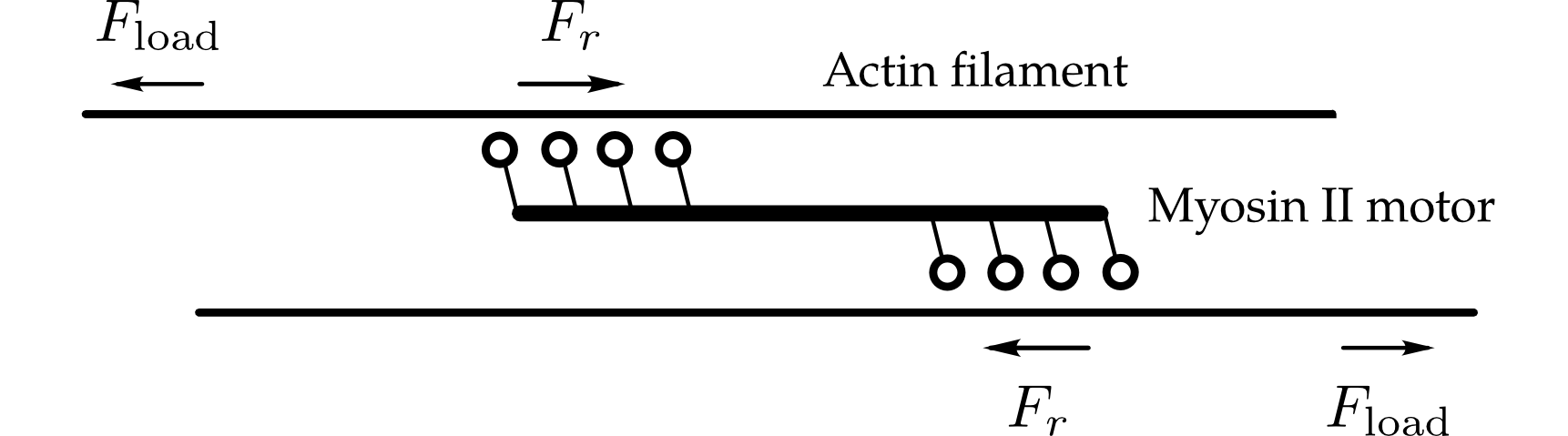}
\caption{
\label{fig:tug_F_illustration}
Setup of a motor interacting with two actin filaments that experience equal but opposite loads.
}
\end{figure}

A third setup that we investigate is a natural extension of the previous setup, but now the myosin motor is interacting with two, rather than one, 
anti-parallel actin filaments. 
The external load is again applied directly to the filaments instead of to the motor, as demonstrated in Fig.~\ref{fig:tug_F_illustration}. 
This corresponds to a situation where the motor is actively cross-linking two polymers, which are subjected to the tension caused by other motors and other external influences.  
Hence the equations of motion read
\begin{equation}
\begin{gathered}
\begin{multlined}[b][.45\textwidth]
\rmd x_\text{M} = 
- \frac{1}{\gamma_\text{M}} \nabla_\text{M} \bigl[ 
V_t(x_{\text{A}_1}(t) - x_\text{M}(t)) 
\\
+ V_t(x_\text{M}(t) - x_{\text{A}_2}(t)) 
\bigr] \; \rmd t 
+ \sqrt{ \frac{2 k_B T}{\gamma_\text{M}} } \; \rmd W_\text{M} ,
\end{multlined}\\
\begin{multlined}[b][.45\textwidth]
\rmd x_{\text{A}_1} = 
- \frac{1}{\gamma_\text{A}} \left[ F_\text{load} + \nabla_{\text{A}_1} V_t(x_{\text{A}_1}(t) - x_\text{M}(t)) \right] \; \rmd t 
\\
+ \sqrt{ \frac{2 k_B T}{\gamma_\text{A}} } \; \rmd W_{\text{A}_1} ,
\end{multlined}\\
\begin{multlined}[b][.45\textwidth]
\rmd x_{\text{A}_2} = 
\frac{1}{\gamma_\text{A}} \left[ F_\text{load} - \nabla_{\text{A}_2} V_t(x_\text{M}(t) - x_{\text{A}_2}(t)) \right] \; \rmd t 
\\
+ \sqrt{ \frac{2 k_B T}{\gamma_\text{A}} } \; \rmd W_{\text{A}_2} ,
\end{multlined}
\end{gathered}
\label{eq:eom_two_actins}
\end{equation}
where subscripts ${\text{A}_1}$ and ${\text{A}_2}$ respectively denote the first and second actin filament 
and where the anti-alignment of polymers is represented by reversing the sign in the argument of the ratchet potential~\eqref{eq:ratchet_interaction} for polymer $\text{A}_1$.

\subsubsection{Elastic environment}
\label{sec:environment}
\begin{figure}[t]
\centering
\includegraphics[width=0.45\textwidth,height=!]{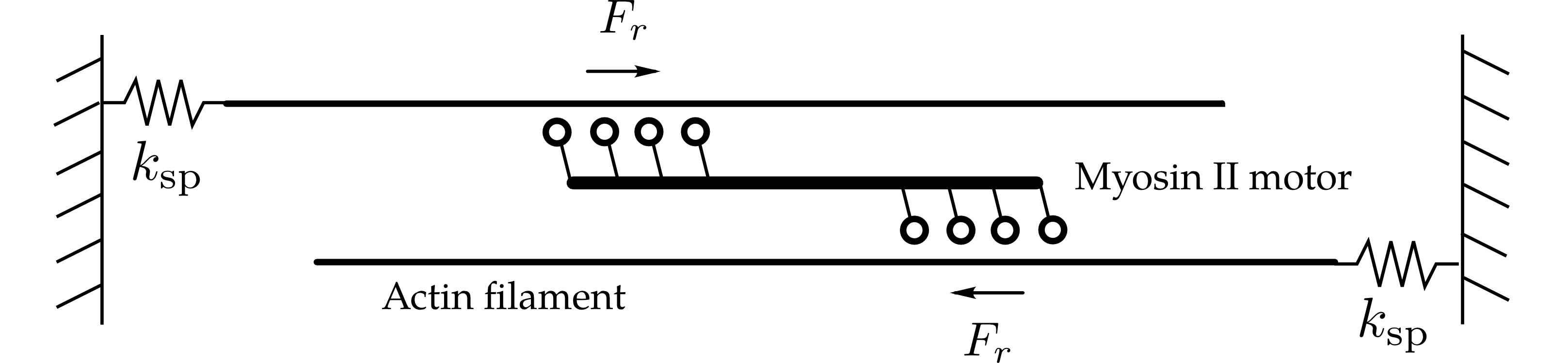}
\caption{
\label{fig:tug}
Illustration of a motor pulling on two actin filament that are each connect to a wall by a spring.
}
\end{figure}

As polymers inside the cytoskeleton are usually thoroughly interconnected rather than freely diffusing \cite{blanchoin2014actin,ennomani2016architecture},
we can opt for an alternative approach.
Rather than applying an external load we attach filaments to fixed points by a spring with constant stiffness $k_\text{sp}$, 
similarly to the approach discussed in \cite{Albert2014}. 
These represent points of interaction with the rest of the network while the spring stiffness $k_\text{sp}$ represents the elastic properties of the full network. 
The myosin motor is again interacting with two filaments that are arranged in an anti-parallel fashion. 
Hence, similar to the previous case, it no longer has a preferred direction of motion 
and it eventually reaches a steady state where its mean velocity will become zero. 
The observable of interest is now the mean force that the motor applies to the filaments, 
which is obtained via the extension of the springs attached to the filaments,
and which represent the active tension in the network produced by the single myosin minifilament. 
The corresponding setup is sketched in Fig. \ref{fig:tug}. 

The equations of motion are just a modification of \eqref{eq:eom_two_actins}, i.e.,   
\begin{equation}
\begin{gathered}
\begin{multlined}[b][.45\textwidth]
\rmd x_\text{M} = 
- \frac{1}{\gamma_\text{M}} \nabla_\text{M} \bigl[ 
V_t(x_{\text{A}_1}(t) - x_\text{M}(t)) 
\\
+ V_t(x_\text{M}(t) - x_{\text{A}_2}(t)) 
\bigr] \; \rmd t 
+ \sqrt{ \frac{2 k_B T}{\gamma_\text{M}} } \; \rmd W_\text{M} ,
\end{multlined}\\
\begin{multlined}[b][.45\textwidth]
\rmd x_{\text{A}_1} = 
- \frac{1}{\gamma_\text{A}} \bigl[ \nabla_{\text{A}_1} V_t( x_{\text{A}_1}(t) - x_\text{M}(t) ) 
\\
+ k_\text{sp} \left( x_{\text{A}_1}(t) - x_{\text{A}_1}(0) \right) \bigr] \; \rmd t 
+ \sqrt{ \frac{2 k_B T}{\gamma_\text{A}} } \; \rmd W_{\text{A}_1} ,
\end{multlined}\\
\begin{multlined}[b][.45\textwidth]
\rmd x_{\text{A}_2} = 
- \frac{1}{\gamma_\text{A}} \bigl[ \nabla_{\text{A}_2} V_t( x_\text{M}(t) - x_{\text{A}_2}(t) )
\\
+ k_\text{sp} \left( x_{\text{A}_2}(t) - x_{\text{A}_2}(0) \right) \bigr] \; \rmd t 
+ \sqrt{ \frac{2 k_B T}{\gamma_\text{A}} } \; \rmd W_{\text{A}_2} ,
\end{multlined}
\end{gathered}
\label{eq:eom_two_actins_no_load}
\end{equation}
where the load is replaced by the force caused by springs and where the reference point for the springs is chosen as their position at some initial time. 

\subsection{Numerical method}
We use an Euler-Maruyama integration scheme to stochastically evolve the systems from a randomized initial configuration \cite{kloeden1989survey,maruyama1955continuous}. 
Waiting times between chemical jumps are generated by a Gillespie algorithm \cite{rao2003stochastic}.
Most results are obtained after evolving the system up to a time of $1\;\mathrm{s}$ with time step $\Delta t = 0.01\;\mathrm{ns}$ and averaging over $100$ independent copies.    
An exception is the data in Fig.~\ref{fig:tug_k}, 
obtained after a time-average every $1\;\mathrm{ms}$ over $10\;\mathrm{s}$ with time step $\Delta t = 0.01\;\mathrm{ns}$ and ensemble average over $1000$ independent copies. Finally, to obtain the empirical distribution in Fig.~\ref{fig:pos_distr}, 
we sampled every $0.5\;\mathrm{ms}$ for a total time of $1\;\mathrm{s}$ with time step $\Delta t = 0.001\;\mathrm{ns}$ and from $100$ independent simulations.  The data were converted into a histogram with a bin width of $0.1\;\mathrm{nm}$.

\section{Model verification}
\label{sec:model_verification}
As our model deviates from the traditionally used ratchet model, as discussed in section~\ref{sec:ratchet},  
we need to verify that we can reproduce the behavior reported in the literature for flashing ratchet models and myosin II motors. 
The mean velocity, force--velocity relation and the stall force are numerically obtained for our model.

\subsection{Mean velocity}
\label{sec:velocity}
We start with a study of the mean velocity of a four-headed motor in the ``load on the motor'' setup,
\begin{equation}
\langle v \rangle 
= \frac{ \left\langle \rmd x_\text{M} \right\rangle }{ \rmd t } 
- \frac{ \left\langle \rmd x_\text{A} \right\rangle }{ \rmd t } ,
\label{eq:mean_velocity} 
\end{equation}
for different values of the scaling factor $c$ and without any external load $F_\text{load} = 0$.
Since the scaling factor $c$ modifies the interaction strength between the motor and the polymer, it effectively models the electrostatic screening of the charges in the ionic solution inside a cell \cite{barterls1993myosin}.
Note that in this particular case, where there is no external load, setups ``load on the motor'' and ``load on the polymer'' coincide. 

\begin{figure}[t]
\centering
\includegraphics[width=0.45\textwidth,height=!]{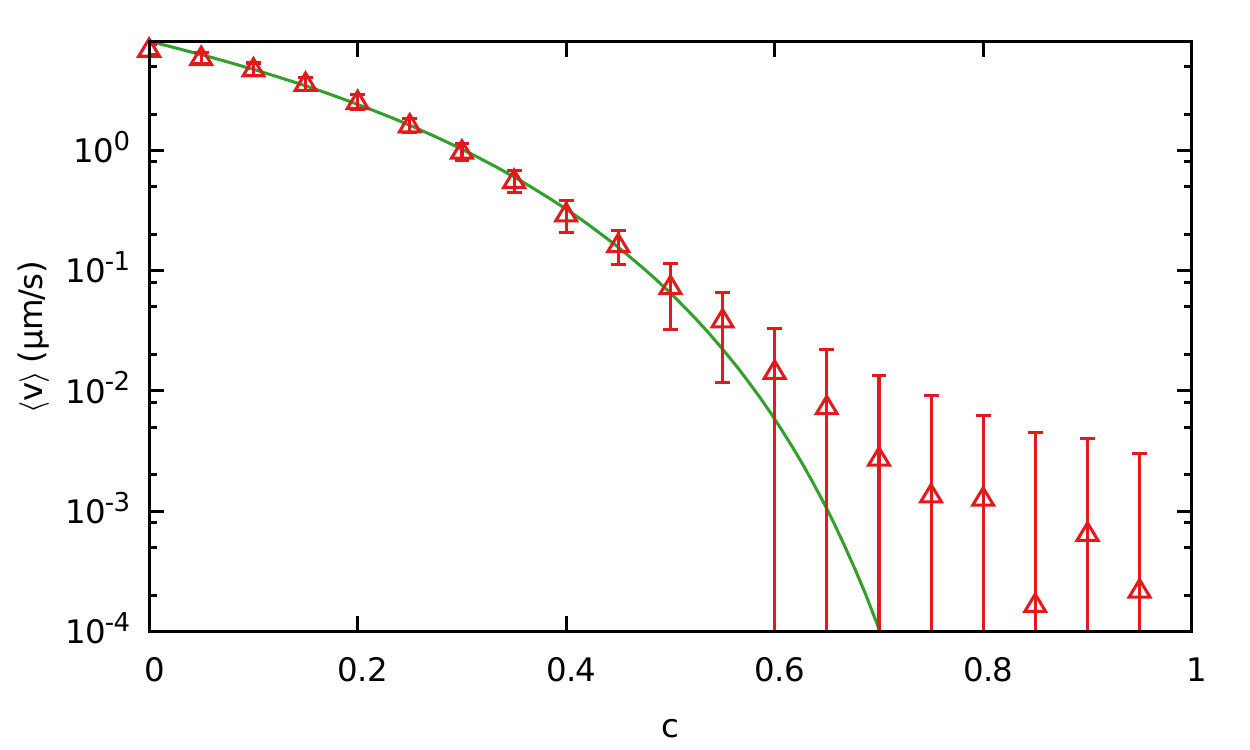}
\caption{
\label{fig:c_v}
Mean velocity $\langle v \rangle$ of the motor relative to the actin filament for $c \in [0,1]$.
The mean velocity rapidly vanishes towards equilibrium, $c \to 1$.  
The solid green line corresponds to the fit $ \left\langle v \right\rangle \propto \exp ( - \frac{\alpha}{1-c} ) $.
} 
\end{figure}

From Fig.~\ref{fig:c_v}
we conclude that the mean velocity has a maximum at $c=0$ and goes to zero when $c$ approaches $1$. 
This result is expected since $c=1$ corresponds to equilibrium, where there is no difference between the two internal states, cf.~\eqref{eq:transition}.
On the other hand, for $c=0$ we have the biggest discrepancy between the two internal states of each head.

\begin{figure}[t]
\centering
\includegraphics[width=0.45\textwidth,height=!]{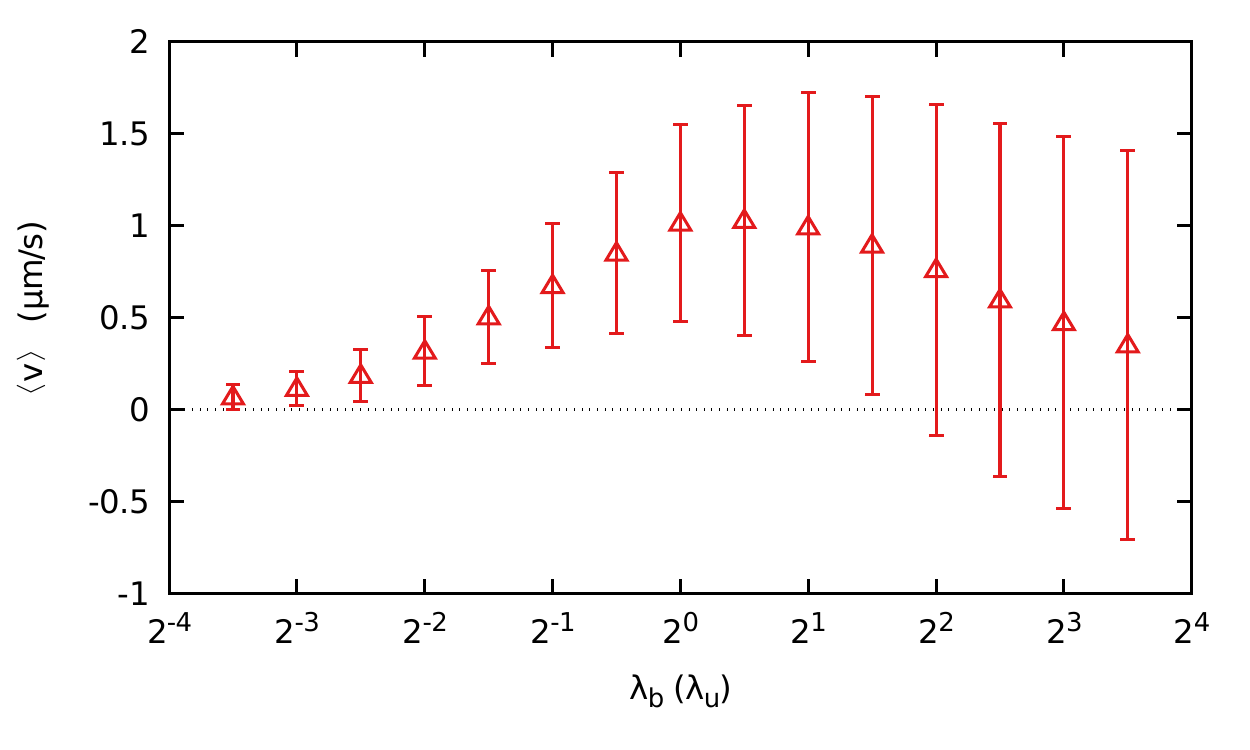}
\caption{
\label{fig:v_k} 
Mean velocity of the motor relative to the actin filament $\langle v \rangle$ for varying $\lambda_\text{b}$.
The velocity decreases to $0$ for large discrepancies between binding and unbinding rate and reaches maximum when they are comparable.
Note that the ATP binding rate $\lambda_\text{b}$ is shown in units of $\lambda_\text{u}$ on a logarithmic scale.
}
\end{figure}

Secondly, the mean velocity~\eqref{eq:mean_velocity} also depends on the concentration of ATP through the binding rate $\lambda_\text{b}$. 
In this setup, we fix the ATP unbinding rate $\lambda_\text{u}$ and scaling factor $c$, cf. table~\ref{tab:parameters}, and vary the ATP binding rate $\lambda_\text{b}$. 
The resulting dependency is shown in Fig.~\ref{fig:v_k}. 
We observe that the maximal velocity is reached for rates close to $\lambda_\text{b} = \lambda_\text{u}$. 
Deviating far from this ratio means that one of the states become dominant and thus the system effectively approaches equilibrium.  
The main difference between $\lambda_\text{b}/\lambda_\text{u}$ being low or high is the mean amplitude of the interaction with actin. 
In the ATP bound state the interaction is weaker, 
which is reflected in the fact that the system is trapped for shorter times close to the local minima and diffuses more. 
Hence, the variance of the relative velocity is higher when compared to the ATP unbound state.

\subsection{Force--velocity relation}
\label{sec:force-velocity}
In the previous section, we reported on the mean relative velocity for a motor in the ``load on the motor'' setup without any external load. 
Here we focus on the dependence of the mean relative velocity \eqref{eq:mean_velocity} on an external load force $F_\text{load}$ applied on the motor anti-parallel to the preferred direction, 
cf. Fig.~\ref{fig:ratchet_setup}.

\begin{figure}[t]
\centering
\includegraphics[width=0.45\textwidth,height=!]{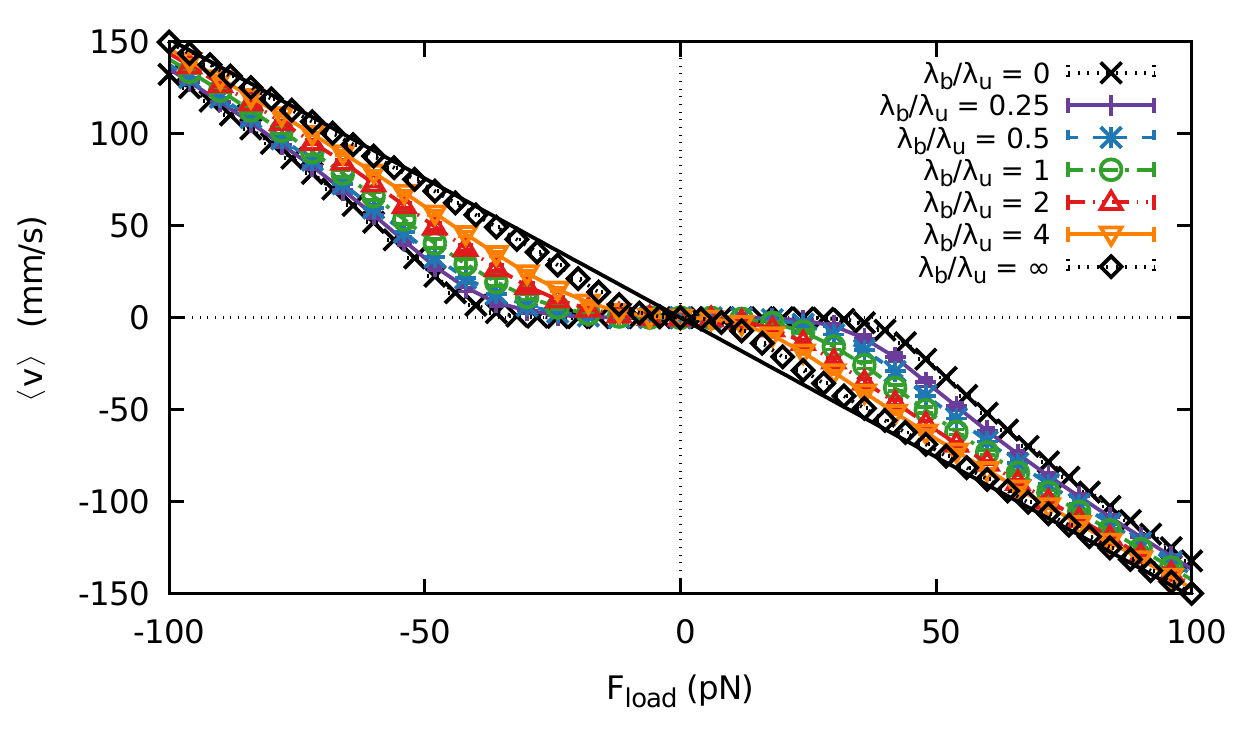}
\caption{
\label{fig:F_v} 
Mean velocity $\langle v \rangle$ of the motor as a function of the load force $F_\text{load}$ for $c=0.3$ and for various ATP binding rate $\lambda_\text{b}$.
The solid black line corresponds to a situation where there will be no interaction with the actin filament, i.e. $\langle v \rangle = - F_\text{load} / \gamma_\text{M}$. 
}
\end{figure}

Fig.~\ref{fig:F_v} shows a typical force -- mean velocity relations for various ATP binding rates $\lambda_\text{u}$, which represents the dependency on ATP concentration.
For small load forces, the interaction with an actin polymer provides an additional friction force countering the load force, 
which results in small response of the relative velocity $\langle v \rangle$.
For larger forces the system looses its resilience and the contribution from the load will dominate. 
The larger the load, the more the velocity will approach the velocity of a non-interacting motor, 
\[
\langle v \rangle \asymp - \frac{1}{\gamma_\text{M}} F_\text{load}.
\]
We prove this dependency in appendix~\ref{sec:ind_velo}. 
Curves for higher ATP binding rates are closer to the curve for the completely unbound motor.
As the motor is more loosely bound to the polymer, it is more submissive to the external load.
Such a force-velocity relation was already presented in 
 \cite{reimann2002brownian} for a flashing ratchet, representing a motor with a single head.

\begin{figure}[t]
\centering
\includegraphics[width=0.45\textwidth,height=!]{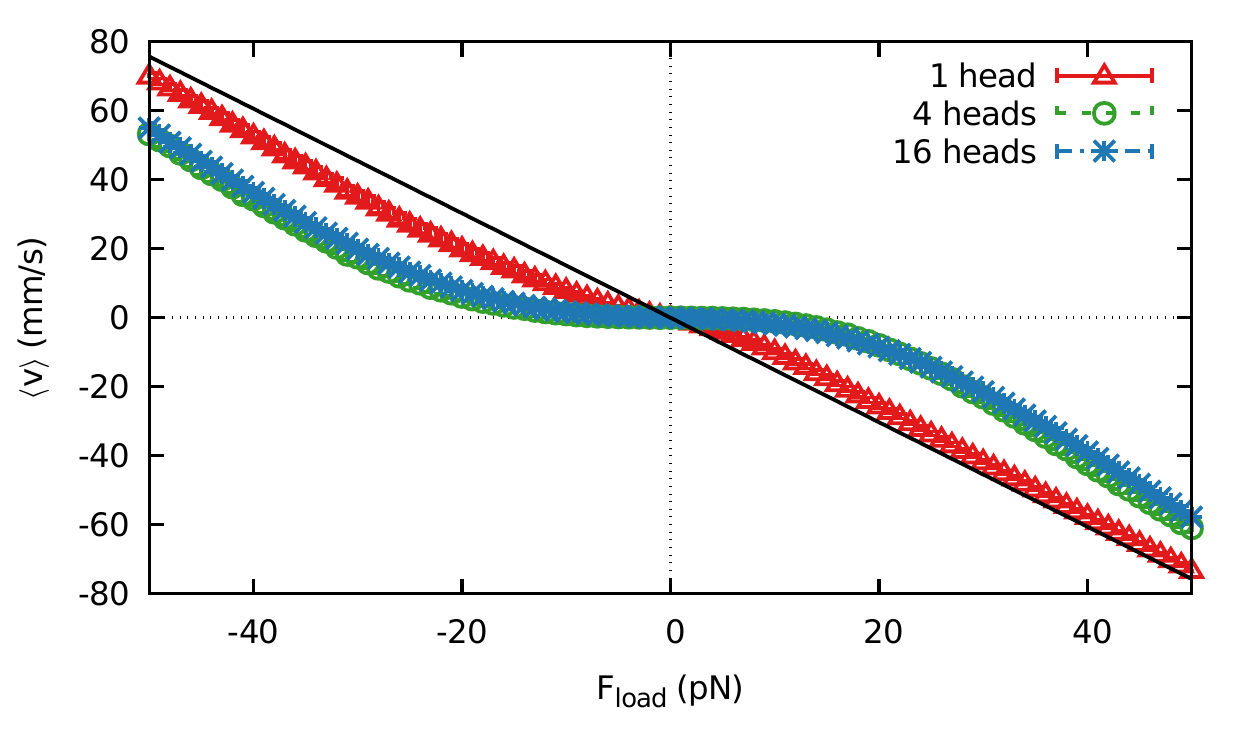}
\caption{
\label{fig:F_v_heads} 
Mean velocity $\langle v \rangle$ of the motor as a function of the load force $F_\text{load}$ for $c=0.3$ and for motors with different number of heads $N$.
The solid black line corresponds to a non-interacting motor, i.e. $\langle v \rangle = - F_\text{load} / \gamma_\text{M}$. 
}
\end{figure}

A similar plot is shown in Fig. \ref{fig:F_v_heads}, where instead of varying ATP binding rate $\lambda_\text{b}$ we look at motors with different numbers of heads $N$. 
Motors with only one head are very susceptible to external loads and the force from the ratchet offers little resistance. 
On the other hand, the motors with four and sixteen heads have almost identical force--velocity relations. 
However, it would be naive to consider these motors identical as they behave very differently regarding their energetics,  
which we will discuss in section~\ref{sec:efficiency}.

A comparison of the force--velocity relation for alternative setups with the ``load on the motor'' setup is provided in appendix~\ref{sec:force_velocity_other}.

\subsection{Stall force}
\begin{figure}[t]
\centering
\includegraphics[width=0.45\textwidth,height=!]{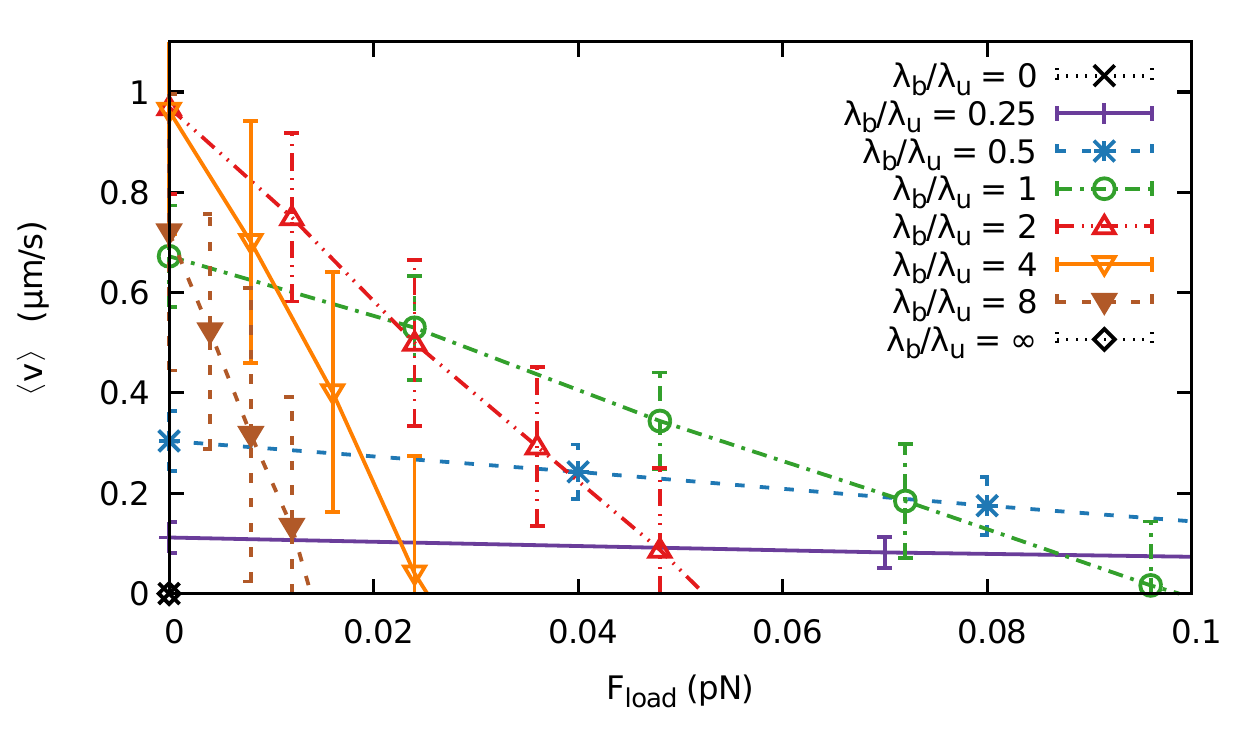}
\caption{
\label{fig:F_v_zoom} 
Mean velocity of the motor $\langle v \rangle$ with a small load force $F_\text{load}$ applied in the direction opposing the movement for different ATP binding rates $\lambda_\text{b}$.
Intersections with the $x$ axis correspond to stall forces $F_\text{stall}$ for given detach rates. 
With increasing $\lambda_\text{b}/\lambda_\text{u}$  the slope increases and the full curve shifts in the low force region, 
causing a persistent decrease in stall force. 
Note that both $\lambda_\text{b} = 0 $ and $\lambda_\text{b} = \infty$ correspond to equilibrium where the mean velocity for zero load and the stall force are zero. 
}
\end{figure}

\begin{figure}[t]
\centering
\includegraphics[width=0.45\textwidth,height=!]{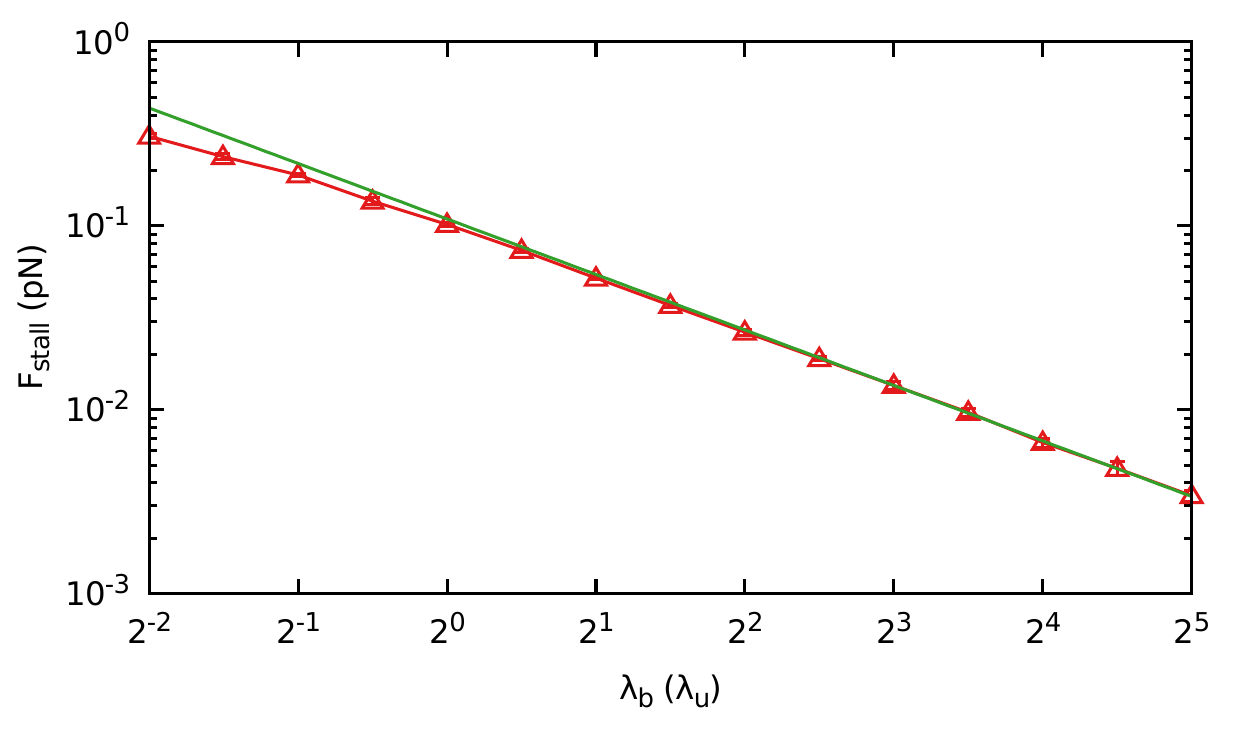}
\caption{
\label{fig:k_Fstall} 
Stall force $F_\text{stall}$ for different ATP binding rates $\lambda_\text{b}$.
The power-law decay $F_\text{stall} \propto \lambda_\text{b}^{-\alpha}$ with exponent $\alpha = 1.002 \pm 0.014$,
which is in good agreement with the dependence $F_\text{stall} \propto 1 / \lambda_\text{b}$ in the simple ratchet model of~\cite{astumian1996mechanochemical}. 
}
\end{figure}

From Fig.~\ref{fig:F_v} it is apparent that when applying a force against the preferred direction of the motor, 
the motor first slows down until it stops before it reverses its direction. 
The force magnitude that is needed to stop the motor's movement relative to the actin filament is called the stall force $F_\text{stall}$,  
\[
\left\langle v ( F_\text{load} \equiv F_\text{stall} ) \right\rangle = 0 .
\]
Fig.~\ref{fig:F_v_zoom} shows a zoomed in version of Fig.~\ref{fig:F_v} for small loads, opposite to the motor's preferred direction of motion. 
The intersections with the horizontal axis correspond to stall forces,  
while intersections of the curves with the vertical axis correspond to the mean velocities for system without any external load, compare Fig.~\ref{fig:v_k}. 
We see that for small loads the mean velocity decreases linearly with increasing load. 
Moreover, the slope of the force -- mean velocity relation increases with the binding rate $\lambda_\text{b}$,
that is responsible for the monotonous decrease of the stall force $F_\text{stall}$ with respect to the ATP binding rate $\lambda_\text{b}$, see Fig.~\ref{fig:k_Fstall}.
When the ATP binding rate increases, so does the fraction of time in which the motor is weakly interacting with the actin filament. 
Hence, on average it takes a smaller load to halt a motor since the amplitude of its confining potential is typically lower.
Ultimately, when the system is found only in the ATP unbound state the relative mean velocity is zero, $\langle v \rangle = 0$, as the system reached an equilibrium 
and thus the stall force reaches zero as well, $F_\text{stall} \to 0$.

\section{Tension}
\label{sec:tension}
While the force--velocity relation captures the non-linear nature of the motor's transport, 
it alone does not truly capture its behavior inside polymer networks.
There, the myosin II minifilament typically crosslinks two polymers being not necessarily displaced on average.
This is demonstrated in setups ``tug of war'' and ``elastic environment''. 
However, as we show in this section, the motor exerts an active tension in the system.
This tension is further propagated through actin polymers over the full network up to the cell membrane and largely contributes to the overall stiffness of the cell \cite{chugh2017actin,ma2012nonmuscle,monier2010actomyosin}.

\subsection{Ratchet force} 
In order to define a tension generated by the motor, we first need to evaluate all forces applied on the motor in the steady state. 
In the equation of motion for the ``load on the motor'' setup \eqref{eq:eom_M} we identify the force generated by the ratchet potential as
\begin{equation}
F_\text{r} = - \nabla_\text{M} V_t( x_\text{M} - x_\text{A} ) = \nabla_\text{A} V_t(x_\text{M} - x_\text{A} ) . 
\label{eq:ratchet_force}
\end{equation}
Without the ratchet interaction the motion of the myosin, relative to the actin filament, would satisfy $ \gamma_\text{M}\langle v \rangle = -F_\text{load}$. 
The discrepancy in the relation for the free motor, up to a pre-factor, equals the mean force exerted via the flashing ratchet $\langle F_\text{r}\rangle$, 
see also equation~\eqref{eq:mean_velocity_Fr_Fl} in appendix~\ref{sec:ind_velo}.
By taking the mean value of equation~\eqref{eq:eom_M}, 
we can extract the mean ratchet force in terms of external load and mean relative velocity as 
\begin{equation}
\langle F_\text{r} \rangle = \frac{\gamma_\text{A}\gamma_\text{M}}{\gamma_\text{A} + \gamma_\text{M} } \left(\langle v \rangle + \frac{F_\text{load}}{\gamma_\text{M}}\right) .
\label{eq:mean_ratchet_force}
\end{equation}
Note that this is the same, up to a scaling factor, as the distance between the force-velocity curves and the ratchet-free curve in Fig.~\ref{fig:F_v}. 
Also note that, using the re-scaled parameters \eqref{eq:rescale_parameters} introduced in appendix~\ref{sec:perturb}, 
we can write the mean ratchet force as $\langle F_\text{r} \rangle = \gamma^* \langle v \rangle - F^*$.
We will discuss the ratchet force in other setups later on in the context of the tension for those setups.
The typical behavior of the mean ratchet force in the ``load on the polymer'' setup is discussed in appendix~\ref{sec:mean_ratchet_force}.

\subsection{Tension in the motor}
\label{sec:tension_motor}
Since both the load and ratchet force are directly applied on the respective ends of the motor in the ``load on the motor'' setup, 
the tension applied to the motor $\tau_\text{M}$ is proportional to their difference.
Here we consider the contribution from the thermal force and from the drag negligible in the steady state as they are typically applied at the center of the mass of the motor. 
We choose the sign of the tension in such a way 
that a positive value corresponds to a pair of forces applied on each end of the motor in the direction away from the motor,
i.e.,  
\begin{align}
\left\langle \tau_\text{M} \right\rangle 
&= \langle F_\text{r} \rangle + F_\text{load} \nonumber \\
&= \frac{\gamma_\text{A}\gamma_\text{M}}{\gamma_\text{A}+\gamma_\text{M}}\langle v\rangle 
+ \left(1 + \frac{\gamma_\text{A}}{\gamma_\text{A}+\gamma_\text{M}} \right) F_\text{load},
\label{eq:tension}
\end{align}
where we will consider the contribution from the ratchet potential as an active component of the tension, i.e. $\tau_\text{M}^\text{(A)} \equiv F_\text{r}$, in this particular setup. 
Note, that we omit the normalization by the length of the motor as it is just a division by a constant due to the infinite rigidity of the motor in our model.   
In a similar fashion we can define the tension on the polymer as
\begin{align}
\left\langle \tau_\text{A} \right\rangle 
&= - \langle F_\text{r} \rangle \nonumber \\
&= - \frac{\gamma_\text{A}\gamma_\text{M}}{\gamma_\text{A}+\gamma_\text{M}}\langle v\rangle 
- \frac{\gamma_\text{A}}{\gamma_\text{A}+\gamma_\text{M}} F_\text{load}.
\label{eq:tension_on_polymer}
\end{align}

\begin{figure}
\centering
\includegraphics[width=.45\textwidth,height=!]{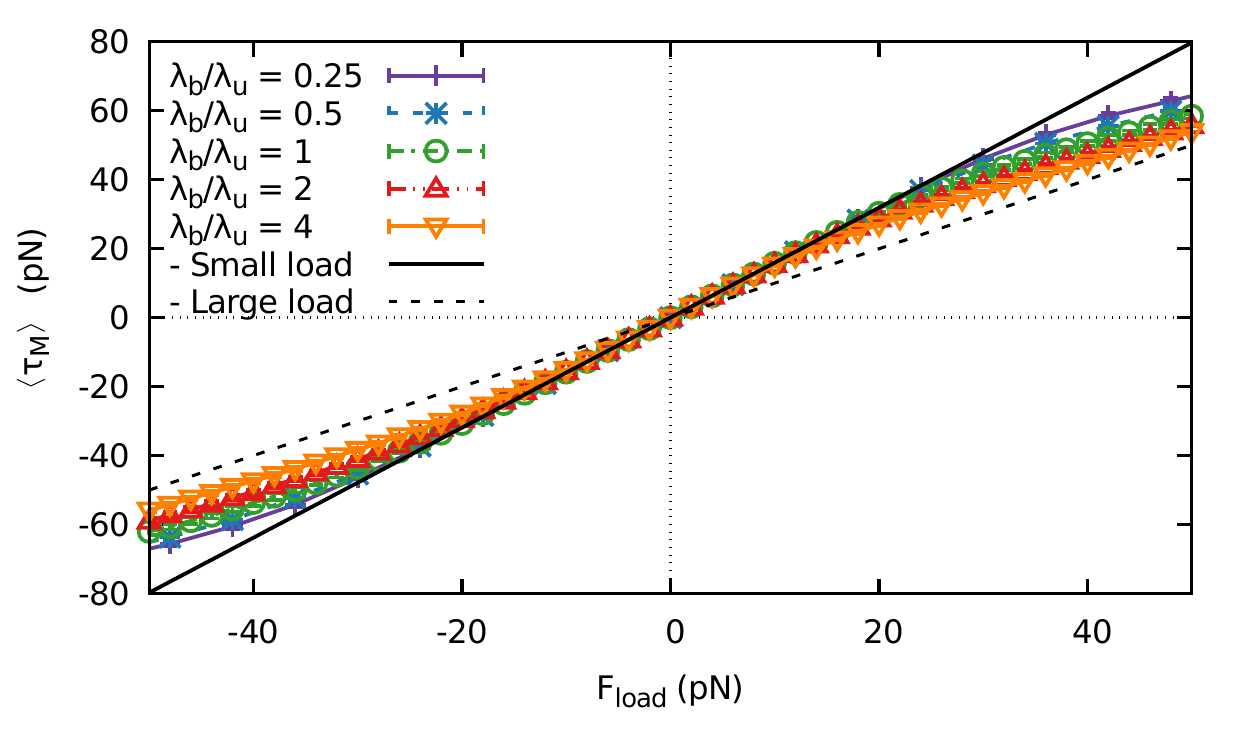}
\caption{\label{fig:tension}
The dependence of the tension~\eqref{eq:tension} $\langle \tau_\text{M}\rangle$ on the load force $F_\text{load}$ for different ATP binding rates $\lambda_\text{b}$.
We can see that with increasing ATP binding rate there is a faster departure from the small load regime towards the large load asymptotic behavior. 
}
\end{figure}

\begin{figure}
\centering
\includegraphics[width=.45\textwidth,height=!]{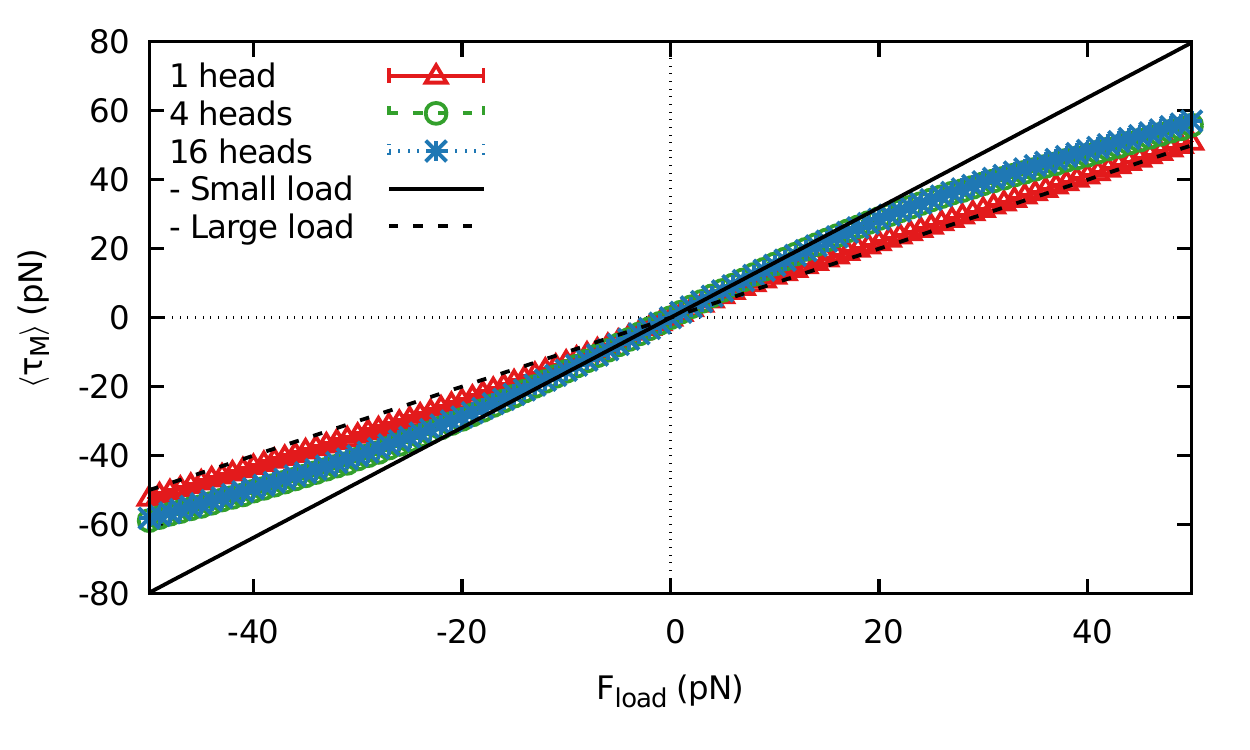}
\caption{\label{fig:tension_heads}
The dependence of the tension~\eqref{eq:tension} $\langle \tau_\text{M}\rangle$ on the load force $F_\text{load}$ for different number of heads. 
We observe a similar behavior as in Fig.~\ref{fig:tension} for the motor with four and sixteen heads which seems to be indistinguishable, 
while the motor with a single head departs very rapidly from the low force regime. 
}
\end{figure}

We plot the tension on the motor as a function of the load $F_\text{load}$ in Fig.~\ref{fig:tension} for various ATP binding rates 
and in Fig.~\ref{fig:tension_heads} as a function of the amount of heads on a motor $N$. 
We observe that for small loads the tension is linear in load, in good agreement with the observation of the plateau in the force velocity relation;
see Fig.~\ref{fig:F_v}.
In that regime, the velocity is small compared to the initial load as most of the load is countered by the ratchet interaction. 
Hence only the second term in~\eqref{eq:tension} contributes to the tension, 
\begin{equation}
\left\langle \tau_\text{M} \right\rangle
\approx
\left(1 + \frac{\gamma_\text{A}}{\gamma_\text{A}+\gamma_\text{M}} \right) F_\text{load}.
\label{eq:tension_low_load} 
\end{equation}

For large forces, the motor slips over the potential and thus the mean ratchet force $\langle F_\text{r}\rangle$ approaches zero,
as can be seen in Figs.~\ref{fig:ratchet_force} and~\ref{fig:ratchet_force_decay}.
Hence, the large force asymptotic is given solely by the load
\begin{equation}
\left\langle \tau_\text{M} \right\rangle
\asymp F_\text{load} 
\qquad \text{for} \qquad F_\text{load} \to \pm \infty. 
\label{eq:tension_large_load}
\end{equation}
That is further supported by calculations of the mean velocity for a motor with a single head, provided in appendix~\ref{sec:perturb}, cf. equation~\eqref{eq:mean_velocity_large}.

\subsubsection{Load on the polymer}
A similar analysis can be made for alternative setups presented in Section~\ref{sec:other_setups}. 
In the case where the load is applied directly to the polymer instead of the motor, cf. the setup ``load on the polymer,'' 
the motor only experiences a force from the ratchet interaction with the polymer and not from a load. 
Therefore, the tension on the motor is given only by the mean ratchet force,  
cf. equation~\eqref{eq:mean_ratchet_force},  
where the load must be divided by $\gamma_\text{A}$ since the load is exerted on the actin filament instead of the motor,
i.e.
\begin{equation}
\left\langle \tau_\text{M} \right\rangle 
= \langle F_\text{r} \rangle 
= \frac{\gamma_\text{A}\gamma_\text{M}}{\gamma_\text{A} + \gamma_\text{M} } \left(\langle v \rangle + \frac{F_\text{load}}{\gamma_\text{A}}\right) . 
\label{eq:tension_load_on_polymer}
\end{equation}
Following the same argumentation as in the ``load on the motor'' setup, for small loads, the tension is given by
\begin{equation}
\left\langle \tau_\text{M} \right\rangle
\approx \frac{\gamma_\text{M}}{\gamma_\text{A}+\gamma_\text{M}} F_\text{load}.
\label{eq:tension_low_load_on_polymer}
\end{equation}
This is similar to~\eqref{eq:tension_low_load}, where the motor experiences the load directly, up to an exchange of friction coefficients of the motor and actin filament. 

Furthermore, since the load is not directly applied to the motor in this setup, the overall term $F_\text{load}$ is not present. 
Therefore, the tension vanishes for large load in the same way that the relative velocity approaches $-F_\text{load} / \gamma_\text{A}$; see appendix~\ref{sec:mean_ratchet_force},
\begin{equation}
\langle \tau_\text{M} \rangle 
\propto
F_\text{load}^{-1}
\qquad\text{for}\qquad
F_\text{load} \to \infty 
.
\label{eq:tension_load_on_polymer_large}
\end{equation}

\subsubsection{Tug of war}
In the case of Section~\ref{sec:tug_of_war}, the motor is aligned anti-parallel between two actin polymers. 

We again start from the mean velocities for all components of the system in this setup \eqref{eq:eom_two_actins},
\begin{align*}
\langle v_{\text{A}_1} \rangle &= - \frac{1}{\gamma_\text{A} } \left( \langle F_{\text{r}_1} \rangle + F_\text{load} \right) , \\
\langle v_\text{M} \rangle &= \frac{1}{\gamma_\text{M} } \left( \langle F_{\text{r}_2} \rangle + \langle F_{\text{r}_1} \rangle \right) , \\
\langle v_{\text{A}_2} \rangle &= \frac{1}{ \gamma_\text{A} } \left( F_\text{load} - \langle F_{\text{r}_2} \rangle \right) ,
\end{align*}
where ${\text{A}_1}$ and ${\text{A}_2}$ denote the two different actin filaments 
and $F_{\text{r}_1}$ and $F_{\text{r}_2}$ are the ratchet forces~\eqref{eq:ratchet_force} between the motor and respectively filament ${\text{A}_1}$ and ${\text{A}_2}$,
\begin{align*}
F_{\text{r}_1} &= - \nabla_\text{M} V( x_{\text{A}_1} - x_\text{M} ) , 
\\
F_{\text{r}_2} &= - \nabla_\text{M} V( x_\text{M} - x_{\text{A}_2} ) .
\end{align*} 

\begin{figure}[t]
\centering
\includegraphics[width=0.45\textwidth,height=!]{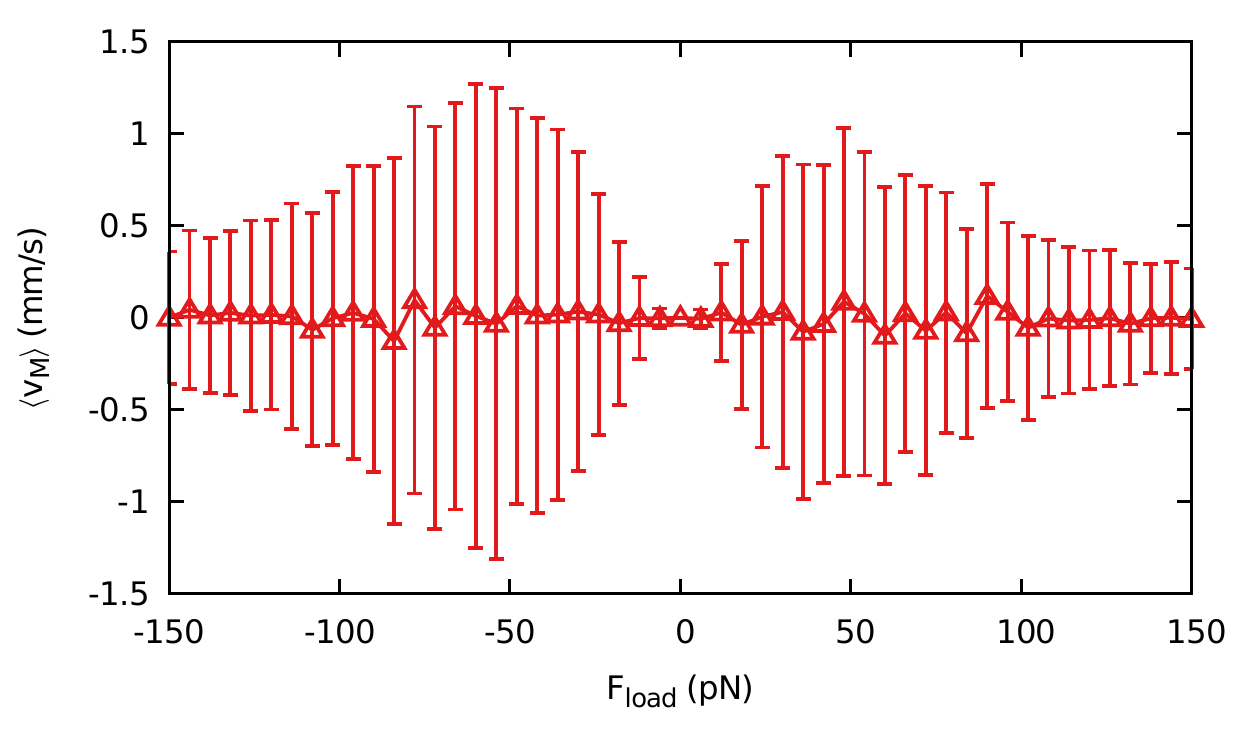}
\caption{\label{fig:tug_F_motor}
Mean velocity of the frustrated motor  between the two filaments under load. 
The variance of the motor's velocity grows as the load increases. 
For even larger loads the motor and filaments start slipping and the variance decreases again.
}
\end{figure}

Note, that due to the symmetry of the setup in the steady state the mean velocity of the myosin is inherently zero,
\begin{multline*}
\left\langle v_\text{M} \right\rangle 
= - \frac{1}{\gamma_\text{M}} \bigl[ 
\left\langle \nabla_\text{M} V_t( x_{\text{A}_1}(t) - x_\text{M}(t) ) \right\rangle
\\ 
+ \left\langle \nabla_\text{M} V_t( x_\text{M}(t) - x_{\text{A}_2}(t) ) \right\rangle 
\bigr] 
\\
= - \frac{\gamma_\text{A}}{\gamma_\text{M}} \left[ 
\left\langle v_{\text{A}_1} \right\rangle 
+ \left\langle v_{\text{A}_2} \right\rangle 
\right]
\equiv 0 , 
\end{multline*}
which is demonstrated in Fig.~\ref{fig:tug_F_motor}.

\begin{figure}[t]
\centering
\includegraphics[width=0.45\textwidth,height=!]{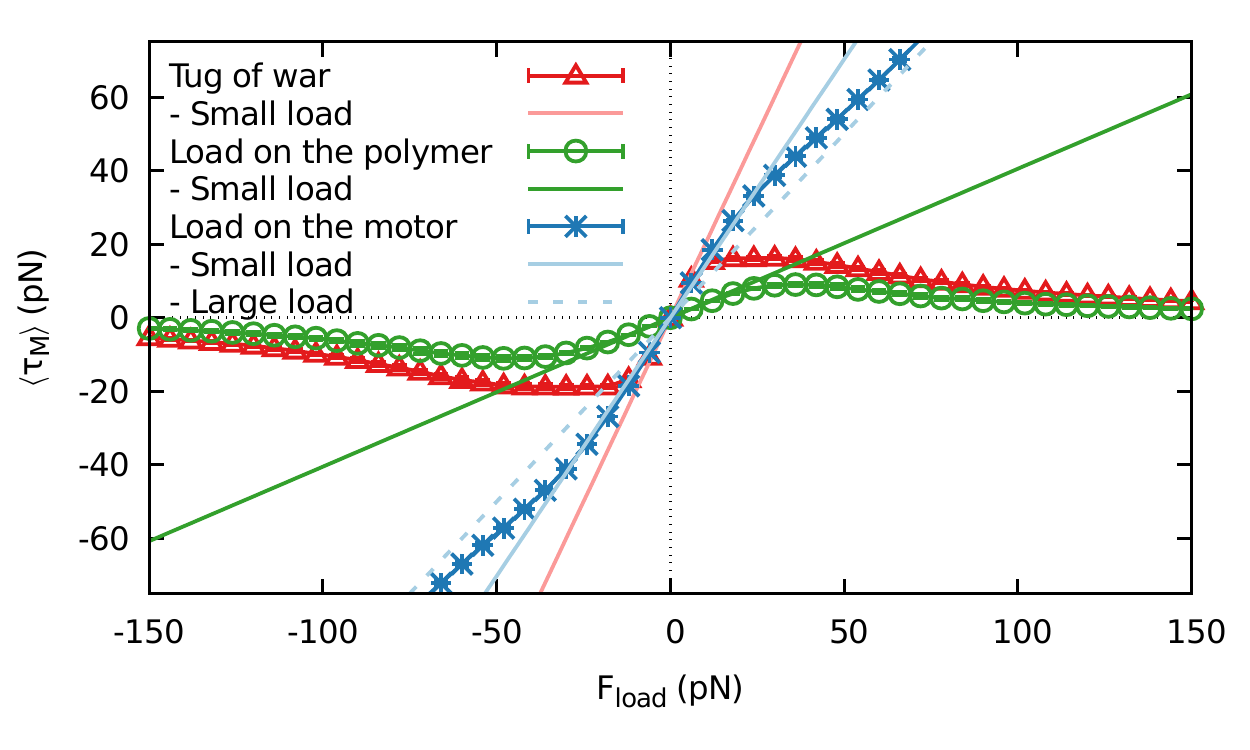}
\caption{
\label{fig:tug_F}
Tension in the motor $\langle \tau_\text{M}\rangle$, comparison between setup \ref{sec:tug_of_war} with two loaded filaments, setup \ref{sec:load_on_polymer} with one loaded filament and the basic setup with the load on the motor. 
The tension for small loads follows the same trend for the basic setup and ``tug of war'', whereas for large loads the tensions agree for the setups without a load on the motor. 
That is because, only in the case when the load is applied directly on the motor, it remains under tension even for large loads.
}
\end{figure}

Although the load is not applied to the motor directly, 
the motor is subjected to a tension due to interaction with both actin filaments. 
This tension is again given by the difference in the forces applied to the respective ends of the motor,
\begin{equation}
\langle \tau_\text{M} \rangle
= \langle F_{\text{r}_2}\rangle - \langle F_{\text{r}_1}\rangle 
= 2 F_\text{load} - \gamma_\text{A} \left(\langle v_{\text{A}_2}\rangle - \langle v_{\text{A}_1}\rangle \right) .
\label{eq:tension_tug}
\end{equation}
The tension in the motor for the different setups is shown in Fig.~\ref{fig:tug_F}, 
compared to the tension in the setups with only one actin filament.

For small loads, ratchet forces balance the load applied on both filaments.  The tension is thus approximately proportional to twice the load 
\[
\langle \tau_\text{M} \rangle \approx 2 F_\text{load} .
\] 
This behavior is similar to the tension in the previously discussed setups  ``load on the motor'' and ``load on the polymer'' setup; cf. equations~\eqref{eq:tension_low_load} and~\eqref{eq:tension_load_on_polymer}, 
where the proportionality constants are $ 1 + \gamma_\text{A} / (\gamma_\text{A} + \gamma_\text{M} )$ and $\gamma_\text{M} / (\gamma_\text{A} + \gamma_\text{M} )$ respectively. 
Note that the proportionality constant for the ``load on the motor'' setup, which lies in the closed interval $[1,2]$, 
is always bigger than or equal to the one in the ``load on the polymer'', which is in the interval $[0,1]$, and smaller than or equal to the one in the ``tug of war'' setup. 
In the ``load on the polymer'' setup, the tension in the motor does not build up as fast with increasing load because the motor experiences a ratchet force only from one actin filament.
The initial tension there is less than half of the tension in the setup with two loaded filaments.
This suggests that the ratchet force, per motor head, is better transmitted to the motor in the ``tug of war'' setup.
That is confirmed by Fig.~\ref{fig:tug_F} and represents our first result.

As described earlier, the motor starts to slip when the load is increased.
Consequently, the mean ratchet force decreases, cf. Fig.~\ref{fig:ratchet_force} and~\ref{fig:ratchet_force_decay}, and the increase in tension diminishes, 
as shown in Fig.~\ref{fig:tug_F}.
This behavior underlines the main difference between the ``tug of war'' and ``load on the polymer'' setups and the ``load on the motor'' setup, 
where the load is applied directly to the motor. 
Furthermore, as the ratchet force vanishes for large load, a residual tension from the load remains.
However, in the ``tug of war'' and ``load on the polymer'' setup, this residual tension is not transmitted further to the motor due to the vanishing ratchet interaction.

Finally, the variance of the motor velocity, cf. Fig.~\ref{fig:tug_F_motor}, seems to behave similarly to the tension on the motor in Fig.~\ref{fig:tug_F}. 
While the mean velocity stays close to zero, the variance increases as the tension builds up in the motor. 
As the motor begins to slip over the filaments, the variance appears to decrease again.
The asymmetry with respect to reversal of the load comes again from the inherent asymmetry of the ratchet potential.

\subsection{Mechanosensing}
\label{sec:mechanosensing}
\begin{figure}[t]
\centering
\includegraphics[width=0.45\textwidth,height=!]{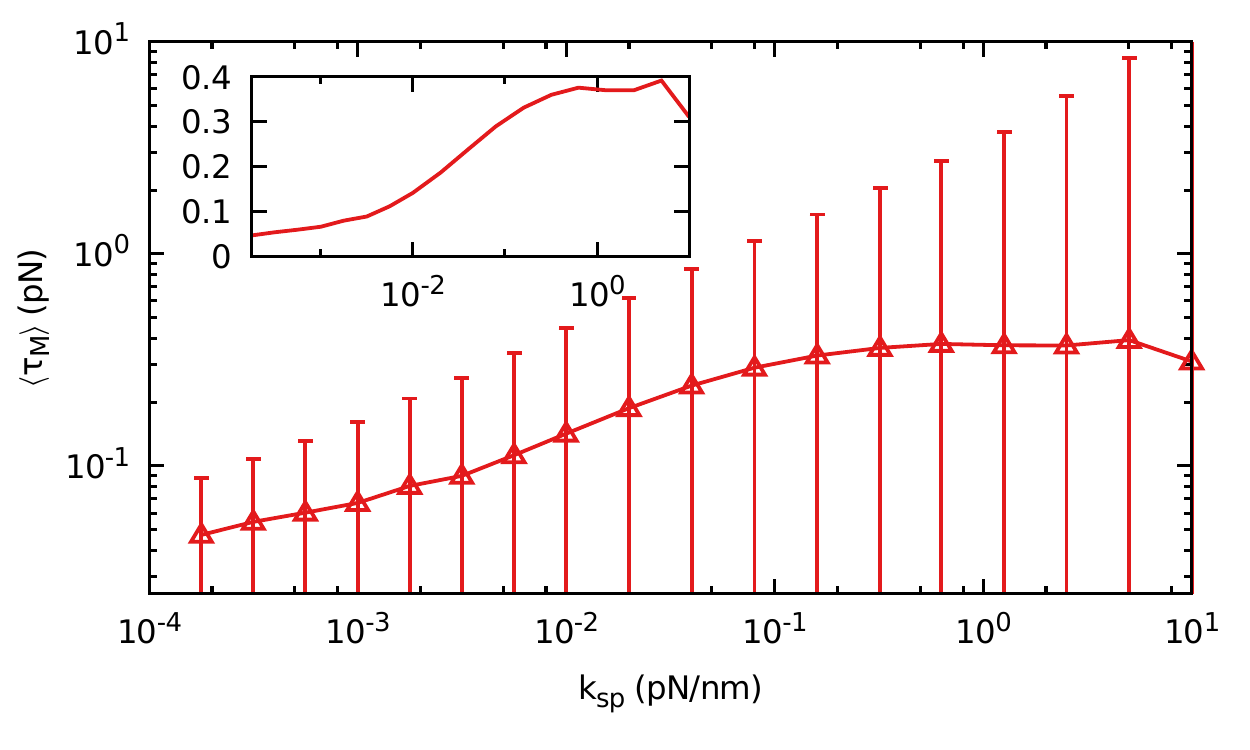}
\caption{
\label{fig:tug_k}
Mean tension $\langle \tau_\text{M} \rangle$ on the motor for varying spring stiffness $k_\text{sp}$.
Lower stiffness enables the polymers to find a local minimum of the ratchet potential more easily, 
thus the exerted force is lower as the contribution of ratchet potential is minimized. 
On the other hand, high stiffness leads to saturation of the exerted force as the main contribution is given by the ratchet potential. 
The insert depicts the mean tension as a function of the spring stiffness on semi-log scale. 
Note that the exact trend for large stiffness remains undetermined due to the quickly growing variance of the tension. 
}
\end{figure}

The last setup, introduced in Section~\ref{sec:other_setups}, is the case where the motor is situated between two anti-parallel actin filaments, which are attached to a wall via springs, see sketch in Fig.~\ref{fig:tug}. 
Similarly as in the previous setup, the tension in the motor is given by the difference of the ratchet forces applied on it.
However, in this setup, there is no external load on actin polymers and thus in the steady state we have a simple balance between the ratchet interaction and springs. 
That means that the tension on the motor is given by
\[
\tau_\text{M} 
= \langle F_{\text{r}_2}\rangle - \langle F_{\text{r}_1} \rangle 
= k_\text{sp} \left( \left\langle \Delta x_2 \right\rangle - \left\langle \Delta x_1 \right\rangle \right) ,
\]
where $k_\text{sp}$ is the spring stiffness and $\Delta x_1 = x_1(t) - x_1(0)$ ($\Delta x_2 = x_2(t) - x_2(0)$) is the extension of the spring attached to first (second) actin. 
The tension is plotted in Fig.~\ref{fig:tug_k} for different stiffnesses $k_\text{sp}$ that represent the varying elasticity of the surrounding network, 
over a biologically relevant range \cite{schwarz2006focal,prager2011fibroblast}.
The tension that the motor generates depends strongly on this elasticity $k_\text{sp}$ and saturates for large $k_\text{sp}$. 
This suggests that the motor is not only able to respond to external forces, but also to sense mechanical properties of its environment.

Similar results were obtained for different motor models \cite{stam2015isoforms,Albert2014}. 
However, the models used in those studies exhibit rates that depend on the strain or tension in the myosin motor. 
In our study, we show that this behavior also occurs in a simple Brownian ratchet model with constant chemical rates.
It is important to include the fluctuations in this stochastic model to preserve the mechanosensing feature.
In mean field models, where the phenomenological force-velocity relation would be implemented deterministically, 
the motor would essentially stall in this given setup and cease to exert a net force on the filaments. 
Therefore mechanosensing will not occur spontaneously in those models.
This concludes the main result of the first part of our study. 

It has been shown that mechanosensing takes place on the cellular scale, 
where the cytoskeleton is coupled to the environment through focal adhesions \cite{geiger2009environmental,etienne2015cells}.

\section{Energy balance}
\label{sec:energie}
Up to this point we have discussed mechanical properties of myosin II inside various setups mimicking the acto-myosin cortex. 
Most of these mechanical properties emerge from the motor activity, 
which is driven by a chemical energy turnover into active tension and active movement of polymers and the motor itself inside the network. 
This leads to a natural question: ``How efficient is the motor?''. 

Presently we formulate a relation for the excess heat produced by the ATP cycle, that demonstrates its dependency on tension in the system.
This relation will be important in our interpretation of the efficiency in section~\ref{sec:chemical_efficiency}.

We start our discussion from the energy balance on the level of individual trajectories \cite{Pesek2013},
which can be symbolically summarized as 
\begin{equation}
\rmd E = \dbar {\mathcal Q}_\text{M}^\text{hb} + \dbar {\mathcal Q}_\text{A}^\text{hb} + \dbar {\mathcal W}^\text{ext}_\text{M} + \dbar {\mathcal Q}^\text{chem}, 
\label{eq:energy_balance}
\end{equation}
where the energy of the full system consists of the interaction with the ratchet potential~\eqref{eq:ratchet_interaction}
and the energy stored in the ATP molecules that are bound to the motor heads,
\begin{equation}
E( x_\text{M}, x_\text{A}, \{ \zeta_i \}) = V_t( x_\text{M} - x_\text{A} ) + \Delta E \sum\limits_{i=0}^{N-1} \chi( \zeta_i = c ) ,
\label{eq:internal_energy}
\end{equation}
where $\Delta E$ represents the actual internal energy stored by in a single ATP and $\chi$ is a characteristic function \footnote{
$\chi(\text{condition}) = 1$ if condition is fulfilled otherwise $0$.
}.

The energy of the system is subject to changes driven by the interaction with the environment.
In the case of overdamped diffusion, the environment represents a heath bath
and thus every interaction is associated with a heat exchange \cite{Pesek2013}
\begin{align}
{\mathcal Q}_\text{M}^\text{hb} 
&= \int \rmd x_\text{M} \circ \left[ - \gamma_\text{M} \frac{\rmd x_\text{M} }{\rmd t} + \sqrt{ 2 k_B T \gamma_\text{M} } \, \frac{ \rmd W_\text{M} }{ \rmd t } \right] , 
\label{eq:heat_hb_M} \\
{\mathcal Q}_\text{A}^\text{hb} 
&= \int \rmd x_\text{A} \circ \left[ - \gamma_\text{A} \frac{\rmd x_\text{A} }{\rmd t} + \sqrt{ 2 k_B T \gamma_\text{A} } \, \frac{ \rmd W_\text{A} }{ \rmd t } \right] ,
\label{eq:heat_hb_A}
\end{align} 
where the first term is the dissipation of energy due to the friction and the second term is the energy absorbed from the thermal activity of the environment.
Note that those integrals are evaluated in the Stratonovich sense.
The energy is also changed by the work of the external load 
\begin{equation}
{\mathcal W}^\text{ext}_\text{M} = - \int \rmd x_\text{M} \circ F_\text{load} = - \Delta x_\text{M} F_\text{load}, 
\label{eq:work_load}
\end{equation}
where we use the fact that the external load $F_\text{load}$ is constant. 

The last term in the energy balance \eqref{eq:energy_balance} is the contribution from the chemical cycle. 
The chemical cycle of a motor head, in the most basic form \cite{astumian1996mechanochemical,Bierbaum2011,Bierbaum2013,Albert2014}, is a sequence of binding an ATP molecule, its hydrolysis, release of phosphor and finally release of ADP.
The first step provides the chemical energy to the system, stored in ATP, while the next steps are associated with dissipation of the remaining energy to the surroundings.
Thus the total chemical energy consists of two parts ${\mathcal Q}^\text{chem} = {\mathcal Q}^\text{chem}_\text{in} + {\mathcal Q}^\text{chem}_\text{out}$,
\begin{align}
{\mathcal Q}_\text{in}^\text{chem} 
&= \sum_{t_\text{U} \leq t_\text{fin} } \left[ \Delta E + (c-1) V_\text{r}(x_{t_\text{U}}) \right] , 
\label{eq:q_in} \\
{\mathcal Q}_\text{out}^\text{chem} 
&= -\sum_{t_\text{B} \leq t_\text{fin} } \left[ \Delta E + (c-1) V_\text{r}(x_{t_\text{B}}) \right] 
\label{eq:q_out}
\end{align}
where we sum over the jump times $t_\text{B}$($t_\text{U}$) of all ATP binding (ADP unbinding) events.
$t_\text{fin}$ denotes the duration of the process. 

\begin{figure}
\centering
\includegraphics[width=.45\textwidth,height=!]{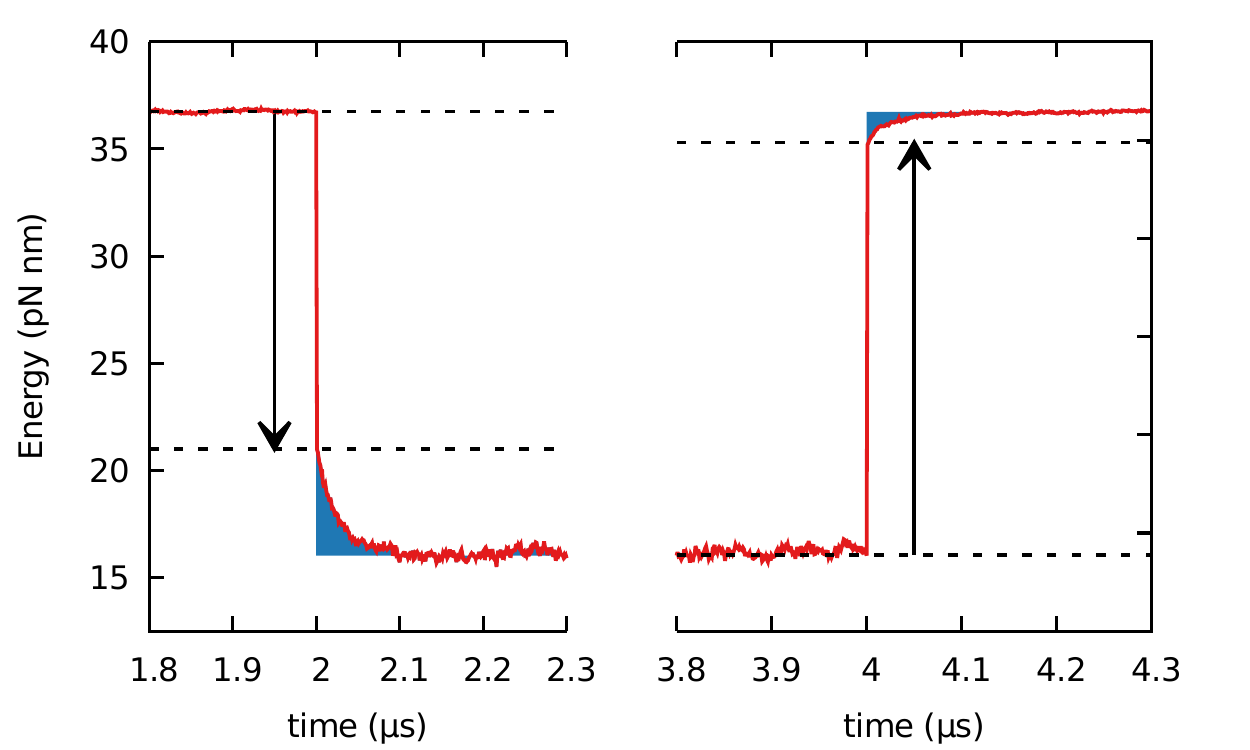}
\caption{\label{fig:energy_evolution}
Illustration of the time-evolution of the energy of the system over a single trajectory for $F_\text{load} = 0$.
The red line depicts the internal energy $E$ of the system over the time~\eqref{eq:internal_energy},
black arrows denote the dissipated and obtained chemical heat~\eqref{eq:q_out}, resp.~\eqref{eq:q_in},
and the blue area denote the total heat dissipated to the heat bath ${\mathcal Q}_\text{M}^\text{hb} + {\mathcal Q}_\text{A}^\text{hb}$, 
see~\eqref{eq:heat_hb_M} and~\eqref{eq:heat_hb_A}.
}
\end{figure}

Fig.~\ref{fig:energy_evolution} depicts a time evolution of the internal energy~\eqref{eq:internal_energy} for a single trajectory of the system without any external load, $F_\text{load} = 0$.
Note, that while we can identify the chemical input and output by the direction of the jump in the total energy, 
without any additional information it is impossible to distinguish which part of the heat dissipated to the heat bath corresponds to myosin or to actin.
For non-zero load, the heat dissipated to the heat bath will contain the work of the external forces as well,
which makes the separation in individual components even more difficult.

Also note that while expressions~\eqref{eq:heat_hb_M}, \eqref{eq:heat_hb_A}, \eqref{eq:q_in} and~\eqref{eq:q_out} are independent of the setup, 
the total energy balance~\eqref{eq:energy_balance} and the internal energy itself~\eqref{eq:internal_energy} are not. 

\subsection{Load on the motor}
In the ``load on the motor'' setup, we can further simplify the expressions~\eqref{eq:q_in} and~\eqref{eq:q_out} 
by using the equation of motion~\eqref{eq:eom_M}, 
which will allow us to interpret the excess chemical heat in terms of the ratchet force~\eqref{eq:ratchet_force} and thus link it to the tension.  
We start with the heat exchange of the molecular motor with the surrounding, i.e. \eqref{eq:heat_hb_M} expressed as 
\begin{multline}
{\mathcal Q}_\text{M}^\text{hb} 
= \int \rmd x_\text{M} \circ \left[ \nabla_\text{M} V_t(x_\text{M} - x_\text{A} ) + F_\text{load} \right] 
\\
= \int \rmd x_\text{M} \circ \nabla_\text{M} V_t(x_\text{M} - x_\text{A} ) - {\mathcal W}_\text{M}^\text{ext} ,
\label{eq:heat_hb_M_expl}
\end{multline}
where we recognize two terms. 
The first one is the work done by the ratchet potential, while the second one is the work done by an external load~\eqref{eq:work_load}. 
Similarly for the heat dissipated by the polymer's interaction with the environment we get 
\begin{multline}
{\mathcal Q}_\text{A}^\text{hb} 
= \int \rmd x_\text{A} \circ \nabla_\text{A} V_t(x_\text{M} - x_\text{A} )
\\
= - \int \rmd x_\text{A} \circ \nabla_\text{M} V_t(x_\text{M} - x_\text{A} ) .
\label{eq:heat_hb_A_expl}
\end{multline}
Consequently, the total energy balance equation~\eqref{eq:energy_balance} simplifies to 
\[
\rmd E 
= \rmd x_\text{M} \circ \nabla_\text{M} V_t  
+ \rmd x_\text{A} \circ \nabla_\text{A} V_t  
+ {\mathcal Q}_\text{in}^\text{chem}
+ {\mathcal Q}_\text{out}^\text{chem} , 
\]
which does not explicitly depend on the load $F_\text{load}$ anymore. 
Note that the total differential here is taken in the Stratonovich sense. 

Finally, in the steady state, 
we can see that the excess of the chemical cycle in the steady state is dissipated as the work of ratchet potential~\eqref{eq:ratchet_potential} over the separation of motor and filament 
\begin{multline}
\left\langle 
{\mathcal Q}_\text{in}^\text{chem} + {\mathcal Q}_\text{out}^\text{chem} 
\right\rangle 
\\
= 
\left\langle 
\int \rmd \left( x_\text{M} - x_\text{A} \right) \circ \left[ - \nabla_\text{M} V_t( x_\text{M} - x_\text{A} ) \right]
\right\rangle ,
\label{eq:chemical_excess}
\end{multline} 
which can be further interpreted in terms of tension on the polymer $\tau_\text{A}$, cf.~\eqref{eq:tension_on_polymer}, and the relative velocity $v$; see~\eqref{eq:mean_velocity}, 
\begin{equation}
\left\langle 
{\mathcal Q}_\text{in}^\text{chem} + {\mathcal Q}_\text{out}^\text{chem} 
\right\rangle 
= \left\langle \int \rmd t \; v \, \tau_\text{A} \right\rangle .
\label{eq:excess_heat} 
\end{equation}

\subsection{Heat fluxes}
\label{sec:heat_flux}
The internal energy of the system~\eqref{eq:internal_energy} is bounded.
Hence, if we interpret the energy balance equation~\eqref{eq:energy_balance} in terms of mean dissipation rates defined as 
\begin{equation}
q = \lim_{\Delta t \to \infty} \frac{1}{\Delta t} \left\langle {\mathcal Q}(\Delta t) \right\rangle ,
\label{eq:heat_flux}
\end{equation}
we obtain an equation representing the conservation of fluxes 
\begin{equation}
0 = q_\text{M}^\text{hb} + q_\text{A}^\text{hb} + w^\text{ext}_\text{M} + q^\text{chem} 
\label{eq:heat_flux_balance}
\end{equation}
for arbitrary initial condition. 
Such equality is trivially true for arbitrarily small times when the initial condition corresponds to the steady state.

\subsection{Alternative setups} 
In section \ref{sec:ratchet} we introduced three alternative models, 
more suitable for capturing the behavior of myosin motors inside the polymer network known as the actomyosin cortex.

\subsubsection{Load on the polymer}
As the ``load on the polymer'' setup is similar to the ``load on the motor'' setup, cf. section~\ref{sec:load_on_polymer}, 
the energy balance~\eqref{eq:energy_balance} is preserved 
with the for heat dissipation~\eqref{eq:heat_hb_M} and~\eqref{eq:heat_hb_A} kept intact 
and external work on the motor ${\mathcal W}_\text{M}^\text{ext}$ being replaced by the external work on the polymer
\[
{\mathcal W}_\text{A}^\text{ext} = \int \rmd x_\text{A} \circ F_\text{load} = \Delta x_\text{A} \, F_\text{load} . 
\]
Note, that in this particular setup the load is applied in the opposite direction, as can be seen from the equations of motion~\eqref{eq:eom_A_lp}. 

Similarly expressions~\eqref{eq:heat_hb_M_expl} and~\eqref{eq:heat_hb_A_expl} are modified to
\begin{align*}
{\mathcal Q}_\text{M}^\text{hb} 
&= \int \rmd x_\text{M} \circ \nabla_\text{M} V_t(x_\text{M} - x_\text{A} ) ,
\\
{\mathcal Q}_\text{A}^\text{hb} 
&= - \int \rmd x_\text{A} \circ \nabla_\text{M} V_t(x_\text{M} - x_\text{A} ) - {\mathcal W}_\text{A}^\text{ext} .
\end{align*} 
Consequently the excess of the chemical cycle~\eqref{eq:chemical_excess} changes to, cf.~\eqref{eq:excess_heat}, 
\begin{multline*}
\left\langle 
{\mathcal Q}_\text{in}^\text{chem} + {\mathcal Q}_\text{out}^\text{chem} 
\right\rangle 
\\
= 
\left\langle 
\int \rmd \left( x_\text{M} - x_\text{A} \right) \circ \left[ - \nabla_\text{M} V_t( x_\text{M} - x_\text{A} ) \right]
\right\rangle ,
\end{multline*} 
which can be again interpreted in terms of the relative velocity $v$, cf.~\eqref{eq:mean_velocity}, 
and the tension on the motor $\tau_\text{M}$, cf.~\eqref{eq:tension_load_on_polymer},  
\begin{equation}
\left\langle 
{\mathcal Q}_\text{in}^\text{chem} + {\mathcal Q}_\text{out}^\text{chem} 
\right\rangle 
= \left\langle \int \rmd t \; v \, \tau_\text{M} \right\rangle .
\label{eq:chemical_excess_poly}
\end{equation}
An argumentation similar to the one used in the previous setup can be also made here.

\subsubsection{Tug of war}
In the ``tug of war'' setup, see section~\ref{sec:tug_of_war} and Fig.~\ref{fig:tug_F_illustration},
the motor is placed between two polymers on which the load is applied.
Therefore the energy balance~\eqref{eq:energy_balance} is modified to  
\begin{multline}
\rmd E = \dbar {\mathcal Q}_\text{M}^\text{hb} + \dbar {\mathcal Q}_{\text{A}_1}^\text{hb} + \dbar {\mathcal Q}_{\text{A}_2}^\text{hb} 
\\
+ \dbar {\mathcal W}^\text{ext}_{\text{A}_1} + \dbar {\mathcal W}^\text{ext}_{\text{A}_2} + \dbar {\mathcal Q}^\text{chem}, 
\label{eq:energy_balance_two_filaments}
\end{multline}
where the energy now contains contributions from both polymers
\begin{multline*}
E( x_\text{M}, x_{\text{A}_1}, x_{\text{A}_2}, \{ \zeta_i \}) 
= V_t( x_\text{M} - x_{\text{A}_2} ) 
\\
+ V_t( x_{\text{A}_1} - x_\text{M} ) 
+ \Delta E \sum\limits_{i=0}^{N-1} \chi( \zeta_i = c ) .
\end{multline*}
Note, that the last sum is over all heads of the motor, $N = N_{\text{A}_1} + N_{\text{A}_2}$.
Contributions from the heat exchange with the environment~\eqref{eq:heat_hb_A} remain the same, even though in this case we have two contributions from both filaments
\begin{align*}
{\mathcal Q}_{\text{A}_1}^\text{hb} 
&= \int \rmd x_{\text{A}_1} \circ \left[ - \gamma_\text{A} \frac{\rmd x_{\text{A}_1} }{\rmd t} + \sqrt{ 2 k_B T \gamma_\text{A} } \, \frac{ \rmd W_{\text{A}_1} }{ \rmd t } \right] , \\ 
{\mathcal Q}_{\text{A}_2}^\text{hb} 
&= \int \rmd x_{\text{A}_2} \circ \left[ - \gamma_\text{A} \frac{\rmd x_{\text{A}_2} }{\rmd t} + \sqrt{ 2 k_B T \gamma_\text{A} } \, \frac{ \rmd W_{\text{A}_2} }{ \rmd t } \right] ,
\end{align*}
and now the external load is not applied on the motor but on each actin polymer.
Hence, equation~\eqref{eq:work_load} is replaced by
\begin{align*}
{\mathcal W}^\text{ext}_{\text{A}_1} &= - \int \rmd x_{\text{A}_1} \circ F_\text{load} = - \Delta x_{\text{A}_1} F_\text{load}, 
\\
{\mathcal W}^\text{ext}_{\text{A}_2} &= \int \rmd x_{\text{A}_2} \circ F_\text{load} = \Delta x_{\text{A}_2} F_\text{load}. 
\end{align*}

Note, that the interpretation of the energy balance \eqref{eq:energy_balance_two_filaments} in terms of the heat fluxes \eqref{eq:heat_flux}
while following the same arguments as for \eqref{eq:heat_flux_balance} is possible and leads to   
\[
0 = q_\text{M}^\text{hb} + q_{\text{A}_1}^\text{hb} + q_{\text{A}_2}^\text{hb} + w^\text{ext}_{\text{A}_1} + w^\text{ext}_{\text{A}_2} + q^\text{chem} .
\]

Similarly to the ``load on the motor'' setup, we can express the excess from the chemical cycle in the steady state as sum of work by the ratchet forces with respect to their respective displacements,
\begin{multline}
\left\langle 
{\mathcal Q}_\text{in}^\text{chem} + {\mathcal Q}_\text{out}^\text{chem} 
\right\rangle 
\\
= 
\left\langle 
\int \rmd \left( x_\text{M} - x_{\text{A}_1} \right) \circ \left[ - \nabla_\text{M} V_t( x_{\text{A}_1} - x_\text{M} ) \right]
\right\rangle 
\\
+
\left\langle 
\int \rmd \left( x_\text{M} - x_{\text{A}_2} \right) \circ \left[ - \nabla_\text{M} V_t( x_\text{M} - x_{\text{A}_2} ) \right]
\right\rangle .
\label{eq:excess_heat_tug}
\end{multline}
 
If we naively assume that relative velocities and ratchet force are well localized random Gaussian variables, 
we can further approximate the excess chemical heat as 
\begin{multline*}
\left\langle 
{\mathcal Q}_\text{in}^\text{chem} + {\mathcal Q}_\text{out}^\text{chem} 
\right\rangle 
\approx T \left[ 
\left\langle v_1 \right\rangle \left\langle F_{\text{r}_1} \right\rangle  
+
\left\langle v_2 \right\rangle \left\langle F_{\text{r}_2} \right\rangle  
\right]
\\
= \frac{T}{2} \left\langle v \right\rangle \left\langle \tau_\text{M} \right\rangle
, 
\end{multline*}
where the last equality follows from the anti-symmetry of the mean relative velocities of the motor with respect to individual polymers in the steady state. 

As the previous assumption is far from reality, such straightforward interpretation of the excess heat is invalid. 
However, if we assume that the displacements in the respective integrals are typically anti-symmetric, due to the anti-symmetry of the setup, 
we can interpret the excess chemical heat as a covariance of the tension on the motor~\eqref{eq:tension_tug} with the relative velocity of the motor with respect to a single polymer \eqref{eq:mean_velocity} integrated over time, 
as in equation~\eqref{eq:excess_heat}. 
Note, that in this particular setup the interpretation in terms of the tension is less straightforward in comparison with the ``load on the motor'' or the ``load on the polymer'' setup, 
because the individual displacements $x_\text{M} - x_{\text{A}_1}$ and $x_\text{M} - x_{\text{A}_2}$ are subjected to thermal fluctuations on short time scale. 
Only on the long time scale they are anti-symmetric and the tension relaxes to a steady value.

\subsubsection{Elastic environment}
\label{sec:power_elastic}
For the model described in paragraph~\ref{sec:environment} and depicted in Fig.~\ref{fig:tug}, 
the energy balance~\eqref{eq:energy_balance}, resp.~\eqref{eq:energy_balance_two_filaments} is further modified,
\begin{equation*}
\rmd E = \dbar {\mathcal Q}_\text{M}^\text{hb} + \dbar {\mathcal Q}_{\text{A}_1}^\text{hb} + \dbar {\mathcal Q}_{\text{A}_2}^\text{hb} + \dbar {\mathcal Q}^\text{chem}, 
\end{equation*}
as there is no external load but all the accumulated energy is now stored in harmonic potentials 
\begin{multline*}
E( x_\text{M}, x_{\text{A}_1}, x_{\text{A}_2}, \{ \zeta_i \}) 
= V_t( x_\text{M} - x_{\text{A}_2} ) 
+ V_t( x_{\text{A}_1} - x_\text{M} ) 
\\
+ \frac{1}{2} k_\text{sp} \left[ x_{\text{A}_1} - x_{\text{A}_1}(0) \right]^2 
+ \frac{1}{2} k_\text{sp} \left[ x_{\text{A}_2} - x_{\text{A}_2}(0) \right]^2 
\\
+ \Delta E \sum\limits_{i=0}^{N-1} \chi( \zeta_i = c ) .
\end{multline*}
Contrary to the previous setup, the total energy of the system is no longer bound, hence the heat fluxes~\eqref{eq:heat_flux} are only balanced in the steady state.

Finally we look again at the excess output from the chemical cycle
\begin{multline*}
\left\langle 
{\mathcal Q}_\text{in}^\text{chem} + {\mathcal Q}_\text{out}^\text{chem} 
\right\rangle 
\\
= 
\left\langle 
\int \rmd \left( x_\text{M} - x_{\text{A}_1} \right) \circ \left[ - \nabla_\text{M} V_t( x_{\text{A}_1} - x_\text{M} ) \right]
\right\rangle 
\\
+
\left\langle 
\int \rmd \left( x_\text{M} - x_{\text{A}_2} \right) \circ \left[ - \nabla_\text{M} V_t( x_\text{M} - x_{\text{A}_2} ) \right] 
\right\rangle 
\\
+ k_\text{sp} \left\langle \int \rmd x_{\text{A}_1} \circ \left( x_{\text{A}_1} - x_{\text{A}_1}(0) \right) \right\rangle
\\ 
+ k_\text{sp} \left\langle \int \rmd x_{\text{A}_2} \circ \left( x_{\text{A}_2} - x_{\text{A}_2}(0) \right) \right\rangle
,
\end{multline*} 
which further simplifies to 
\begin{multline*}
\left\langle 
{\mathcal Q}_\text{in}^\text{chem} + {\mathcal Q}_\text{out}^\text{chem} 
\right\rangle 
\\
= 
\left\langle 
\int \rmd \left( x_\text{M} - x_{\text{A}_1} \right) \circ \left[ - \nabla_\text{M} V_t( x_{\text{A}_1} - x_\text{M} ) \right]
\right\rangle 
\\
+
\left\langle 
\int \rmd \left( x_\text{M} - x_{\text{A}_2} \right) \circ \left[ - \nabla_\text{M} V_t( x_\text{M} - x_{\text{A}_2} ) \right]
\right\rangle 
\\
+ \left\langle 
\Delta E^\text{sp}_1 
\right\rangle
+ \left\langle 
\Delta E^\text{sp}_2 
\right\rangle ,
\end{multline*} 
where $E^\text{sp}$ denotes the energy stored in the spring;, i.e.,
\[
\left\langle \Delta E^\text{sp}_\alpha \right\rangle 
= 
\frac{1}{2} k_\text{sp}
\left\langle
\left( 
x_{\text{A}_\alpha}(t) 
- 
x_{\text{A}_\alpha}(0)  
\right)^2 
\right\rangle
.
\]
The interpretation of this excess chemical energy in terms of the tension is even more difficult than in the ``tug of war'' setup~\eqref{eq:excess_heat_tug} due to multiple factors.
First, there are additional terms related to the change of the energy of the springs.
Secondly, the mean velocity of all elements in the steady state is zero due to the anti-symmetry of the setup and no additional external force breaking this symmetry. 

\section{Efficiency} 
\label{sec:efficiency}
In previous years, a lot of effort was directed towards the assessment of the efficiency of molecular motors \cite{Schmiedl2008,boksenbojm2009entropy,parmeggiani1999energy,parrondo1998efficiency,zhang2009efficiency}. 
The main focus of these works was towards the transport properties of the aforementioned motors. 
However the main role of myosin II is to generate tension in a cytoskeletal network \cite{ma2012nonmuscle,chugh2017actin,monier2010actomyosin}.

Efficiency, in general, relates a ``useful'' output to the input.
As what is considered ``useful'' heavily depends on the considered setup, there is a plethora of possible choices.
Here we discuss a common choice as well as one particular choice which is to the best of our knowledge not discussed previously in the literature. 
First, we briefly discuss the most common definition of efficiency, well discussed in the literature \cite{Schmiedl2008,boksenbojm2009entropy,parmeggiani1999energy,zhang2009efficiency}, which is related to the transport properties of molecular motors. 
Then we provide an alternative definition of the chemical efficiency, which can be related to the tension generated by myosin~II in networks. 

\subsection{Transport efficiency} 
As we have stated, the first efficiency we discuss is related to transport properties. 
In this respect, we identify the heat dissipated to the environment by the motor alone as the output work 
\[
{\mathcal W}^\text{out}_\text{M} = - {\mathcal Q}^\text{hb}_\text{M} 
\] 
and chemical work $\mathcal Q^\text{chem}$ and the work done by the load $\mathcal W^\text{ext}_\text{M}$.
Hence the transport efficiency can be defined as 
\[
\eta_\text{tr} = \frac{ \left\langle {\mathcal W}^\text{out}_\text{M} \right\rangle }{ \left\langle {\mathcal Q}^\text{chem} \right\rangle + \left\langle {\mathcal W}^\text{ext}_\text{M} \right\rangle } .
\]
Note that for a large load ($F_\text{load} \to \infty$) such efficiency approaches $\eta_\text{tr} \to 1$, 
while in absence of a load $F_\text{load} = 0$ it captures how much of the chemical energy is dissipated by the motor's motion  
\[
\eta_\text{tr}( F_\text{load} = 0 ) = \frac{ \left\langle \int \rmd x_\text{M} \circ \left[ - \nabla_\text{M} V_t \right] \right\rangle }{ \left\langle {\mathcal Q}^\text{chem} \right\rangle } ,
\]
i.e., how much of the chemical energy provided to the system is dissipated by the motor and not by the polymer. 

Note that this particular definition of the efficiency has several shortcomings.
First, the output work has two components, one from the chemical source and the second from the external driving, 
thus in principle the efficiency can be greater than one. 
Secondly, 
only for long time scales, in situations where the motor displacement and ratchet force are to a great extent uncorrelated, 
one can relate the output work to the heat dissipated by the drag force to the medium.
This means that in systems where the motor is frustrated and does not move on average, 
like in the ``tug of war'' and ``elastic environment'' setups,
one cannot distinguish between various sources of the output work. 
Finally, the output work depends on the details of the ratchet interaction, which renders it practically immeasurable.

\subsection{Chemical efficiency} 
\label{sec:chemical_efficiency}
An alternative approach is to identify the ``useful'' output by looking at the chemical energy, supplied by ATP, that is transformed to other forms of energy, 
i.e. either it is dissipated by myosin and actin through their diffusion or it is stored in the interaction potential and thus increases the tension in the system. 
This consideration leads to a novel chemical efficiency defined as 
\begin{equation}
\eta_\text{chem} = \frac{\langle\mathcal Q^\text{chem}_\text{in}\rangle+\langle\mathcal Q^\text{chem}_\text{out}\rangle}{\langle\mathcal Q^\text{chem}_\text{in}\rangle} .
\label{eq:efficiency}
\end{equation}
Note that in the steady state all excess chemical energy is dissipated to the environment, see section~\ref{sec:heat_flux}, 
and thus can in principle be measured by calorimetric techniques~\cite{Maskow2015}.  

\subsubsection{Load on the motor}
\begin{figure}[t]
\centering
\includegraphics[width=0.45\textwidth,height=!]{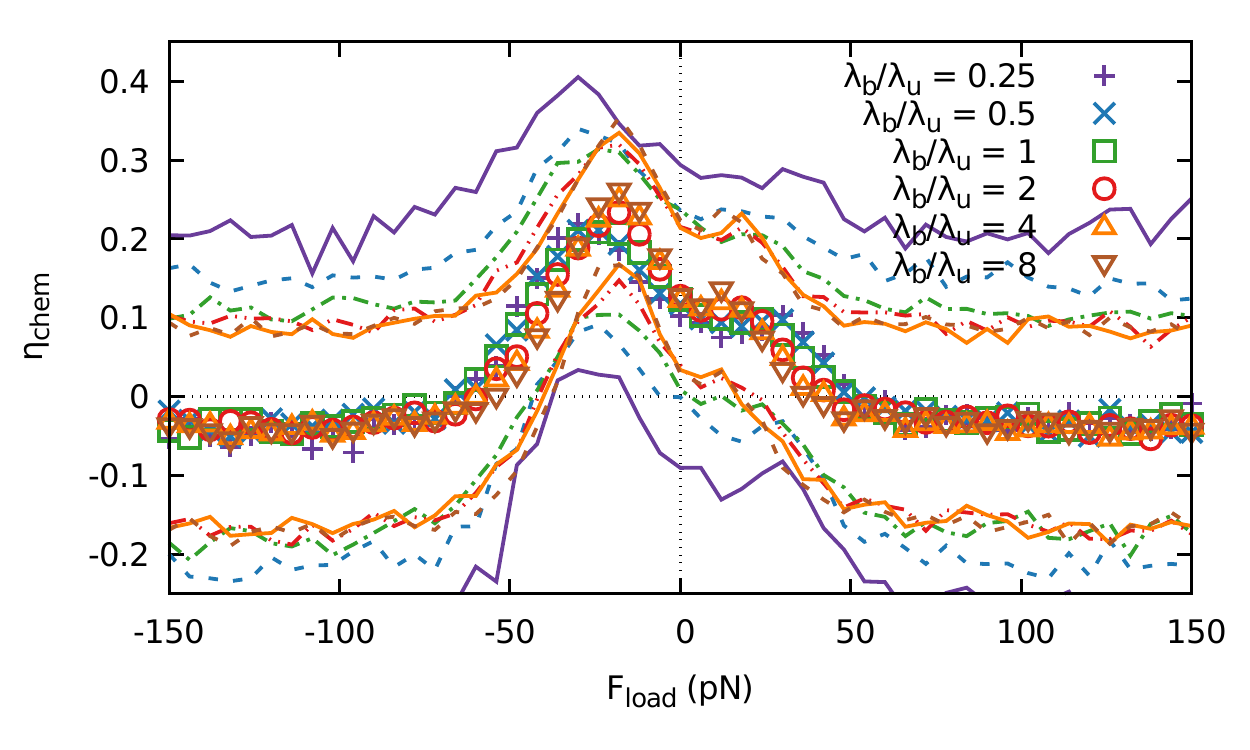}
\caption{
\label{fig:chem}
Chemical efficiency $\eta_\text{chem}$ as a function of load force $F_\text{load}$.
Solid lines are plotted at the level of $1\sigma$. 
By increasing the $\lambda_\text{b}$ the local maxima of chemical efficiencies are getting closer to the origin,
and their fluctuations decrease. 
}
\end{figure}

We start our discussion again with the ``load on the motor'' setup with a four headed motor, see Fig.~\ref{fig:chem}.
We observe that the chemical efficiency has two local maxima. 
We show later that these two maxima correspond to a frustrated system, 
in which the load is close to the point where it is able to overcome any force induced by the ratchet potential. 
Another observation, which we discuss later in more detail, is that the efficiency reaches negative values for large loads, 
which we interpret as work done by the load $F_\text{load}$ that is converted back to chemical energy. 

In order to properly understand these features, 
we try to evaluate the chemical efficiency \eqref{eq:efficiency} in a steady state. 
We start by evaluating the mean heat \eqref{eq:q_in} and \eqref{eq:q_out} produced over a time interval of length $\Delta t$ 
by transitions of a single head $i$ displaced with respect to the end of the motor by $i d$,  
\begin{gather*}
\begin{multlined}[b][.42\textwidth]
\langle \mathcal Q^\text{chem}_\text{in}(i) \rangle 
= \Delta t \, \lambda_\text{u} 
\\ \times 
\!\!\!\! \sum\limits_{\{\zeta_j\} : \zeta_i = 1 } \!\!\!\! \rho(\{ \zeta_j \} ) 
\left[ \Delta E - (1-c) \left\langle V_\text{r}(x - i \, d ) \middle| \{ \zeta_j \} \right\rangle \right] 
\end{multlined}
\\
\begin{multlined}[b][.42\textwidth]
\langle \mathcal Q^\text{chem}_\text{out}(i) \rangle 
= - \Delta t \, \lambda_\text{b} 
\\ \times 
\!\!\!\! \sum\limits_{\{\zeta_j\} : \zeta_i = c } \!\!\!\! \rho(\{ \zeta_j \} ) 
\left[ \Delta E - (1-c) \left\langle V_\text{r}(x - i \, d ) \middle| \{ \zeta_j \} \right\rangle \right] 
\end{multlined}
\end{gather*}
where $\langle V | X \rangle$ is the steady state conditional mean value with respect to the fixed internal state $X$ 
and $\rho(X) \in [0,1]$ denotes the probability of the occupation of the corresponding internal state $X$ in the steady state, 
which does not depend on the position due to the independence of the internal state dynamics~\eqref{eq:transition} on position. 
Moreover, due to the independence of individual heads the probability occupation of a given internal state can be written as 
\[
\rho( \{ \zeta_i \} ) = \prod_{i=0}^{N-1} \rho_1( \zeta_i ), 
\]  
where 
\[
\rho_1(\zeta) = \begin{cases} 
\frac{ \lambda_\text{b} }{ \lambda_\text{b} + \lambda_\text{u} } & \zeta = 1 , \\
\frac{ \lambda_\text{u} }{ \lambda_\text{b} + \lambda_\text{u} } & \zeta = c . 
\end{cases} 
\]

\begin{figure}[t]
\centering
\subfigure[ATP unbound state]{ 
\centering
\includegraphics[width=.45\textwidth,height=!]{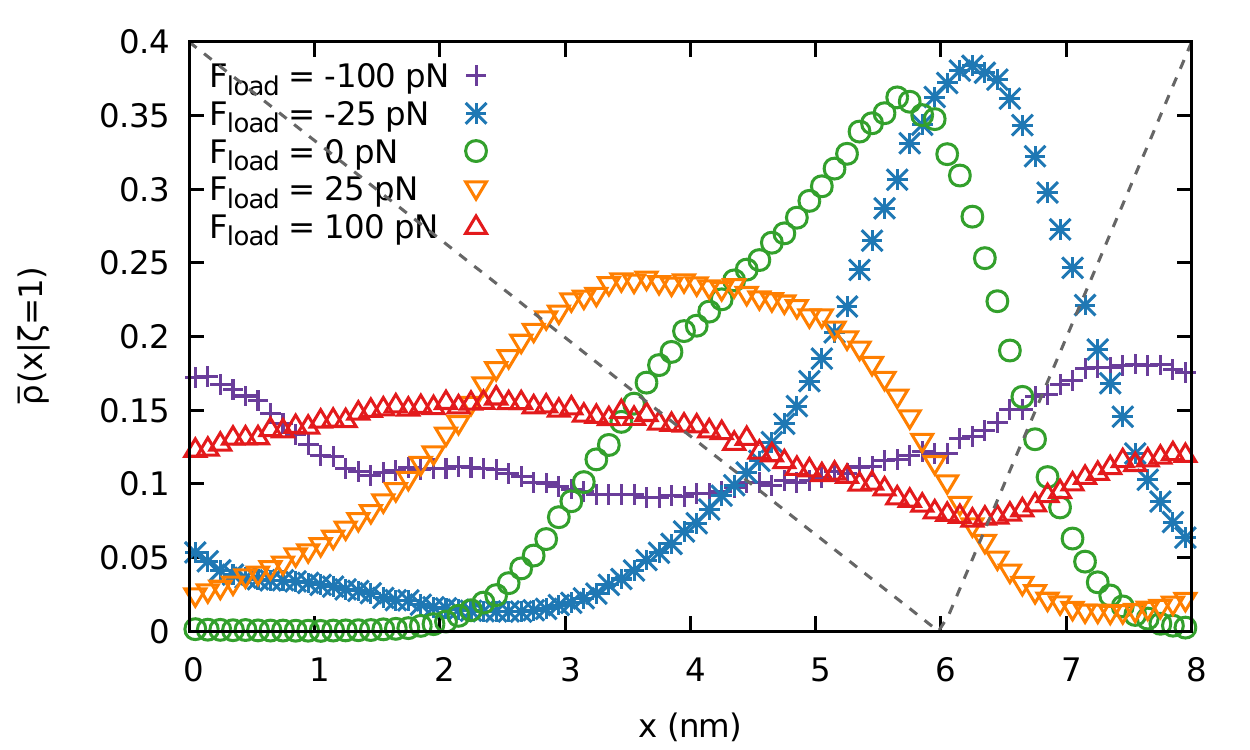}
}
\\
\subfigure[ATP bound state]{
\centering
\includegraphics[width=.45\textwidth,height=!]{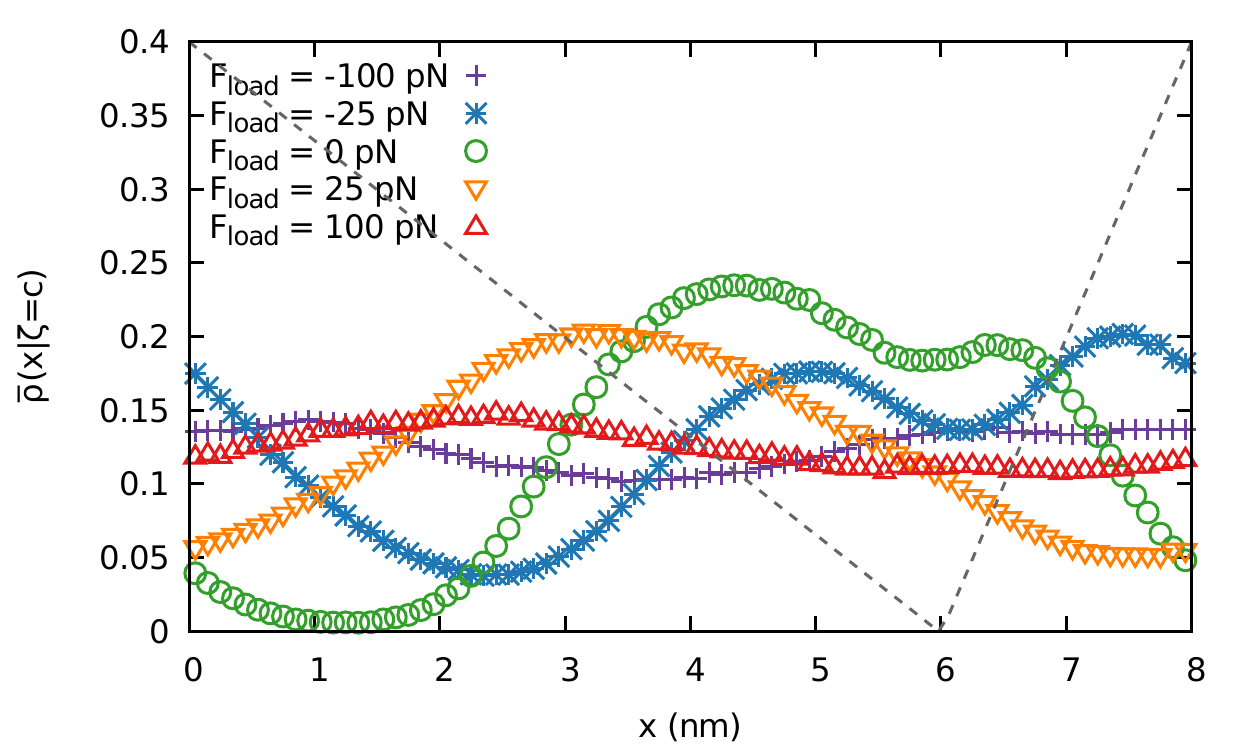}
}
\caption{
\label{fig:pos_distr}
Spatial distributions $\rho$ of the motor heads with respect to the ratchet potential (denoted by dashed black line) for different load forces $F_\text{load}$.
The spatial distribution is taken as the average over all heads of a given motor in a given internal state.
We observe that heads in the ATP unbound state follow the potential closely while those in the bound state do not. 
Also by increasing the external load, all distributions universally become flat. 
}
\end{figure}

Another consequence of the independence of the individual heads 
is that we can express the total mean heat, produced in the chemical cycle, as a sum of contributions from individual heads
\begin{align*}
\langle \mathcal Q_\text{in}^\text{chem} \rangle 
=& \sum\limits_{i=0}^{N-1} \left\langle \mathcal Q_\text{in}^\text{chem}(i) \right\rangle 
= \Delta t \frac{ \lambda_\text{b} \lambda_\text{u} }{ \lambda_\text{b} + \lambda_\text{u} } 
\\ &\times
\left[ N \, \Delta E 
- (1-c) \sum\limits_{i=0}^{N-1} \left\langle V_\text{r}(x - i d ) \middle| \zeta_i = 1 \right\rangle 
\right] ,  
\\
\langle \mathcal Q^\text{chem}_\text{out} \rangle 
=& \sum\limits_{i=0}^{N-1} \left\langle \mathcal Q_\text{out}^\text{chem}(i) \right\rangle 
= - \Delta t \frac{ \lambda_\text{b} \lambda_\text{u} }{ \lambda_\text{b} + \lambda_\text{u} } 
\\ &\times
\left[ N \, \Delta E 
- (1-c) \sum\limits_{i=0}^{N-1} \left\langle V_\text{r}(x - i d ) \middle| \zeta_i = c \right\rangle  
\right] , 
\end{align*}
which leads to a more explicit expression for the efficiency \eqref{eq:efficiency} 
\begin{multline*}
\eta_\text{chem} = \frac{1-c}{N}
\\ \times
\sum\limits_{i=0}^{N-1} \frac{ \left\langle V_\text{r}(x-id) \middle| \zeta_i = c \right\rangle - \left\langle V_\text{r}( x -id ) \middle| \zeta_i = 1 \right\rangle }
{ \Delta E - \frac{1-c}{N} \sum\limits_{j=0}^{N-1} \left\langle V_\text{r}(x-jd) \middle| \zeta_j = 1 \right\rangle } 
. 
\end{multline*}
We can further introduce an average probability density over all heads 
\begin{equation}
\bar{\rho}(x,\zeta) = \frac{1}{N} \sum\limits_{i=0}^{N-1} \sum\limits_{ \{ \zeta_j \} : \zeta_i = \zeta } \rho( x + i d, \{ \zeta_j \} ) ,
\label{eq:eff_distribution}
\end{equation}
and consequently the conditional average probability density
\begin{equation}
\bar{\rho}(x|\zeta) = \bar{\rho}(x,\zeta) \left[ \int\limits_0^\ell \rmd y \; \bar{\rho}(y,\zeta) \right]^{-1}
\label{eq:cond_prob}
\end{equation}
which allow us to further simplify the expression for the chemical efficiency to 
\begin{equation}
\eta_\text{chem} = (1-c)
\frac{ \left\langle V_\text{r}(x) \middle| \zeta = c \right\rangle_{\bar{\rho}} - \left\langle V_\text{r}(x) \middle| \zeta = 1 \right\rangle_{\bar{\rho}} }
{ \Delta E - (1-c) \left\langle V_\text{r}(x) \middle| \zeta = 1 \right\rangle_{\bar{\rho}} } 
. 
\label{eq:eta}
\end{equation}
We notice that the numerator is given by the average difference between mean potentials in the ATP attached and detached states over all heads. 
As the potential is known in our model, the only missing information is thus the steady state distribution 
or to be more precise the average steady state density for all heads \eqref{eq:eff_distribution}.  
These distributions are empirically obtained by collecting the positions of the heads over all times in which the head was in the ATP bound/unbound state.
They are shown in Fig.~\ref{fig:pos_distr} for different load forces. 

\begin{figure}[t]
\centering
\includegraphics[width=0.45\textwidth,height=!]{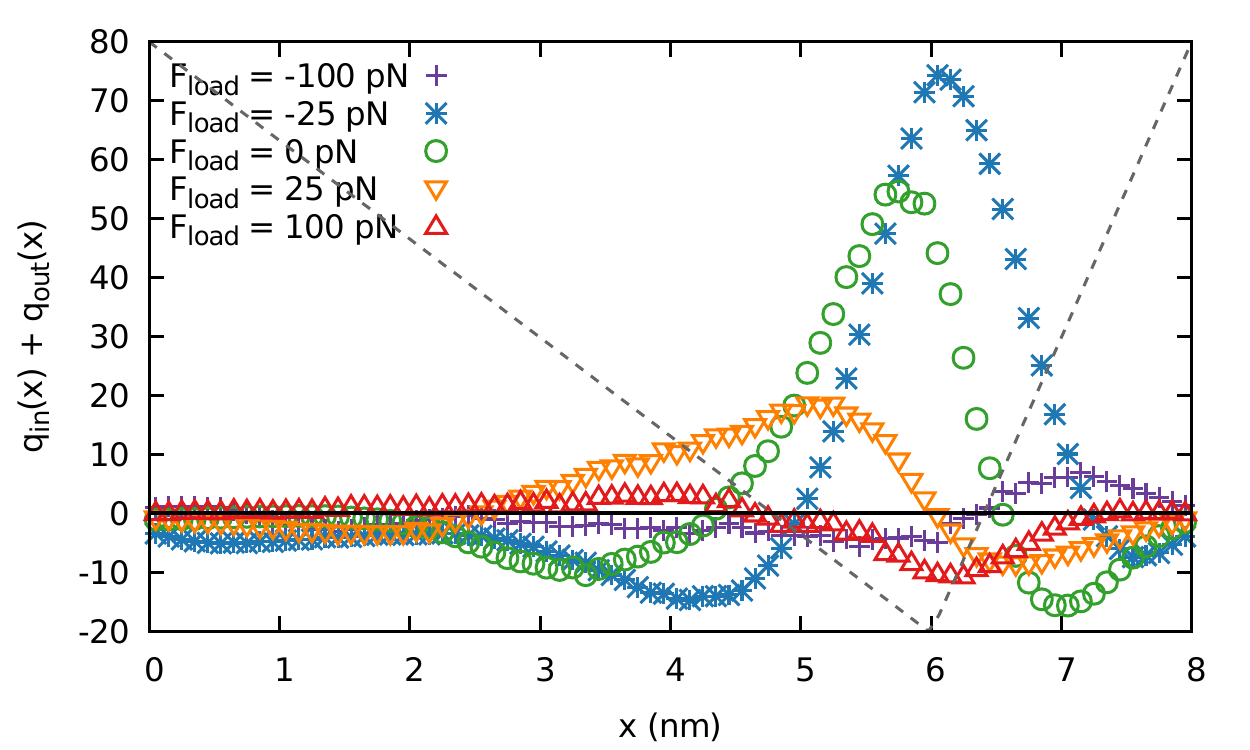}
\caption{
\label{fig:chem_energy_distr}
Heat flux~\eqref{eq:heat_flux_explicit} density, i.e. $\int \rmd x \; q(x) = q$, for various loads.  
We can see that the largest heat flux is around the ratchet potential~\eqref{eq:ratchet_potential} minimum, where the gap is largest. 
However, the region which contributes to the chemical efficiency~\eqref{eq:eta} the most is the close to the ratchet potential maximum;
see Fig.~\ref{fig:chem_efficiency_distr} where the heat flux is very small. 
}
\end{figure}

\begin{figure}[t]
\centering
\includegraphics[width=0.45\textwidth,height=!]{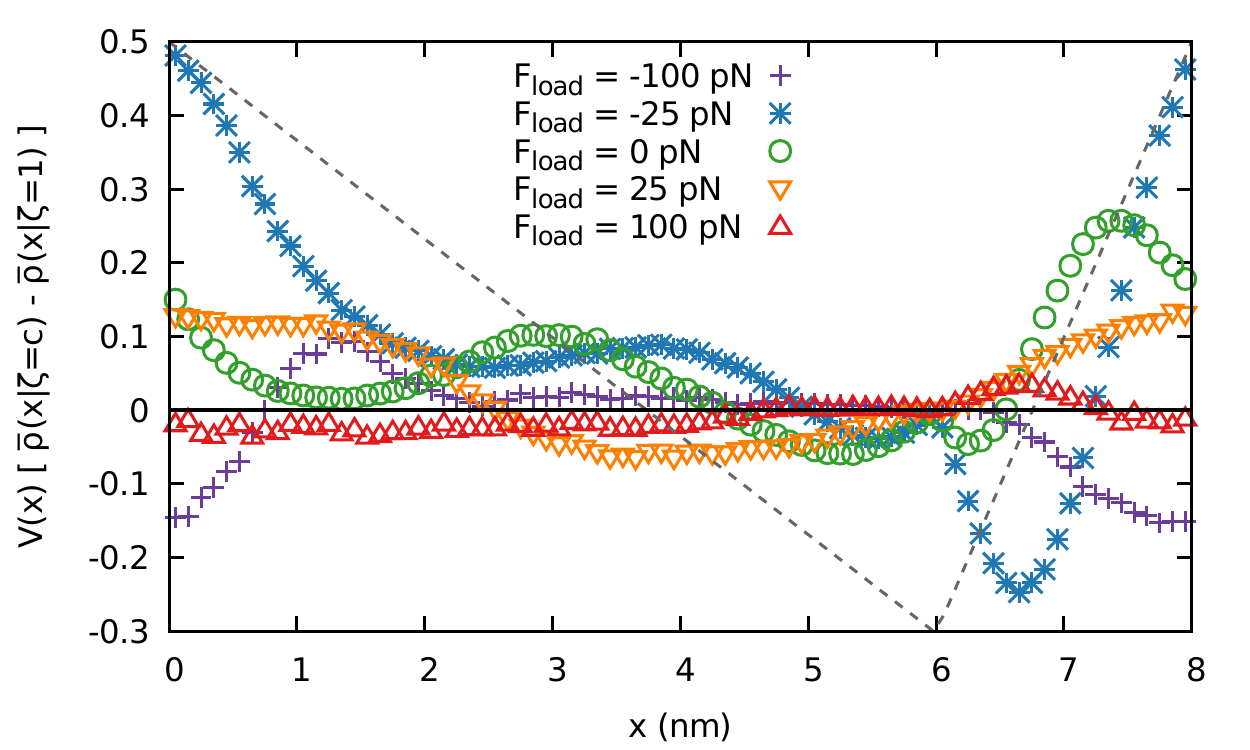}
\caption{
\label{fig:chem_efficiency_distr}
Space resolved numerator of the efficiency \eqref{eq:eta} for various load forces $F_\text{load}$. 
We observe that the optimal output is achieved when there is the biggest discrepancy between spatial distributions; compare with Fig.~\ref{fig:pos_distr} and \ref{fig:chem}. 
}
\end{figure}

The difference between the average distribution \eqref{eq:eff_distribution} in the ATP bound state and ATP unbound state is directly translated to the mean heat flux \eqref{eq:heat_flux} out of the chemical system
\begin{multline}
q_\text{in}^\text{chem} + q_\text{out}^\text{chem} 
= (1-c) \frac{ \lambda_\text{b} \lambda_\text{u} }{ \lambda_\text{b} + \lambda_\text{u} } 
\\ \times
\left[ \left\langle V_\text{r}(x) \middle| \zeta = c \right\rangle_{\bar{\rho}} - \left\langle V_\text{r}(x) \middle| \zeta = 1 \right\rangle_{\bar{\rho}} \right] ,
\label{eq:heat_flux_explicit}
\end{multline} 
which is depicted in Fig.~\ref{fig:chem_energy_distr}, 
where we show the spatial dependency of the heat fluxes.
We can see that a major flux is close to the ratchet potential's minimum \eqref{eq:ratchet_potential} where both the probability density is localized and the gap is the largest. 
On the other hand, close to the ratchet potential's maximum the heat flux is close to zero.
However, if we compare it to the numerator of the chemical efficiency~\eqref{eq:eta}; see Fig.~\ref{fig:chem_efficiency_distr}, 
we can see that a major contribution to the chemical efficiency actually comes from the region around the ratchet potential's maximum.
This discrepancy can be interpreted in the following sense: 
While most transitions occur close to the ratchet potential's minimum due to the large probability density, thus providing a large flux, 
those particles do not diffuse far before the next transition, 
hence the contribution to the chemical efficiency, associated with the discrepancy between chemical input and output, is relatively small. 
On the other hand, a particle that switched its internal state close to the potential maximum relaxes relatively fast, compared to the transition rates, 
to the vicinity of the ratchet potential's minimum and thus creates a larger discrepancy leading to high contribution to the efficiency. 
However these transitions are rare due to the low probability density and hence the total flux density is relatively small.  

In absence of an external load, $F_\text{load}=0$, 
the maximum of the distribution, see Fig.~\ref{fig:pos_distr}, is close to the minimum of the potential,  
where the discrepancy in the position comes from the averaging over multiple heads with a fixed distance between them, 
which does not match with the period of the ratchet potential. 
Also the fact that the distributions seem to exhibit multiple peaks, can be attributed to the motor having four heads. 
Both in the bound and unbound state, the head can be found predominantly on the ``soft'' slope of the ratchet potential.

When the load is increased to $F_\text{load} = - 25 \, \mathrm{pN}$, the average steady state distribution shifts to the right in both the ATP bound and unbound state. 
While the distribution for the ATP unbound state remains peaked around the potential minimum, cf. Fig.~\ref{fig:pos_distr},
the distribution in the ATP bound state flattens out and increases around the potential maximum,
which is a trademark of a frustrated system.  
This is due to the potential being less steep, meaning that heads in that particular state are more susceptible to external forces,  
which in this particular case leads to an increase of $\eta$ due to the increase in discrepancy between the average distributions, 
see Figs~\ref{fig:chem} and \ref{fig:chem_energy_distr}. 

For a load in the opposite direction $F_\text{load}= 25 \, \mathrm{pN}$ the effect on average densities is similar but less pronounced, 
see Fig.~\ref{fig:chem_energy_distr}, 
since the slope of the ratchet potential on the side is less steep and the distributions get more spread out, cf. Fig.~\ref{fig:pos_distr}. 

For larger loads, distributions shift and flatten even more, Fig.~\ref{fig:pos_distr}. 
In the extreme situation, it leads to completely flat distribution and thus zero efficiency, Fig.~\ref{fig:chem}.
For high loads we also observe a population inversion; see Fig.~\ref{fig:pos_distr}, 
where the potential maximum is more occupied than the potential minimum. 
This is due to the effective friction caused by the potential. 
As the motor slows down when dragged along the potential gradient, 
 it spends more time on the up-slope than on the down-slope of the potential.
Such behavior is associated with a negative efficiency. 
In Appendix~\ref{sec:perturb} we have proven this hypothesis on a system with a single head and we have shown that the efficiency decays with the force 
\[
\eta \asymp - \frac{1}{ F_\text{load}^2 } \qquad \text{for} \quad F_\text{load} \to \pm \infty .
\]

\subsubsection{Dependency on number of heads}
\begin{figure}[t]
\centering
\includegraphics[width=0.45\textwidth,height=!]{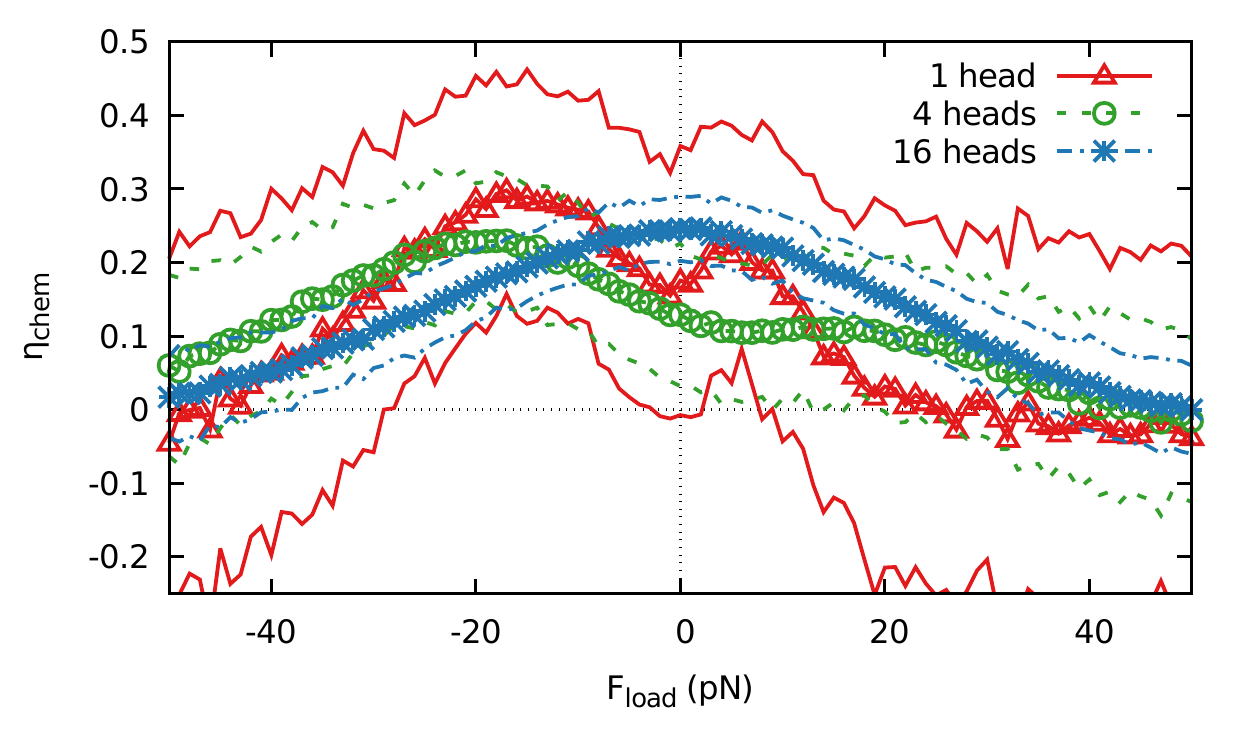}
\caption{
\label{fig:chem_eff_1head}
Chemical efficiency of a motor under load with a variable number of motor heads.
The ``solid'' lines are drawn on the level of one sigma. 
We observe that with an increasing number of heads the efficiency \eqref{eq:efficiency} distribution gets broader and it's variance decreases.
Moreover, for sixteen heads it became centered without two local maxima.
}
\end{figure}

If we interpret the dependency of the chemical efficiency \eqref{eq:eta} on the load as a probabilistic distribution with finite variance and assume that the main contribution to the shape, Fig.~\ref{fig:chem}, is from the numerator, 
by the central limit theorem the efficiency-load dependency approached a Gaussian distribution with increasing number of heads. 
This is confirmed by Fig.~\ref{fig:chem_eff_1head}, where indeed the distribution is broadening with an increasing number of heads and has only a single maximum.
It is also apparent that there is no qualitative difference in chemical efficiency between the system with four heads and one head.
Moreover we can see that the empirical efficiency has a smaller variance with an increasing number of heads. 
From these, we can conclude that there is a significant qualitative difference between motors with low and high numbers of heads.
While the motors with a high number of heads perform better without any external load, motors with a low number of heads perform the best when they are frustrated.
Thus in-vitro experiments with large myosin II motors \cite{brown2009cross-correlated} do not necessarily capture the behavior of the motors inside a cytoskeleton.
That is the first main result of the second part of this study.

\subsubsection{Load on the polymer}
\begin{figure}[t]
\centering
\includegraphics[width=.45\textwidth,height=!]{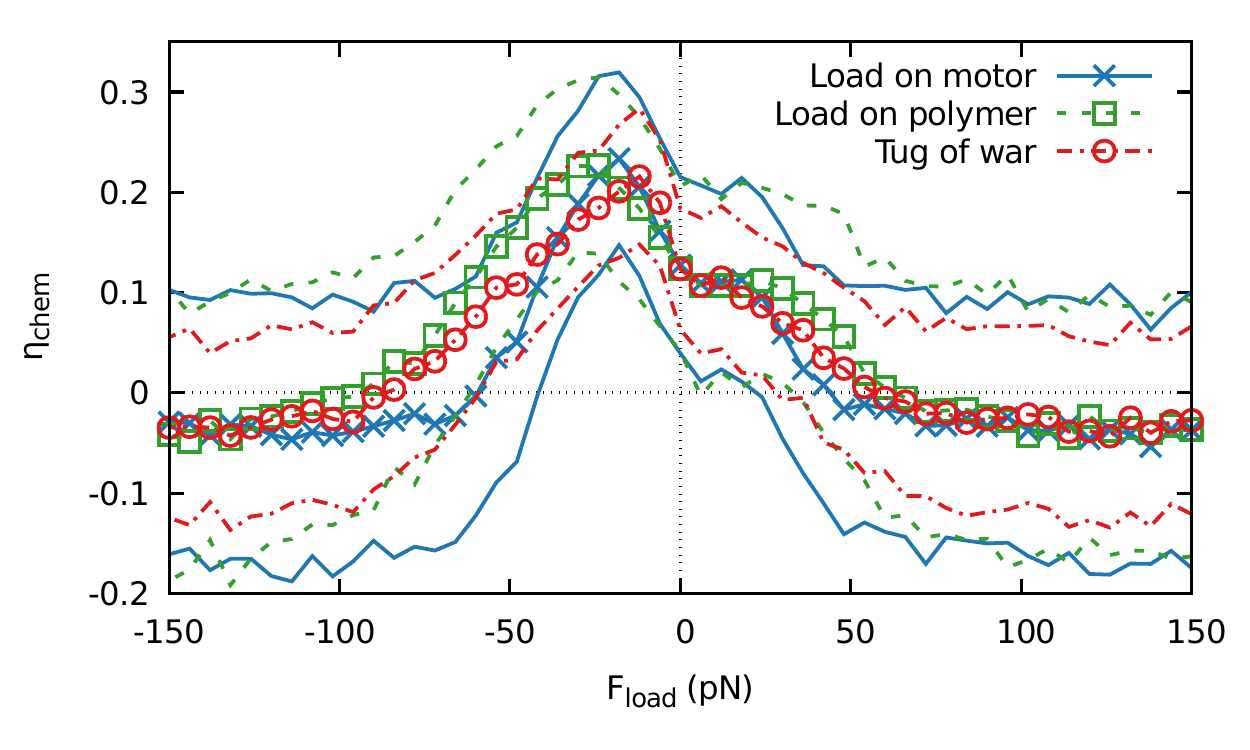}
\caption{\label{fig:load_efficiency_setups}
Comparison of the chemical efficiency~\eqref{eq:eta} as a function of the load $F_\text{load}$ for the ``load on the motor'', ``load on the polymer'' and the ``tug of war'' setups.
We observe that these efficiency profiles differ in the width of the distribution, while it's general shape remains the same.   
}
\end{figure}

\begin{figure}[t]
\centering
\includegraphics[width=.45\textwidth,height=!]{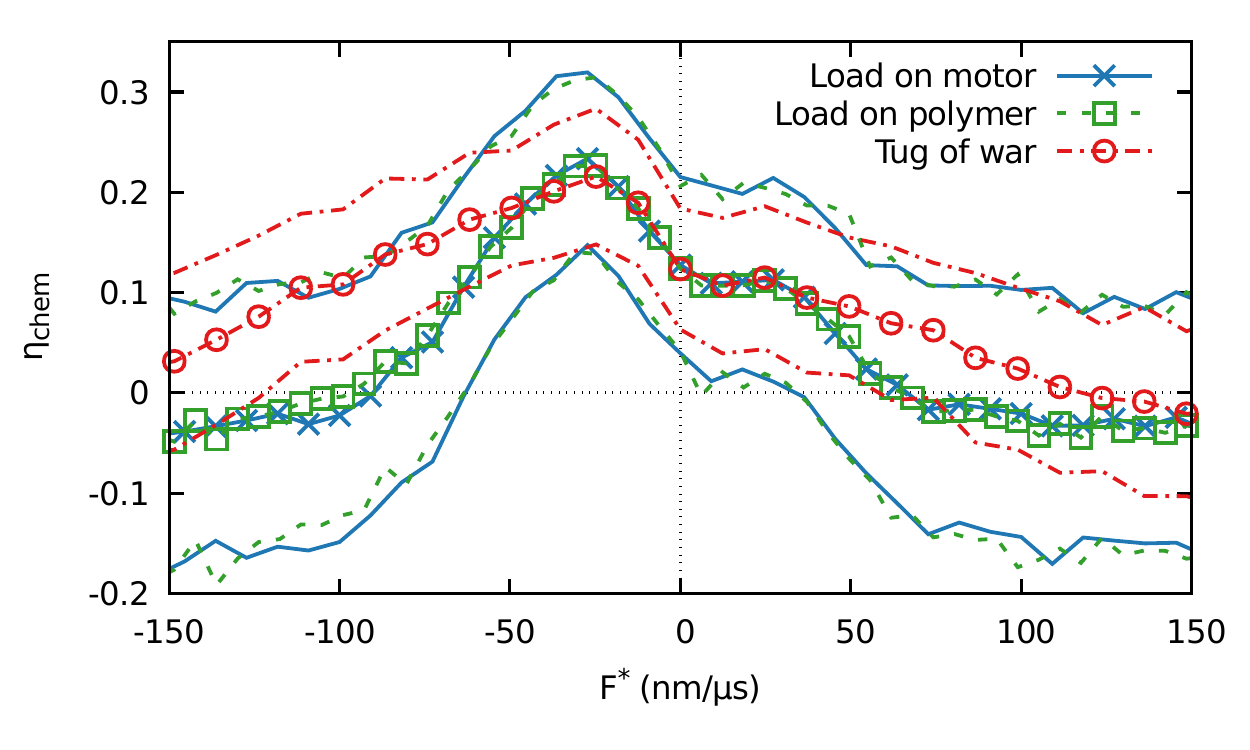}
\caption{\label{fig:load_efficiency_setups_rescaled}
Comparison of the chemical efficiency~\eqref{eq:eta} as a function of the re-scaled load $F^*$ for the ``load on the motor'', ``load on the polymer'' and the ``tug of war'' setups.
For the ``load on the motor'' setup the re-scaled load is $F^* = F_\text{load} / \gamma_\text{M}$, 
while for the ``load on the polymer'' the re-scaled load is defined as $F^* = F_\text{load} / \gamma_\text{A}$.
Finally, for the ``tug of war'' setup we define the re-scaled load as $F^* = 2 F_\text{load} / \gamma_\text{A}$,
where the factor $2$ comes from the fact that the load is applied once on each side of the system.
We observe that these efficiency profiles for ``load on the motor'' and ``load on the polymer'' collapse on each other.    
}
\end{figure}

As was already discussed, the situation with the load on the polymer is almost identical to the ``load on the motor'' setup.
The main difference is in the notion of what we perceive as motor or as polymer, cf. section~\ref{sec:load_on_polymer}.
Hence, it is not surprising that the chemical efficiency \eqref{eq:eta} as a function of the load $F_\text{load}$ 
resembles the situation with the load on the motor, see Fig.~\ref{fig:load_efficiency_setups}. 
The main difference between the curve for the ``load on the motor'' and ``load on the polymer'' setup is the width of the distribution. 
This can be understood from equations~\eqref{eq:chemical_excess} and~\eqref{eq:chemical_excess_poly}, 
where we see that the excess of the chemical energy entering the system is proportional to the relative velocity.
In Fig.~\ref{fig:F_v_setups} we observe a different asymptotic behavior of the relative velocity by a factor $\gamma_\text{M} / \gamma_\text{A}$, Hence we expect that the same factor applies in the steady state for the rate of the excess chemical heat production. 
That hypothesis is further confirmed in Fig.~\ref{fig:load_efficiency_setups_rescaled},
where we plot the chemical efficiency $\eta_\text{chem}$ with respect to re-scaled load. 

\subsubsection{Tug of war}
Fig.~\ref{fig:load_efficiency_setups} and~\ref{fig:load_efficiency_setups_rescaled} also depict the chemical efficiency $\eta$ dependency on the load $F_\text{load}$ for the ``tug of war'' setup, cf. Section~\ref{sec:tug_of_war}.
The general shape is preserved.
From Fig.~\ref{fig:load_efficiency_setups}, we see that the large load asymptotic behavior corresponds to the ``load on the polymer'' setup.
That can be explained by the same asymptotic behavior for the relative velocity $\langle v\rangle$, cf. Fig.~\ref{fig:F_v_setups}, 
and for the tension on the motor $\tau_\text{M}$, cf. Fig~\ref{fig:tug_F}, while having equation~\eqref{eq:excess_heat_tug} in mind.

For small loads, see Fig.~\ref{fig:load_efficiency_setups_rescaled}, the chemical efficiency for the ``tug of war'' setup corresponds to the ``load on polymer'' setup only when the load is scaled by a factor~$2$. 
This points to the fact that, rather than the external load $F_\text{load}$, the mean tension on the motor $\langle \tau_\text{M} \rangle$ generated by the ratchet interaction is important,
which for small loads, when the polymer does not slip, is close to twice the external load. 
We discuss this connection later, see Fig~\ref{fig:tension_efficiency}.

\subsubsection{Elastic environment}
\begin{figure}[t]
\centering
\includegraphics[width=.45\textwidth,height=!]{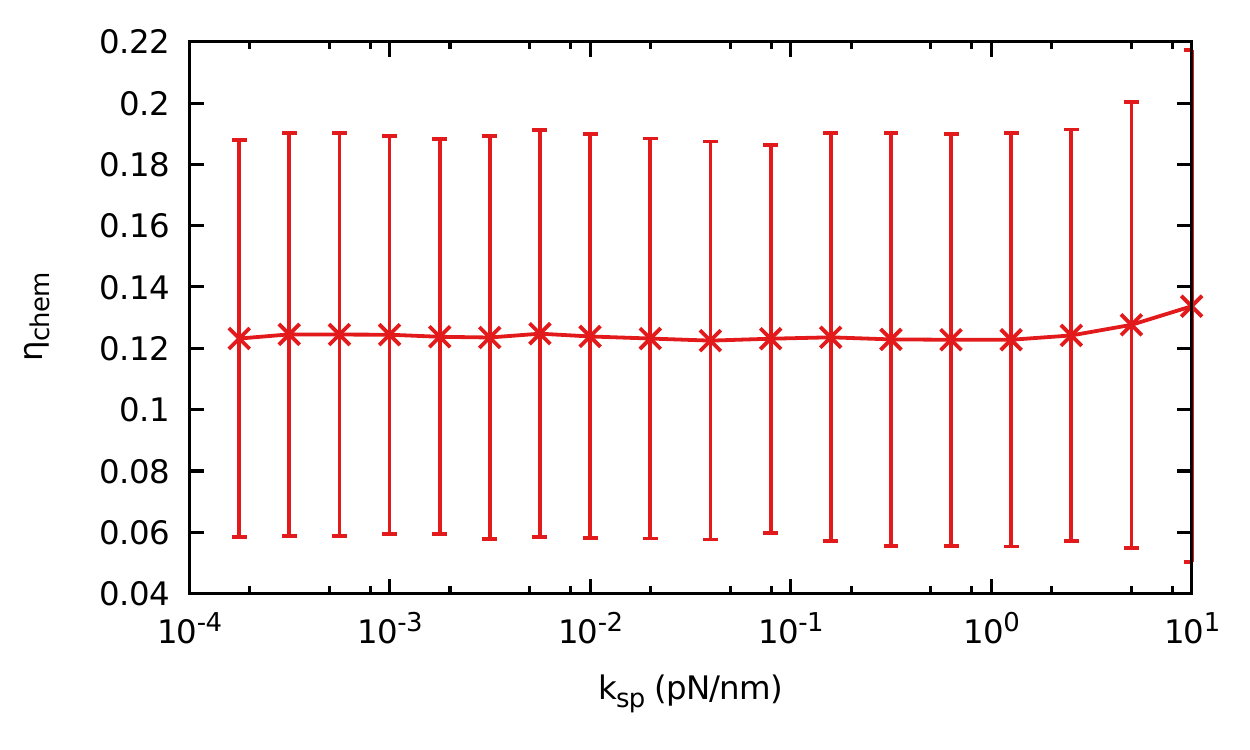}
\caption{\label{fig:stiffness_efficiency}
The chemical efficiency~\eqref{eq:eta} as a function of the spring constant $k_\text{sp}$.
We observe a flat profile.
}
\end{figure}

In the case of the  ``elastic environment'' setup; see Fig.~\ref{fig:stiffness_efficiency}, we observe that the chemical efficiency~\eqref{eq:eta} is practically independent of the stiffness of the environment, $k_\text{sp}$. 
Unfortunately, here we cannot use the same reasoning as for the ``tug of war'' setup, as the mean velocity of the motor in the steady state is zero. 

\subsubsection{Tension and chemical efficiency}
\begin{figure}[t]
\centering
\includegraphics[width=.45\textwidth,height=!]{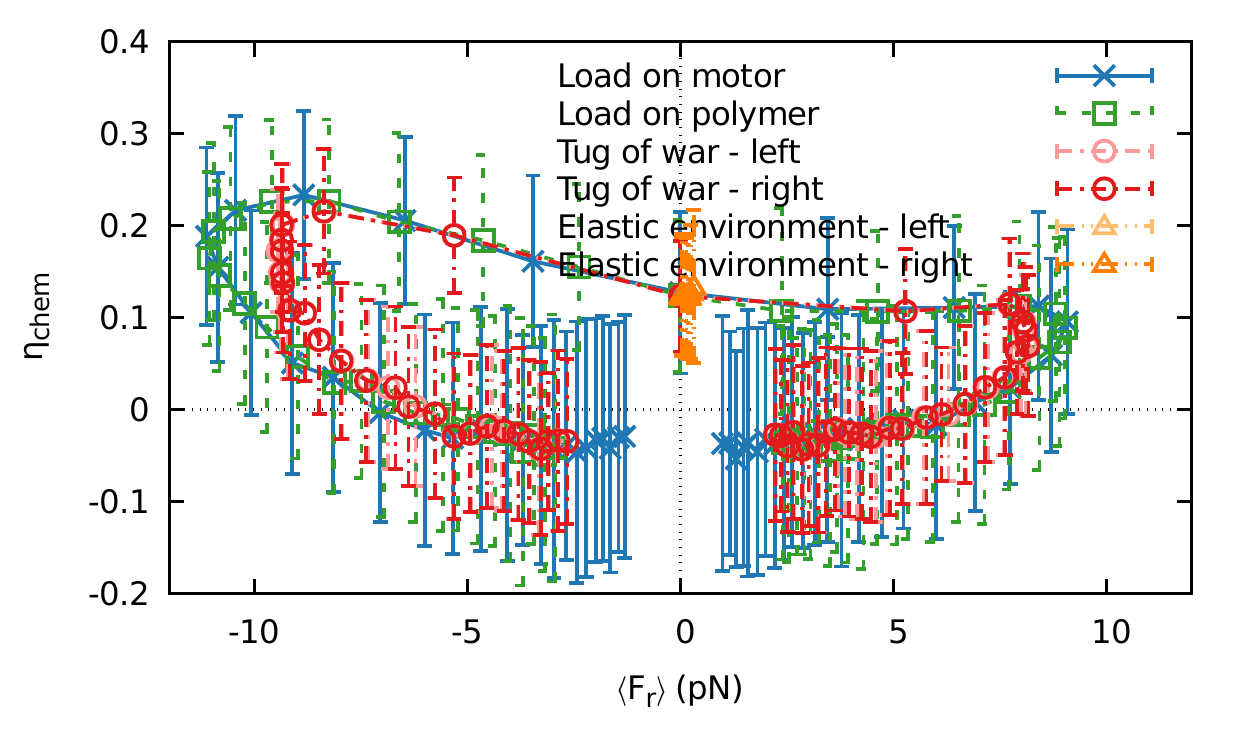}
\caption{\label{fig:ratchet_force_efficiency}
The chemical efficiency $\eta_\text{chem}$ as a function of the mean ratchet force $\langle F_\text{r} \rangle$ (resp. $\langle F_{\text{r}_1} \rangle$ and $-\langle F_{\text{r}_2} \rangle$).
We observe a universal dependency of the efficiency on the mean ratchet force across all setups. 
}
\end{figure}

During the investigation of the chemical efficiency dependency on the external load $F_\text{load}$ for the ``tug of war'' setup, 
we discovered that rather than the load, the tension on the motor $\tau_\text{M}$ might be the intrinsic parameter. 
However, if we take the results of section~\ref{sec:energie} into account, namely equations~\eqref{eq:chemical_excess}, \eqref{eq:chemical_excess_poly} and \eqref{eq:excess_heat_tug}, 
the mean ratchet force~\eqref{eq:ratchet_force} might be a better candidate.
Indeed, as seen in Fig.~\ref{fig:ratchet_force_efficiency}, the chemical efficiency has a universal dependency on the mean ratchet force across all setups. 

For a small mean ratchet force, caused either by a small external load $F_\text{load}$ or a small stiffness of the surrounding network $k_\text{sp}$, 
the chemical efficiency has a relatively flat profile with weak positive response to an increase in tension.
This can be explained by perturbations in the steady state average probability density for the motor's heads 
and was discussed in more details in  section~\ref{sec:chemical_efficiency};
see equations~\eqref{eq:eta} and~\eqref{eq:heat_flux_explicit}, and Figs.~\ref{fig:pos_distr} and~\ref{fig:chem_energy_distr}.  

With increasing load, both positive and negative, the mean ratchet force on the motor in the ``load on the polymer'', ``load on the motor'' and ``tug of war'' setups increase up to a certain threshold.
There, the polymers start to slip with respect to the motor and a further increase in the driving leads only to a decrease in the mean ratchet force and thus to a decrease in the tension~\eqref{eq:tension_load_on_polymer_large} and~\eqref{eq:tension_tug} as well.
This creates a second branch of the dependency for the slipping motor. 
For the ``tug of war'' setup the threshold value is lower than for the ``load on the polymer'' setup due to the interaction mediated by two ratchet potentials, 
as for the complete motor to slip it is sufficient if only one of the ratchet interactions on either side of the motor fails. 

Note that a similar discussion for the ``elastic environment'' setup is highly non-trivial, as shown in section~\ref{sec:power_elastic}.
However, the dependency of the efficiency on the mean ratchet force follows a common pattern and agrees with the results of the other setups. 

\begin{figure}[t]
\centering
\includegraphics[width=.45\textwidth,height=!]{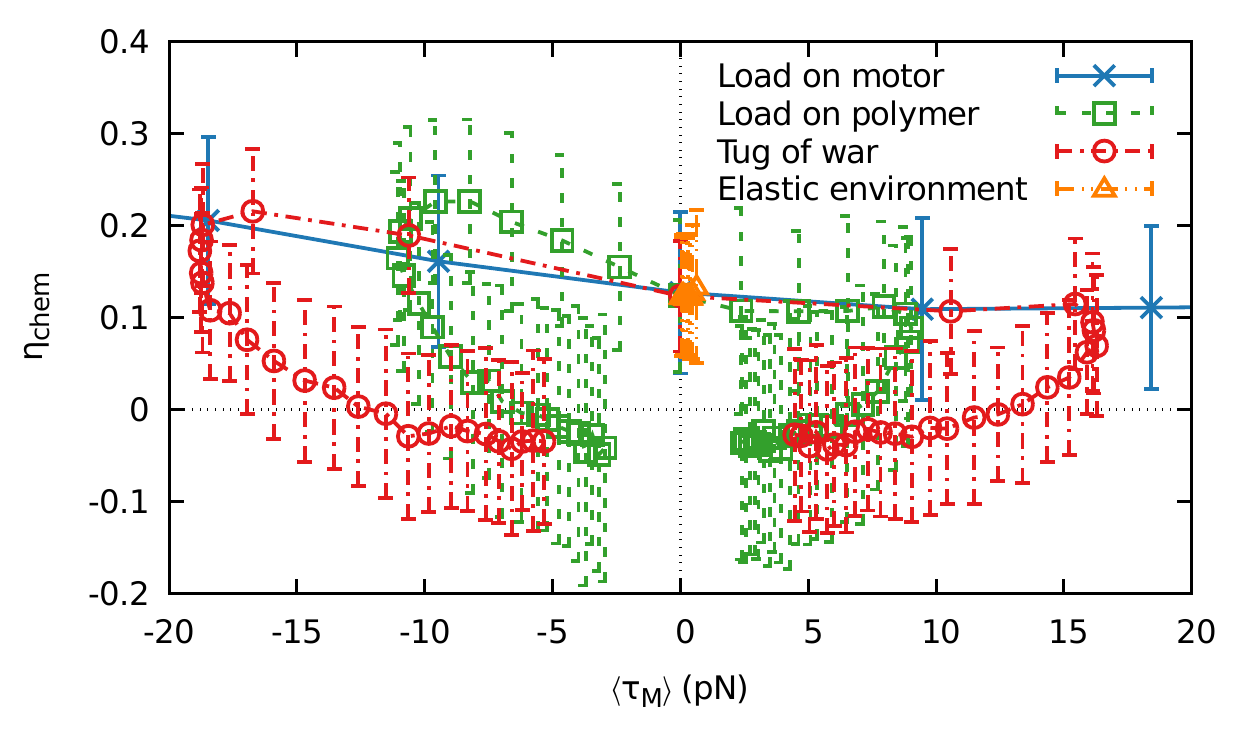}
\caption{\label{fig:tension_efficiency}
The chemical efficiency \eqref{eq:eta} as a function of the tension in the motor $\langle \tau_\text{M} \rangle$.
The efficiencies in all setups agree for small tensions.
The behavior when the motor is not directly loaded is qualitatively very similar.
The discrepancy lies in the different amplitudes of the tension, cf. Fig.~\ref{fig:tug_F}. 
}
\end{figure}

As the mean ratchet force is not directly measurable in comparison with the tension on the motor,
we plot the chemical efficiency as a function of the tension in the motor for all the setups mentioned above;
see Fig.~\ref{fig:tension_efficiency}. 
Here we can see a deviation from a universal behavior, 
most notably for the ``load on the motor'' setup, where there is always a residual tension from the load alone~\eqref{eq:tension_large_load}.
Also note that in the ``tug of war'' setup the threshold value of the tension for motor to slip is larger than in other setups due to two contributions from the mean ratchet force to the tension, one at each respective side of the motor.
These considerations complement the discussion that related the excess chemical energy with the build up tension in the motor, 
or more precisely with the additional tension created by the ratchet interaction, see discussion related to equations~\eqref{eq:excess_heat},~\eqref{eq:chemical_excess_poly} and~\eqref{eq:excess_heat_tug}. 

In the end, we can conclude that the additional tension caused by the ratchet potential, or in other setups representing the tension in the motor itself, is the true intrinsic quantity that determines the chemical efficiency of the molecular motor.

\section{Conclusions}
In this paper we introduced a scaled-ratchet model, which is a variant of a Brownian ratchet.
With this model, we were able to reproduce experimental values for  velocities and stall-forces of the myosin II molecular motor. 
We inspected the dependency of these quantities on the ATP binding rate $\lambda_\text{b}$, which can be linked to the ATP concentration itself, 
and on the scaling factor $c$, that models the screening of the interactions by ionic concentrations in the inter-cellular environment.
We also revisited the force-velocity relation for small non-muscle myosin II motors 
and investigated how it depends on $\lambda_\text{b}$ and on the number of heads of the motor $N$.
From that, we inferred that, from a mechanical point of view, there is hardly any difference between the myosin motor with as little as four heads and a motor with a large number of heads. 

Recently it was found that the main role of myosin II motors is not in active transport, but rather in the generation of tension inside large networks. 
In order to capture the behavior of the motor in such an environment,    
we studied mechanical quantities like the mean ratchet force and the tension in the motor for multiple setups mimicking to various degrees the motor inside such a network.
We observed that for large loads, the motor will slip over the filament track. 
Consequently, the force from the ratchet potential, and the tension generated by it, will vanish.
The decay of the ratchet force corresponds to the theoretical result, obtained by a perturbative expansion around infinitely large loads.

In addition, the motor's ability of mechanosensing naturally emerges from the model without imposing a load-dependence on the chemical rates.
The main result is that all dynamical aspects follow solely from the continuous, mechanical ratchet interaction between the motor and actin filament.

The traditional way of how to evaluate the efficiency of molecular motors is with respect to the efficiency of the transport along the actin filament.
However, such a quantity is unsuitable for motors inside large polymer networks where there is little to no mean displacement of the motor. 
Here we propose an alternative which we call chemical efficiency, that quantifies the fraction of energy, supplied by ATP hydrolysis, 
that is consumed in the diffusion processes of the myosin II motor and actin filament.
This energy is partially used in the movement of the motor along the actin filament or in the build-up of tension in the motor, depending on the setup. 
We have shown that, contrary to the mechanical study, a significant qualitative difference in the behavior of the motor with a high or low number of heads is present. 
The main conclusion is that motors with a small amount of heads perform most efficiently when they are frustrated, a treat which is missing in motors with large number of heads.

We further observed, that for large loads, the chemical efficiency becomes negative, 
which can be interpreted as the energy that is being drained from the external driving to the chemical cycle.
That behavior was confirmed by perturbative calculations, that also predict that the efficiency decays to zero for even bigger loads.

Finally, it seems that this chemical efficiency has a universal dependency on the mean ratchet force, independently of the considered setup,
which is directly related to the tension the motor experiences. 

In summary, we have shown that the scaled ratchet model, introduced in this paper, is naturally capable of covering various biologically relevant treats like the non-linear force velocity relation,  its mechanosensing ability and the fact that the motor's chemical efficiency reaches a maximal value under load. 

\begin{acknowledgments}
The authors are grateful 
to Bart Smeets and Maxim Cuvelier for proofreading of the manuscript.
\end{acknowledgments}

\appendix 
\section{Force -- velocity relation for alternative setups}
\label{sec:force_velocity_other}
\begin{figure}[t]
\centering
\includegraphics[width=0.45\textwidth,height=!]{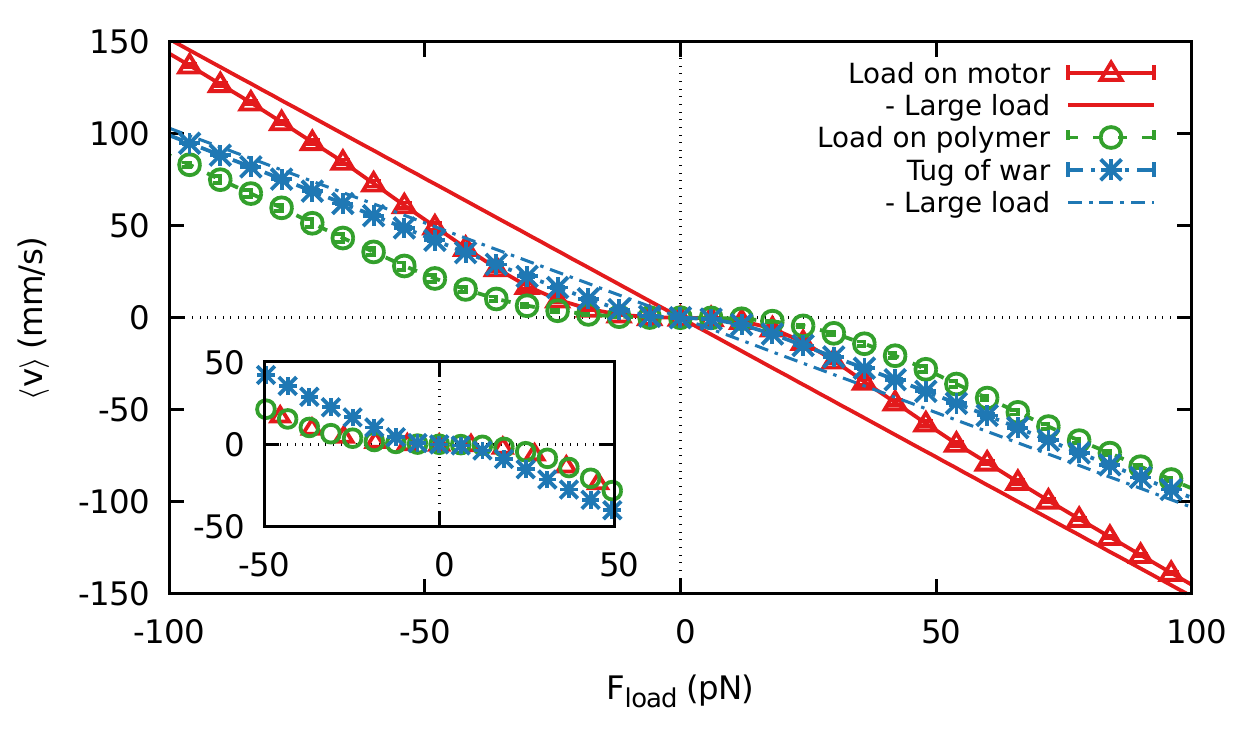}
\caption{
\label{fig:F_v_setups} 
Mean velocity $\langle v \rangle$ of the motor with respect to the filament, as a function of the load force $F_\text{load}$ for different various setups.
For large loads, the curves approach either  $ - F_\text{load} / \gamma_\text{M}$ or $ - F_\text{load} / \gamma_\text{A}$, depending on whether the load is applied to the motor or to the polymer. 
Insert: Force-velocity plotted with rescaled x-axes,
$F_\text{load} \to - F_\text{load} / \gamma_\text{M}$ for the ``load on the motor'' setup 
and $F_\text{load} \to - F_\text{load} / \gamma_\text{A}$ for the ``load on the filament'' and the ``tug of war'' setup. 
}
\end{figure}

In Fig.~\ref{fig:F_v_setups} we compare the force-velocity relation of the ``load on the motor'' setup with alternative setups,
namely with the ``load on the polymer'', cf. Section~\ref{sec:load_on_polymer} and the ``tug of war'', cf. Section~\ref{sec:tug_of_war}.
Note that for the ``tug of war'' setup we have two equivalent choices for the relative velocity,
however due to the symmetry of the setup, cf. Fig.~\ref{fig:tug_F_illustration}, in the steady state these options differ only by sign, 
hence we define the relative velocity~\eqref{eq:mean_velocity} as 
\[
\langle v \rangle 
\equiv \frac{ \left\langle \rmd x_\text{M} \right\rangle }{ \rmd t }
- \frac{ \left\langle \rmd x_{\text{A}_2} \right\rangle }{ \rmd t }
= - \left( \frac{ \left\langle \rmd x_\text{M} \right\rangle }{ \rmd t }
- \frac{ \left\langle \rmd x_{\text{A}_1} \right\rangle }{ \rmd t } \right) .
\]
For small loads, the mean relative velocity $\langle v \rangle$ of the motor with respect to the polymer remains close to each other in all setups.
However, for high loads a clear difference in their asymptotic behavior becomes apparent.
When the load is applied to the motor the relative velocity goes to $ - F_\text{load} / \gamma_\text{M}$.
For the setups where the load is applied to the filaments, the relative velocity goes to $ - F_\text{load} / \gamma_\text{A}$ with increasing load.

In the inset of Fig.~\ref{fig:F_v_setups}, 
the force-velocity relations are plotted on rescaled x-axes for the different setups, 
which takes the difference in their asymptotic behavior into account. 
Thus, the ``load on the motor'' setup is plotted with respect to $ - F_\text{load} / \gamma_\text{M}$ instead of $F_\text{load}$,
while the other setups are plotted with respect to $ - F_\text{load} / \gamma_\text{A}$.
The curve for the simplest setups, with one motor and one polymer, collapse in this plot, 
confirming the fact that they can be mapped onto each other by switching friction coefficients $\gamma_\text{M} \leftrightarrow \gamma_\text{A}$ 
and reversing the relative velocity. 
However, the ``tug of war'' setup manifests a clear difference for mid-range loads due to the more complex interaction with two polymers.

\section{Velocity with respect to medium}
\label{sec:ind_velo}
We have investigated the relative velocity of the myosin and actin polymer \eqref{eq:mean_velocity} in Section~\ref{sec:force-velocity}. 
Here we provide a study whether it is the motor that is moving with respect to the medium or rather the actin polymer that is being pushed back through the medium.
In order to find the dependency between individual velocities and relative velocity in the ``load on the motor'' setup, 
we start by taking expectation values of their equations of motion, cf. equations~\eqref{eq:eom_M} and~\eqref{eq:eom_A},
\begin{align}
\gamma_\text{M} \langle v_\text{M} \rangle &= \frac{ \langle \rmd x_\text{M} \rangle }{\rmd t} = \langle F_\text{r} \rangle - F_\text{load} , 
\label{eq:pre_velocity_M} 
\\
\gamma_\text{A} \langle v_\text{A} \rangle &= \frac{ \langle \rmd x_\text{A} \rangle }{\rmd t} = -\langle F_\text{r} \rangle , 
\label{eq:pre_velocity_A}
\end{align}
where $F_\text{r}$ denote ratchet force~\eqref{eq:ratchet_force}. 
Next, we express the mean relative velocity \eqref{eq:mean_velocity} in terms of the mean ratchet force $\langle F_\text{r} \rangle$ and the external load $F_\text{load}$  
\begin{equation}
\langle v \rangle 
= \langle v_\text{M} - v_\text{A} \rangle 
= \frac{\gamma_\text{A} + \gamma_\text{M}}{\gamma_\text{A} \gamma_\text{M}} \langle F_\text{r} \rangle - \frac{1}{\gamma_\text{M}} F_\text{load} ,
\label{eq:mean_velocity_Fr_Fl}
\end{equation}
which leads to individual velocities expressed in terms of the mean relative velocity and load 
\begin{align}
\langle v_\text{M} \rangle &= \frac{1}{ \gamma_\text{A} + \gamma_\text{M} } \left( \gamma_\text{A} \langle v \rangle - F_\text{load} \right) ,
\label{eq:velocity_M} \\
\langle v_\text{A} \rangle &= -\frac{1}{ \gamma_\text{A} + \gamma_\text{M} } \left( \gamma_\text{M} \langle v \rangle + F_\text{load} \right) .
\label{eq:velocity_A}
\end{align}

Note, that for the load equal the stall force, $F_\text{load} = F_\text{stall}$, both velocities are equal and non-zero.
This is related to the fact that the full system is moving with respect to the liquid, as can be seen from the mean velocity of the geometric center 
\begin{multline*}
\langle v_\text{center} \rangle = \left\langle \frac{ v_\text{M} + v_\text{A} }{2} \right\rangle 
= \\
= \frac{ \gamma_\text{A} - \gamma_\text{M} }{ 2 ( \gamma_\text{A} + \gamma_\text{M} ) } \langle v \rangle - \frac{1}{ \gamma_\text{A} + \gamma_\text{M} } F_\text{load}
, 
\end{multline*}
which is proportional to both the load and the relative velocity and is zero only when $\langle v \rangle = 2 F_\text{load} / ( \gamma_\text{A} - \gamma_\text{M} )$.

\begin{figure}[t]
\centering
\includegraphics[width=0.45\textwidth,height=!]{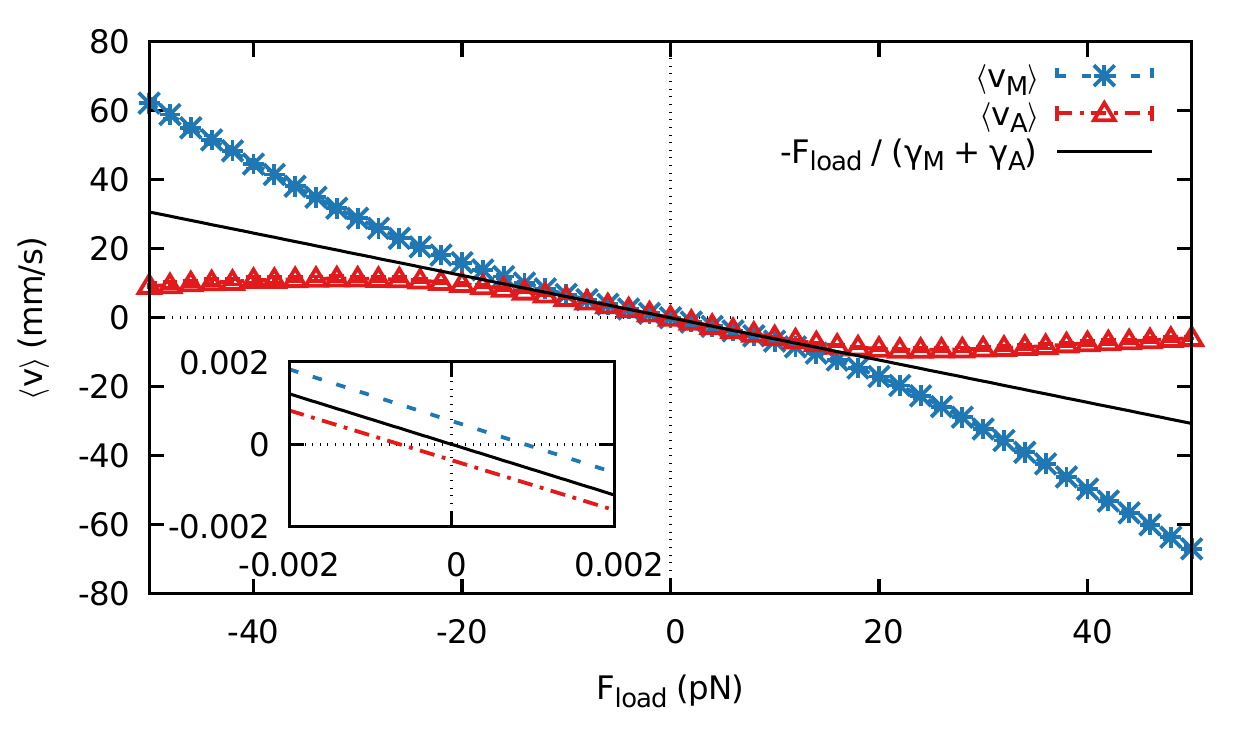}
\caption{
\label{fig:ind_v} 
Mean velocities of the motor and actin filament with respect to an inertial frame of reference for $\lambda_\text{b}/\lambda_\text{u} = 2$.
We observe that only the myosin is moving while the velocity of the dragged actin filament decreases as we increase the load on the myosin motor.
Inset: Zoomed to the linear regime around the origin.
}
\end{figure}

Also note, that the ratchet force is bounded by 
\begin{equation}
- \frac{ N V }{ \ell (1-a) } 
\le 
F_\text{r}
\le
\frac{ N V }{ \ell a } ,
\label{eq:ratchet_force_bounds}
\end{equation}
cf. equations~\eqref{eq:ratchet_potential} and~\eqref{eq:ratchet_interaction}, 
which means that there is a maximal portion of the load that the ratchet interaction can transmit from the motor to the polymer 
and thus there exists an upper and lower bound for the polymer's velocity~\eqref{eq:pre_velocity_A}.   
That is seen in Fig.~\ref{fig:ind_v} depicting the individual velocities as a function of the external load $F_\text{load}$,  
where the polymer's velocity is clearly bounded. 

For small loads, the motor and actin filament are moving together, as the coupling by the ratchet interaction proves to be dominant. 
This hypothesis can be further supported by the linear dependency of equations~\eqref{eq:velocity_M} and~\eqref{eq:velocity_A} on the load, 
which is due to the fact that for small loads also the mean relative velocity \eqref{eq:mean_velocity} is linear in the load; see Fig.~\ref{fig:F_v}. 
Moreover as the mean velocity's response to the external force is typically very small compared to the direct contribution from the external load to the individual velocities,
we can further approximate them by  
\begin{align*}
\left\langle v_\text{M} \right\rangle 
&\approx - \frac{ F_\text{load} }{ \gamma_\text{A} + \gamma_\text{M} } + \frac{ \gamma_\text{A} }{ \gamma_\text{A} + \gamma_\text{M} } \left\langle v \right\rangle_{ F_\text{load} = 0 },
\\
\left\langle v_\text{A} \right\rangle 
&\approx - \frac{ F_\text{load} }{ \gamma_\text{A} + \gamma_\text{M} } - \frac{ \gamma_\text{M} }{ \gamma_\text{A} + \gamma_\text{M} } \left\langle v \right\rangle_{ F_\text{load} = 0 } .
\end{align*}
Such a response is depicted in the insert of Fig.~\ref{fig:ind_v}, where the shift in individual velocities caused by the relative velocity for very small loads is constant.

Since the load is only applied to the motor in this setup, for very high loads, the motor will start slipping over the ratchet potential along the actin filament. 
Consequently, the actin filament's mean velocity will stop increasing and eventually approaches zero, 
while the myosin motor's mean velocity goes to $\langle v_\text{M} \rangle \asymp F_\text{load}/\gamma_\text{M} $
as can be seen in Fig.~\ref{fig:ind_v}. 
This is due to the fact that the asymptotic behavior of the relative velocity  is
\[
\langle v \rangle \asymp - \frac{ F_\text{load} }{\gamma_\text{M} },
\]
which can be obtained directly from equation~\eqref{eq:mean_velocity_Fr_Fl} as the contribution from the ratchet force is bounded.
The asymptotic behavior for the motor and the polymer follows from equations~\eqref{eq:velocity_M} and~\eqref{eq:velocity_A} for $ \left\langle v \right\rangle = - F_\text{load} / \gamma_\text{M} $. 
From there it is also apparent that the relative mean velocity in the limit of infinitely large load $F_\text{load} \to \pm \infty$ will be equal to the motor's velocity $\langle v_\text{M} \rangle$.
That agrees well with the data previously shown in Fig.~\ref{fig:F_v}.

\section{Mean ratchet force}
\label{sec:mean_ratchet_force}
\begin{figure}[t]
\centering
\includegraphics[width=0.45\textwidth,height=!]{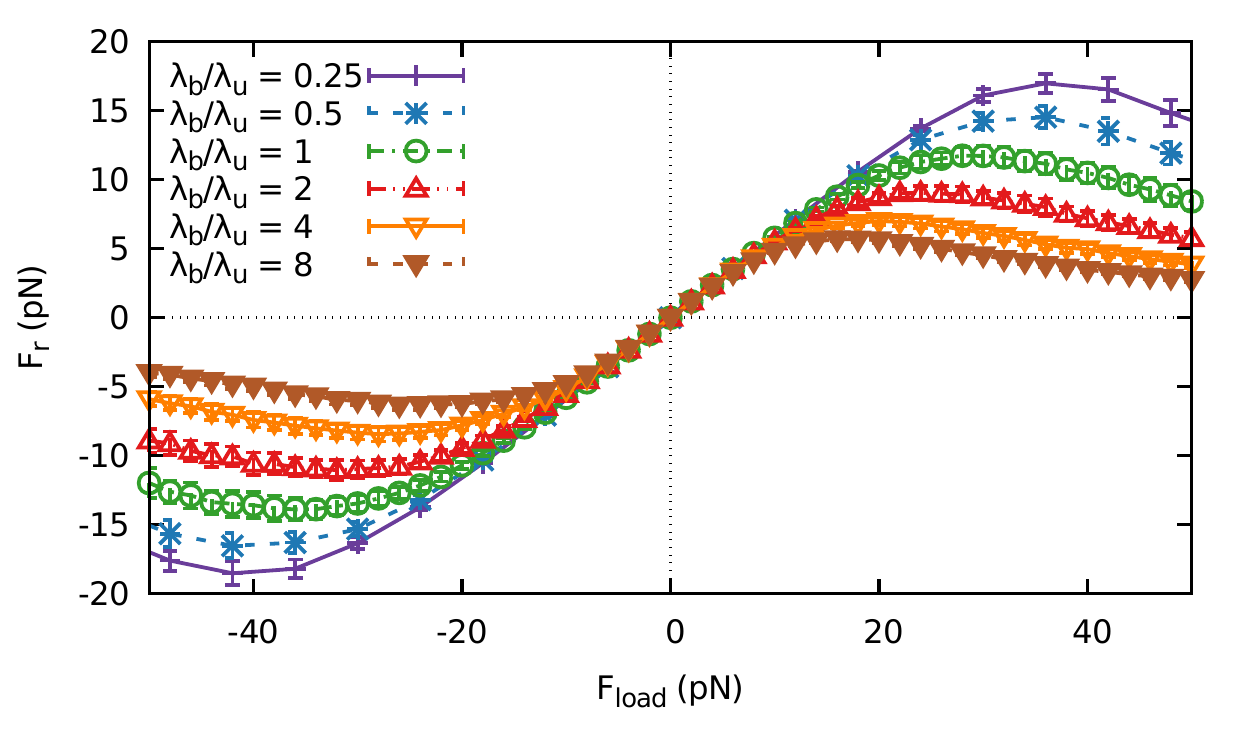}
\caption{
\label{fig:ratchet_force}
Mean force from the ratchet interaction depending on the load force on the motor for several values of $\lambda_\text{b}$. 
For small loads, the ratchet force completely counter to load, hence the linear regime around $F_\text{load}=0$ goes like
}
\end{figure}

\begin{figure}[t]
\centering
\includegraphics[width=0.45\textwidth,height=!]{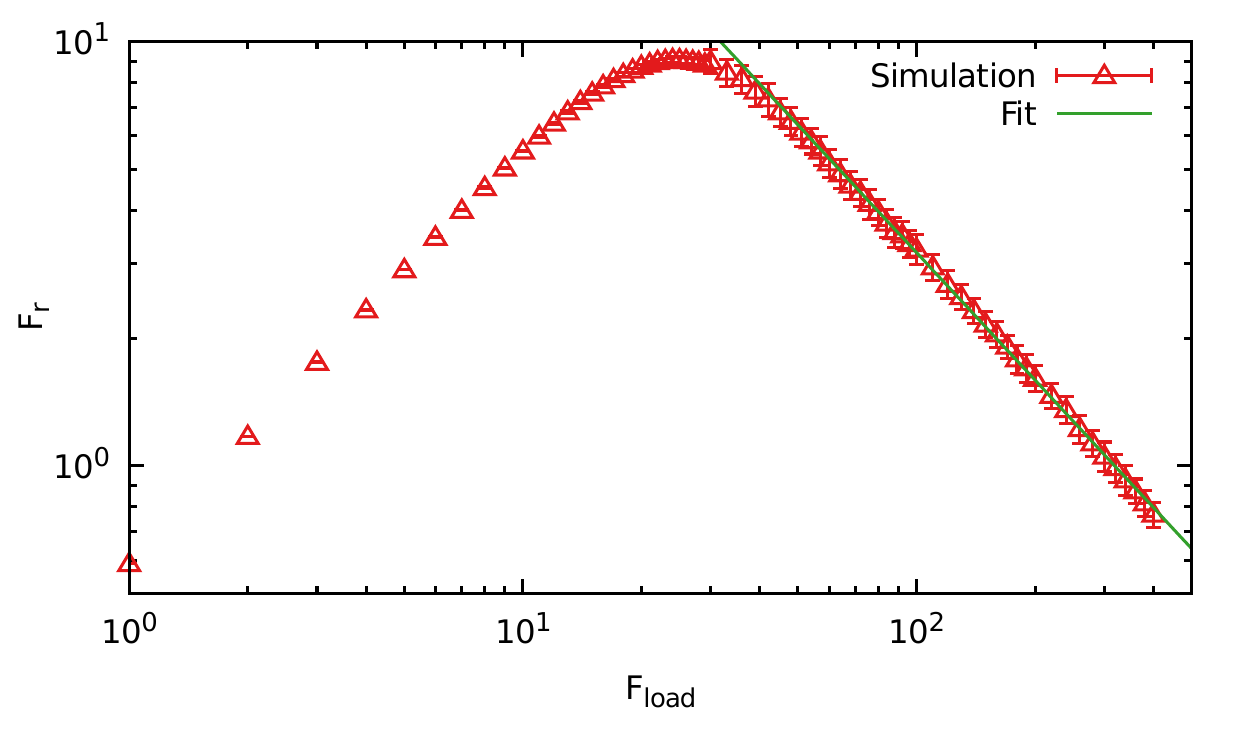}
\caption{
\label{fig:ratchet_force_decay}
The ratchet force, depending on the load, is shown in a logarithmic plot. 
We can see that the ratchet force asymptotically decays according to the power law $F_\text{eff} \propto F_\text{load}^{-\alpha}$, 
with exponent determined from the fit to be $\alpha = 0.9990 \pm 0.0047$, 
which is in good agreement with the theoretical prediction~\eqref{eq:ratchet_vs_load}.
}
\end{figure}

In Fig.~\ref{fig:ratchet_force} we show, 
that the mean ratchet force in the ``load on the polymer'' setup for small loads is linear and independent of the chemical rates and hence of the ATP concentration. 
In this regime the motor hardly moves along the filament and the load is almost completely countered by the mean ratchet force, i.e. it acts as a friction force.
With increasing load, the force generated by the ratchet interaction will reach a maximal value and eventually decay for larger external load forces. 
In appendix~\ref{sec:perturb}, cf. equation~\eqref{eq:mean_velocity_large}, we show that the friction provided by the ratchet will go to zero as the load goes to infinity as 
\begin{equation}
\left| \langle F_\text{r} \rangle_{F_\text{load}\rightarrow\infty} \right| \asymp \frac{1}{F_\text{load}} , 
\label{eq:ratchet_vs_load}
\end{equation}
which is depicted in Fig.~\ref{fig:ratchet_force_decay}. 

Note that while there are natural bounds on the ratchet force due to the shape of the ratchet potential, 
see equation~\eqref{eq:ratchet_force_bounds} in appendix~\ref{sec:ind_velo}, 
here we have shown that the minimal and the maximal ratchet force are comparable to these theoretical bounds. 
Compare this to the maximal tension the motor is able to withstand before it start to slip in section~\ref{sec:tension_motor}.

\section{Local minimum of the chemical efficiency}
\label{sec:local_minimum}
Another trait of the chemical efficiency curves are the local minima around $F_\text{load}=0$. 
This can be understood from equation \eqref{eq:eta}, which is determined by the ratchet potential averaged with the heads' distribution in the attached and detached state. 
For a one headed motor, these distribution will both be peaked around the minimum of the ratchet potential. 
In fact, these distributions will be given by the Boltzmann distribution
\begin{equation*}
\rho_0(x,\zeta) = \frac{1}{Z(\zeta)} e^{-\beta \zeta V_\text{r}(x)}
\end{equation*} 
in the lowest order of approximation, as the binding and unbinding rates operate on much longer time scale than the relaxation governed by the diffusion. 
From this point of view, ATP binding is effectively equivalent to an increasing of the temperature of the heat bath $T\rightarrow\frac{T}{c}$. 
Hence the distributions for the ATP bound state are broader and the expectation value of $V_\text{r}$ will be bigger than in the attached state and $\eta$ will be positive. 

Now if we apply a small external force, those distributions will be slightly perturbed. 
Following the McLennan formula~\cite{Maes2010} we obtain  
\begin{align*}
\rho^*(x) 
&\sim e^{-\beta \left[V_\text{r}(x) + F_\text{load} \left( x - x_\text{min} \right) \right]} \\
&\approx \rho_0(x) \left[1 - \beta F_\text{load} \left( x - x_\text{min} \right) \right]
\end{align*}
close to the potential minimum $x_\text{min}$, i.e. $V(x) \ge V(x_\text{min})$. 
This leads, for small $F_\text{load}$, to a shift of the distribution $\rho$ in the opposite direction of the load
as can be seen from 
\begin{multline}
\left\langle x \right\rangle_{\rho^*} 
= x_\text{min} + \left\langle x - x_\text{min} \right\rangle_{\rho^*}
\\
\approx x_\text{min} + \left\langle x - x_\text{min} \right\rangle_{\rho_0} 
- \beta F_\text{load} \left\langle \left( x - x_\text{min} \right)^2 \right\rangle_{\rho_0} 
\\
= \left\langle x \right\rangle_{\rho_0}
- \beta F_\text{load} \left\langle \left( x - x_\text{min} \right)^2 \right\rangle_{\rho_0} .
\label{eq:mean_position_shift}
\end{multline}
From the previous equation it follows that a more spread-out distribution will be affected more by the force.
This means that in our case the ATP bound state will be more affected.
A larger displacement of the distribution, away from the potential's minimum, 
means that the mean ratchet potential in the ATP bound state $\left\langle V_\text{r} \middle| \zeta = c \right\rangle$ will increase more than the mean ratchet potential in the ATP unbound state $\left\langle V_\text{r} \middle| \zeta = 1 \right\rangle$ and hence the chemical efficiency \eqref{eq:eta} has to increase. 

This argument can be further supported by Jensen's inequality.
The ratchet potential is convex close to its minimum. 
Moreover, we have seen in Fig.~\ref{fig:pos_distr} that the probability distribution for small loads is well localized around it's mean value.
Hence we can concur that the Jensen's inequality will hold for sufficiently small loads  
\begin{equation*}
V_\text{r}(\langle x | \zeta \rangle) \leq \langle V_\text{r}(x) | \zeta \rangle , 
\end{equation*} 
where the lower bound on the expectation value of the ratchet potential~\eqref{eq:ratchet_potential} is given by the value of the ratchet potential at the mean position of the heads in a given state. 
Consequently, any shift in the mean position of the heads due to the external force shifts the lower bound for the mean value of the ratchet potential itself.
Furthermore, in the ATP bound state the mean position will be further away from the minimum of the ratchet potential \eqref{eq:mean_position_shift} so the lower bound is higher than in the ATP unbound state.
Therefore, as the distribution is well localized around the minimum, the shift in the lower bound for the potential represents a shift in the mean value as well. 

For a four headed motor, the distributions of the heads at zero load are not exactly peaked around the minimum of the potential. 
This is due to the fixed distance between each head, which does not correspond to the period of the ratchet potential. 
Therefore the local minimum of $\eta_\text{chem}$ lies not exactly at, but close to, $F_\text{load} =0$; see Fig.~\ref{fig:chem}.

\section{Perturbative treatment}
\label{sec:perturb}
Here we show that the chemical cycle must extract energy from the ratchet potential in the system with one head for large load forces, 
i.e. the chemical efficiency \eqref{eq:eta} must be negative. 
We show this by calculating the perturbative corrections to the probability density of the position of the motor with respect to the actin filament in the regime of large load $F_\text{load} \to \infty$, both in the ATP bound and ATP unbound state. 

First, we assume that the relaxation time of the position in a given potential is much shorter than the typical time between jumps in the chemical state of the heads. 
This effectively leads to time scale separations,
thus the conditional probability distribution \eqref{eq:cond_prob} is given as a steady state solution of the corresponding Fokker-Planck equation in a given state $\zeta$
\begin{multline*}
0 = - \frac{1}{\gamma_\text{M}} \partial_\text{M} \left[ \left( F_\text{load} + \zeta \partial_\text{M} V_\text{r}( x_\text{M} - x_\text{A} ) \right) \mu\left( x_\text{M}, x_\text{A} \right)
\right. \\ \left.
+ k_B T \partial_\text{M} \mu\left( x_\text{M}, x_\text{A} \right) \right] 
\\
+ \frac{1}{\gamma_\text{A}} \partial_\text{A} \left[ 
\zeta \partial_\text{M} V_\text{r} \, \mu( x_\text{M}, x_\text{A} ) 
- k_B T \partial_\text{A} \mu\left( x_\text{M}, x_\text{A} \right) 
\right] .
\end{multline*}
If we change the variables to describe the geometric center $x_C$ and the distance between motor and actin $\Delta x$
\begin{align*}
x_C &= \frac{1}{2} ( x_\text{M} + x_\text{A} ) , \\
\Delta x &= x_\text{M} - x_\text{A}, 
\end{align*}
a more convenient version of the Fokker-Planck is obtained
\begin{align*}
0 &= \begin{multlined}[t][.4\textwidth]
- \partial_\Delta \Biggl[ 
\left( \frac{1}{\gamma_\text{M}} F_\text{load} + \left( \frac{1}{\gamma_\text{M}} + \frac{1}{\gamma_\text{A}} \right) \zeta \partial V_\text{r}( \Delta x ) \right) \mu
\\ 
+ \left( \frac{1}{\gamma_\text{M}} + \frac{1}{\gamma_\text{A}} \right) k_B T \partial_\Delta \mu 
\Biggr] 
\end{multlined}
\\
&- \begin{multlined}[t][.4\textwidth]
\frac{1}{2} \partial_C \Biggl[
\left( \frac{1}{\gamma_\text{M}} F_\text{load} + \left( \frac{1}{\gamma_\text{M}} - \frac{1}{\gamma_\text{A}} \right) \zeta \partial V_\text{r}( \Delta x ) \right) \mu
\\
+ \frac{1}{2} \left( \frac{1}{\gamma_\text{M}} + \frac{1}{\gamma_\text{A}} \right) k_B T \partial_C \mu 
\Biggr] .
\end{multlined}
\end{align*}
If we integrate over all possible position of the center as our focus in this appendix is on the steady state distribution of the relative position of the motor and actin,
which completely determines the mean value of both the ratchet potential and the relative velocity, 
we obtain a reduced Fokker-Planck equation just for the aforementioned displacement ($x \equiv \Delta x$ further on)
\begin{equation}
0 = \frac{1}{\gamma^*} \partial_x \left[ \left( F^* - \zeta \partial_x V_\text{r}(x) \right) \rho(x) - k_B T \partial_x \rho(x) \right] ,
\label{eq:FP_ss}
\end{equation}
where $V_\text{r}$ is the ratchet potential \eqref{eq:ratchet_potential} and 
\begin{equation}
\begin{aligned}[.4\textwidth] 
\gamma^* &= \frac{ \gamma_\text{M} \gamma_\text{A} }{ \gamma_\text{M} + \gamma_\text{A} } , \\ 
F^* &= - \frac{ \gamma_\text{A} }{ \gamma_\text{M} + \gamma_\text{A} } F_\text{load} . 
\end{aligned} 
\label{eq:rescale_parameters}
\end{equation}

The first step is expressing equation \eqref{eq:FP_ss} in terms of $\epsilon = ( F^* )^{-1}$ instead of load force $f_\text{load}$,
which makes it more convenient for the expansion
\begin{equation}
\partial_x \left[ \rho(x) - \epsilon \left( \zeta \partial_x V_\text{r}(x) \rho(x) + k_B T \partial_x \rho(x) \right) \right] = 0 
\label{eq:FP_ss_pert}
\end{equation}
Due to the periodical boundary conditions, the zeroth order approximation correspond to the flat distribution, i.e. $\rho_0 = \frac{1}{\ell}$. 
Note that at the leading order the conditional probability density is independent of the internal state $\zeta$,
thus the chemical efficiency \eqref{eq:eta} is zero. 

For the first order solution we start with an ansatz $\rho_1 = \rho_0 + \delta\rho_1$. 
When inserted into \eqref{eq:FP_ss_pert} we get
\begin{equation*}
\delta\rho_1 = \frac{\epsilon}{\ell} \partial_x V_\text{r} + C
\end{equation*}
where $C$ is an integration constant.
Due to the normalization 
\begin{equation}
1 = \int\limits_0^\ell \rmd x \; \rho_1(x), 
\label{eq:normalization}
\end{equation}
and periodicity of the ratchet potential \eqref{eq:ratchet_potential} the aforementioned integration constant has to be equal to $0$.

Note, that the first order correction does not contribute to calculations of $\langle V_\text{r} | \zeta \rangle_\rho$ 
\begin{multline*}
\langle V_\text{r} | \zeta \rangle_{\delta \rho_1} 
= \int\limits_0^\ell V_\text{r}(x) \, \delta\rho_1(x) \; \rmd x
\\
\propto \int\limits_0^\ell V_\text{r}(x) \, \partial_x V_\text{r}(x) \; \rmd x 
= \frac{1}{2} \int\limits_0^\ell \partial_x V_\text{r}^2(x) \; \rmd x 
= 0
\end{multline*}
where we again used the fact that the ratchet potential $V_\text{r}(x)$ is periodic.

\begin{figure}[t]
\centering
\includegraphics[width=0.45\textwidth,height=!]{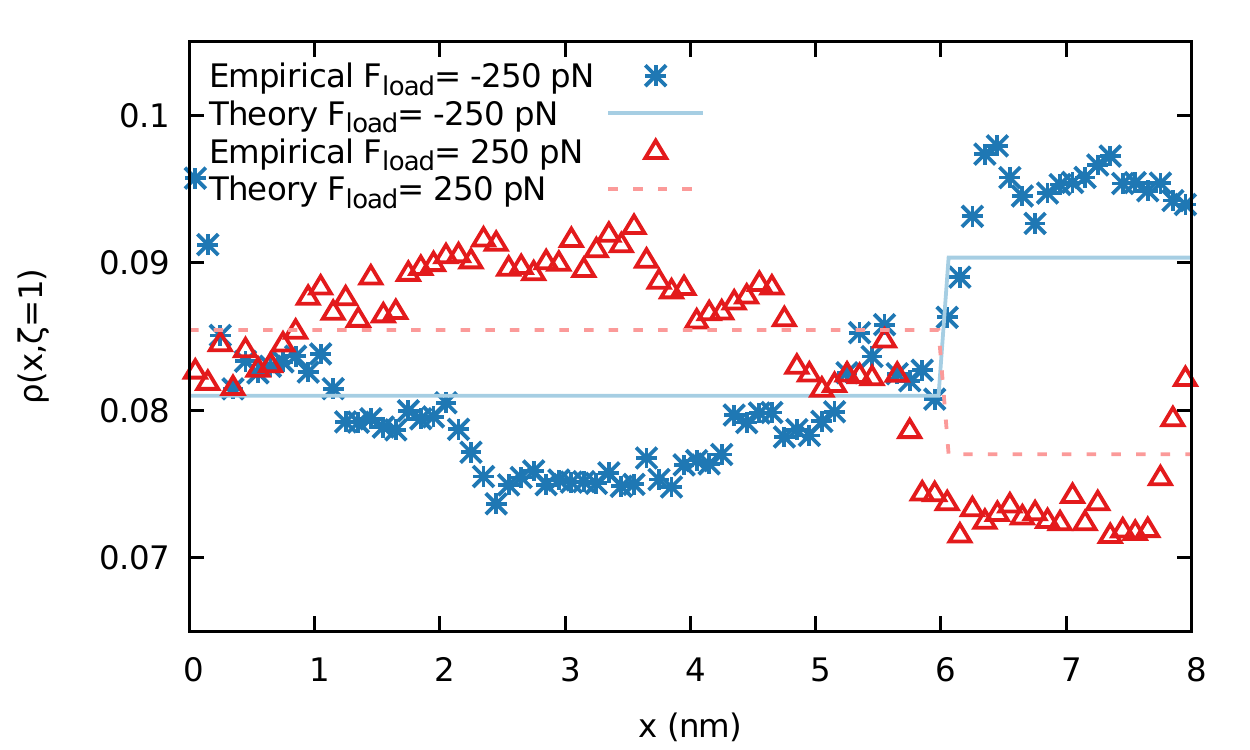}
\caption{\label{fig:prob_density_large}
Empirical probability density for large loads for motor with single head in an ATP unbound state compared with the theoretical prediction \eqref{eq:prob_density_large}.
}
\end{figure}

Consequently, we need to calculate the second order correction $\delta\rho_2$. 
Starting from the ansatz $\rho_2 = \rho_1 + \delta\rho_2$ in \eqref{eq:FP_ss_pert} and keeping up the terms up to quadratic order in $\epsilon$, we find
\begin{equation*}
\delta\rho_2 = \frac{ \epsilon^2 }{ \ell } \left[ \left( \zeta \partial_x V_\text{r} \right)^2 - k_B T \zeta \partial_x^2 V_\text{r} \right] + C^\prime .
\end{equation*}
Again by requiring the normalization~\eqref{eq:normalization}, the integration constant $C^\prime$ can be determined
\begin{multline*}
1 = \int\limits_0^\ell \rho_2(x) \; \rmd x 
\\
= 1  
+ \frac{\epsilon^2}{ \ell } \int\limits_0^\ell \left[ \left( \zeta \partial_x V_\text{r}(x) \right)^2 - k_B T \zeta \partial_x^2 V_\text{r}(x) \right] \; \rmd x 
+ \ell C^\prime ,
\end{multline*}
which leads to
\begin{equation}
C^\prime = - \frac{\epsilon^2 \zeta^2}{ \ell^2 } \int\limits_0^\ell \left( \partial_x V_\text{r}(x) \right)^2 \; \rmd x . 
\end{equation}
Now the density up to second order is given by
\begin{multline}
\rho_2(x)  
= \frac{1}{\ell} \Biggl\{ 1  + \epsilon \zeta \partial_x V_\text{r}(x) \\
+ \epsilon^2 \zeta^2 \left[ \left( \partial_x V_\text{r}(x) \right)^2 - \frac{1}{\ell} \int\limits_0^\ell \left( \partial_y V_\text{r}(y) \right)^2 \; \rmd y \right] 
\\
- \epsilon^2 \zeta k_B T \partial_x^2 V_\text{r}(x)  
\Biggr\} .
\label{eq:prob_density_large}
\end{multline}
Note that up to the quadratic order in $\epsilon$ the conditional probability density depends on the ratchet potential only through a first or second order derivatives. 
Therefore the density will appear as a piece-wise constant function on $x \in [0,\ell]$, with different values for $x < a \ell$ and $x > a \ell$. 
This trend can already be seen in the empirical density obtained from simulations for large load forces $F^* = \pm 250 \, \mathrm{pN}$ in Fig.~\ref{fig:pos_distr},
and is further demonstrated on motor with a single head in Fig.~\ref{fig:prob_density_large}. 

With the second order correction, we can now calculate the change in the expected value of the potential energy
\begin{align*}
\langle V_\text{r} | \zeta \rangle_{\delta\rho_2} 
=& \int\limits_0^\ell V_\text{r}(x) \, \delta\rho_2 \; \rmd x \\
=& \frac{\epsilon^2 \zeta^2}{\ell} \int\limits_0^\ell V_\text{r}(x) \left(\partial_x V_\text{r}(x) \right)^2 \; \rmd x  \\
&- \frac{\epsilon^2 \zeta^2}{\ell^2} \int\limits_0^\ell V_\text{r}(x) \; \rmd x \int\limits_0^\ell \left( \partial_y V_\text{r}(y) \right)^2 \; \rmd y \\
&- \frac{\epsilon^2 \zeta k_B T}{\ell} \int\limits_0^\ell V_\text{r}(x) \, \partial_x^2 V_\text{r}(x) \; \rmd x .
\end{align*}
After tedious but straightforward calculation we found that the first two terms will cancel each other. 
Hence only the last term remains, which after an explicit calculation leads to  
\begin{align*}
\langle V_\text{r} | \zeta = 1 \rangle_{\rho_2} &= \frac{V}{2} + \frac{k_B T}{(F^*)^2} \frac{V^2}{a \left(1-a\right) \ell^2 }, \\
\langle V_\text{r} | \zeta = c \rangle_{\rho_2} &= \frac{V}{2} + \frac{k_B T}{(F^*)^2} \frac{c V^2}{a \left(1-a\right) \ell^2 } .
\end{align*}
By using these expression we get the chemical efficiency~\eqref{eq:eta} up to the second order in $\epsilon$  
\begin{align*}
\eta_\text{chem} 
&= - \frac{ \left(1-c\right)^2 }{ a (1-a) } \, \frac{k_B T} { \Delta E - (1-c) \frac{V}{2} } \left( \frac{V}{\ell F^*} \right)^2 
\\
&= - \frac{ \left(1-c\right)^2 }{ a (1-a) } 
\frac{k_B T} { \Delta E - (1-c) \frac{V}{2} } 
\left[ \frac{ ( \gamma_\text{A} + \gamma_\text{M} ) V }{\gamma_\text{A} \ell F_\text{load}} \right]^2 .
\end{align*}
As was advertised, the zeroth and linear order in $\epsilon$ do not contribute, 
the chemical efficiency decays towards zero as $F_\text{load}^{-2}$ and it is negative for large loads independently of the direction of the load.
The symmetry for of the chemical efficiency $\eta_\text{chem}$ is for large loads confirmed in Fig.~\ref{fig:chem}. 
The fact that the chemical efficiency is negative in this regime, 
means the driven system builds up a chemical energy in the environment, while the external force is performing work on the system. 
Also note that in the expression for the chemical efficiency, 
the ratio between the amplitude of the ratchet potential $V$ and the work done by the external force over a period of the potential $\ell F^*$ appears in the formula, 
as well as the ratio between the thermal energy $k_B T$ and the mean energy gap between the attached and detached state. 

In addition we can compute the steady state probabilistic current in each respective internal state.
The probabilistic current is given by 
\[
\mathcal{J}(x) = \frac{1}{\gamma^*} \left[ \left( F^* - \partial_x V_\text{r}(x) \right) \rho(x) - k_B T \partial_x \rho(x) \right] 
\]
Which simplifies in the second order in the large load regime to
\begin{align*}
\mathcal{J}(x) 
&= \frac{F^*}{\gamma^* \ell} - \frac{\zeta^2}{F^* \gamma^* \ell^2} \int\limits^\ell_0 \left( \partial_x V_\text{r}(x) \right)^2 \; \rmd x \\
&= \frac{F^*}{\gamma^* \ell} - \frac{\zeta^2 V^2}{\gamma^* F^* \ell^3 a (1-a) } ,
\end{align*}
which is independent of the position. 
Note, that the current integrated over the period of the ratchet potential $\ell$, represents the mean velocity of the motor in the given state
\begin{multline}
\left\langle v \middle| \zeta \right\rangle 
= \int\limits_0^\ell \rmd x \; {\mathcal J}(x) 
= \frac{F^*}{\gamma^*} - \frac{\zeta^2 V^2}{\gamma^* F^* \ell^2 a (1-a) } 
\\
= - \frac{F_\text{load}}{\gamma_\text{M}} - \frac{ ( \gamma_\text{M} + \gamma_\text{A} ) \zeta^2 V^2}{\gamma_\text{M} \gamma_\text{A} F_\text{load} \ell^2 a (1-a) } .
\label{eq:mean_velocity_large}
\end{multline}
Indeed, the dominant term for large external forces is $\langle v \rangle_{\rho_0} = - F_\text{load} / \gamma_\text{M}$ independently of the internal state, 
which was already demonstrated in Fig.~\ref{fig:F_v}. 
Furthermore, the force-independent term (i.e. term proportional to $F_\text{load}^0$) vanishes leading to an anti-symmetric form with respect to the load $F_\text{load}$. 
The first correction decay linearly in the external load, which was clearly seen in Fig.~\ref{fig:ratchet_force_decay}, 
which exposes the first correction to the mean velocity, cf. equation~\eqref{eq:mean_ratchet_force}. 
As there is no fundamental difference in the Fokker-Planck equation for one head and for four heads regarding the currents for large driving forces, 
the result obtained for a one-headed motor also holds for the simulations of motors with four heads. 

\bibliography{ratchet}

\begin{thebibliography}{58}%
\makeatletter
\providecommand \@ifxundefined [1]{%
 \@ifx{#1\undefined}
}%
\providecommand \@ifnum [1]{%
 \ifnum #1\expandafter \@firstoftwo
 \else \expandafter \@secondoftwo
 \fi
}%
\providecommand \@ifx [1]{%
 \ifx #1\expandafter \@firstoftwo
 \else \expandafter \@secondoftwo
 \fi
}%
\providecommand \natexlab [1]{#1}%
\providecommand \enquote  [1]{``#1''}%
\providecommand \bibnamefont  [1]{#1}%
\providecommand \bibfnamefont [1]{#1}%
\providecommand \citenamefont [1]{#1}%
\providecommand \href@noop [0]{\@secondoftwo}%
\providecommand \href [0]{\begingroup \@sanitize@url \@href}%
\providecommand \@href[1]{\@@startlink{#1}\@@href}%
\providecommand \@@href[1]{\endgroup#1\@@endlink}%
\providecommand \@sanitize@url [0]{\catcode `\\12\catcode `\$12\catcode
  `\&12\catcode `\#12\catcode `\^12\catcode `\_12\catcode `\%12\relax}%
\providecommand \@@startlink[1]{}%
\providecommand \@@endlink[0]{}%
\providecommand \url  [0]{\begingroup\@sanitize@url \@url }%
\providecommand \@url [1]{\endgroup\@href {#1}{\urlprefix }}%
\providecommand \urlprefix  [0]{URL }%
\providecommand \Eprint [0]{\href }%
\providecommand \doibase [0]{http://dx.doi.org/}%
\providecommand \selectlanguage [0]{\@gobble}%
\providecommand \bibinfo  [0]{\@secondoftwo}%
\providecommand \bibfield  [0]{\@secondoftwo}%
\providecommand \translation [1]{[#1]}%
\providecommand \BibitemOpen [0]{}%
\providecommand \bibitemStop [0]{}%
\providecommand \bibitemNoStop [0]{.\EOS\space}%
\providecommand \EOS [0]{\spacefactor3000\relax}%
\providecommand \BibitemShut  [1]{\csname bibitem#1\endcsname}%
\let\auto@bib@innerbib\@empty
\bibitem [{\citenamefont {Mitchison}\ and\ \citenamefont
  {Cramer}(1996)}]{mitchison1996actin}%
  \BibitemOpen
  \bibfield  {author} {\bibinfo {author} {\bibfnamefont {T.}~\bibnamefont
  {Mitchison}}\ and\ \bibinfo {author} {\bibfnamefont {L.~P.}\ \bibnamefont
  {Cramer}},\ }\href {\doibase 10.1016/S0092-8674(00)81281-7} {\bibfield
  {journal} {\bibinfo  {journal} {Cell}\ }\textbf {\bibinfo {volume} {84}},\
  \bibinfo {pages} {371} (\bibinfo {year} {1996})}\BibitemShut {NoStop}%
\bibitem [{\citenamefont {Pollard}(1982)}]{pollard1982structure}%
  \BibitemOpen
  \bibfield  {author} {\bibinfo {author} {\bibfnamefont {T.~D.}\ \bibnamefont
  {Pollard}},\ }\href {\doibase 10.1083/jcb.95.3.816} {\bibfield  {journal}
  {\bibinfo  {journal} {The Journal of Cell Biology}\ }\textbf {\bibinfo
  {volume} {95}},\ \bibinfo {pages} {816} (\bibinfo {year} {1982})}\BibitemShut
  {NoStop}%
\bibitem [{\citenamefont {Reimann}\ and\ \citenamefont
  {H{\"{a}}nggi}(2002)}]{Reimann2002introduction}%
  \BibitemOpen
  \bibfield  {author} {\bibinfo {author} {\bibfnamefont {P.}~\bibnamefont
  {Reimann}}\ and\ \bibinfo {author} {\bibfnamefont {P.}~\bibnamefont
  {H{\"{a}}nggi}},\ }\href {\doibase 10.1007/s003390201331} {\bibfield
  {journal} {\bibinfo  {journal} {Applied Physics A}\ }\textbf {\bibinfo
  {volume} {75}},\ \bibinfo {pages} {169} (\bibinfo {year} {2002})}\BibitemShut
  {NoStop}%
\bibitem [{\citenamefont {Ross}\ \emph {et~al.}(2008)\citenamefont {Ross},
  \citenamefont {Ali},\ and\ \citenamefont {Warshaw}}]{ross2008cargo}%
  \BibitemOpen
  \bibfield  {author} {\bibinfo {author} {\bibfnamefont {J.~L.}\ \bibnamefont
  {Ross}}, \bibinfo {author} {\bibfnamefont {M.~Y.}\ \bibnamefont {Ali}}, \
  and\ \bibinfo {author} {\bibfnamefont {D.~M.}\ \bibnamefont {Warshaw}},\
  }\href {\doibase 10.1016/j.ceb.2007.11.006} {\bibfield  {journal} {\bibinfo
  {journal} {Current Opinion in Cell Biology}\ }\textbf {\bibinfo {volume}
  {20}},\ \bibinfo {pages} {41} (\bibinfo {year} {2008})}\BibitemShut {NoStop}%
\bibitem [{\citenamefont {Rosenblatt}\ \emph {et~al.}(2004)\citenamefont
  {Rosenblatt}, \citenamefont {Cramer}, \citenamefont {Baum},\ and\
  \citenamefont {McGee}}]{rosenblatt2004myosin}%
  \BibitemOpen
  \bibfield  {author} {\bibinfo {author} {\bibfnamefont {J.}~\bibnamefont
  {Rosenblatt}}, \bibinfo {author} {\bibfnamefont {L.~P.}\ \bibnamefont
  {Cramer}}, \bibinfo {author} {\bibfnamefont {B.}~\bibnamefont {Baum}}, \ and\
  \bibinfo {author} {\bibfnamefont {K.~M.}\ \bibnamefont {McGee}},\ }\href
  {\doibase 10.1016/S0092-8674(04)00341-1} {\bibfield  {journal} {\bibinfo
  {journal} {Cell}\ }\textbf {\bibinfo {volume} {117}},\ \bibinfo {pages} {361}
  (\bibinfo {year} {2004})}\BibitemShut {NoStop}%
\bibitem [{\citenamefont {Vicente-Manzanares}\ \emph
  {et~al.}(2009)\citenamefont {Vicente-Manzanares}, \citenamefont {Ma},
  \citenamefont {Adelstein},\ and\ \citenamefont {Horwitz}}]{vicente2009non}%
  \BibitemOpen
  \bibfield  {author} {\bibinfo {author} {\bibfnamefont {M.}~\bibnamefont
  {Vicente-Manzanares}}, \bibinfo {author} {\bibfnamefont {X.}~\bibnamefont
  {Ma}}, \bibinfo {author} {\bibfnamefont {R.~S.}\ \bibnamefont {Adelstein}}, \
  and\ \bibinfo {author} {\bibfnamefont {A.~R.}\ \bibnamefont {Horwitz}},\
  }\href {\doibase 10.1038/nrm2786} {\bibfield  {journal} {\bibinfo  {journal}
  {Nature Reviews Molecular Cell Biology}\ }\textbf {\bibinfo {volume} {10}},\
  \bibinfo {pages} {778} (\bibinfo {year} {2009})}\BibitemShut {NoStop}%
\bibitem [{\citenamefont {Ma}\ \emph {et~al.}(2012)\citenamefont {Ma},
  \citenamefont {Kovacs}, \citenamefont {Conti}, \citenamefont {Wang},
  \citenamefont {Zhang}, \citenamefont {Sellers},\ and\ \citenamefont
  {Adelstein}}]{ma2012nonmuscle}%
  \BibitemOpen
  \bibfield  {author} {\bibinfo {author} {\bibfnamefont {X.}~\bibnamefont
  {Ma}}, \bibinfo {author} {\bibfnamefont {M.}~\bibnamefont {Kovacs}}, \bibinfo
  {author} {\bibfnamefont {M.~A.}\ \bibnamefont {Conti}}, \bibinfo {author}
  {\bibfnamefont {A.}~\bibnamefont {Wang}}, \bibinfo {author} {\bibfnamefont
  {Y.}~\bibnamefont {Zhang}}, \bibinfo {author} {\bibfnamefont {J.~R.}\
  \bibnamefont {Sellers}}, \ and\ \bibinfo {author} {\bibfnamefont {R.~S.}\
  \bibnamefont {Adelstein}},\ }\href {\doibase 10.1073/pnas.1116268109}
  {\bibfield  {journal} {\bibinfo  {journal} {Proceedings of the National
  Academy of Sciences}\ }\textbf {\bibinfo {volume} {109}},\ \bibinfo {pages}
  {4509} (\bibinfo {year} {2012})}\BibitemShut {NoStop}%
\bibitem [{\citenamefont {Chugh}\ \emph {et~al.}(2017)\citenamefont {Chugh},
  \citenamefont {Clark}, \citenamefont {Smith}, \citenamefont {Cassani},
  \citenamefont {Dierkes}, \citenamefont {Ragab}, \citenamefont {Roux},
  \citenamefont {Charras}, \citenamefont {Salbreux},\ and\ \citenamefont
  {Paluch}}]{chugh2017actin}%
  \BibitemOpen
  \bibfield  {author} {\bibinfo {author} {\bibfnamefont {P.}~\bibnamefont
  {Chugh}}, \bibinfo {author} {\bibfnamefont {A.~G.}\ \bibnamefont {Clark}},
  \bibinfo {author} {\bibfnamefont {M.~B.}\ \bibnamefont {Smith}}, \bibinfo
  {author} {\bibfnamefont {D.~A.~D.}\ \bibnamefont {Cassani}}, \bibinfo
  {author} {\bibfnamefont {K.}~\bibnamefont {Dierkes}}, \bibinfo {author}
  {\bibfnamefont {A.}~\bibnamefont {Ragab}}, \bibinfo {author} {\bibfnamefont
  {P.~P.}\ \bibnamefont {Roux}}, \bibinfo {author} {\bibfnamefont
  {G.}~\bibnamefont {Charras}}, \bibinfo {author} {\bibfnamefont
  {G.}~\bibnamefont {Salbreux}}, \ and\ \bibinfo {author} {\bibfnamefont
  {E.~K.}\ \bibnamefont {Paluch}},\ }\href {\doibase 10.1038/ncb3525}
  {\bibfield  {journal} {\bibinfo  {journal} {Nature Cell Biology}\ }\textbf
  {\bibinfo {volume} {19}},\ \bibinfo {pages} {689} (\bibinfo {year}
  {2017})}\BibitemShut {NoStop}%
\bibitem [{\citenamefont {Monier}\ \emph {et~al.}(2010)\citenamefont {Monier},
  \citenamefont {P{\'{e}}lissier-Monier}, \citenamefont {Brand},\ and\
  \citenamefont {Sanson}}]{monier2010actomyosin}%
  \BibitemOpen
  \bibfield  {author} {\bibinfo {author} {\bibfnamefont {B.}~\bibnamefont
  {Monier}}, \bibinfo {author} {\bibfnamefont {A.}~\bibnamefont
  {P{\'{e}}lissier-Monier}}, \bibinfo {author} {\bibfnamefont {A.~H.}\
  \bibnamefont {Brand}}, \ and\ \bibinfo {author} {\bibfnamefont
  {B.}~\bibnamefont {Sanson}},\ }\href {\doibase 10.1038/ncb2005} {\bibfield
  {journal} {\bibinfo  {journal} {Nature Cell Biology}\ }\textbf {\bibinfo
  {volume} {12}},\ \bibinfo {pages} {60} (\bibinfo {year} {2010})}\BibitemShut
  {NoStop}%
\bibitem [{\citenamefont {Reimann}(2002)}]{reimann2002brownian}%
  \BibitemOpen
  \bibfield  {author} {\bibinfo {author} {\bibfnamefont {P.}~\bibnamefont
  {Reimann}},\ }\href {\doibase 10.1016/S0370-1573(01)00081-3} {\bibfield
  {journal} {\bibinfo  {journal} {Physics Reports}\ }\textbf {\bibinfo {volume}
  {361}},\ \bibinfo {pages} {57} (\bibinfo {year} {2002})},\ \Eprint
  {http://arxiv.org/abs/0010237} {arXiv:0010237 [cond-mat]} \BibitemShut
  {NoStop}%
\bibitem [{\citenamefont {Yogurtcu}\ \emph {et~al.}(2012)\citenamefont
  {Yogurtcu}, \citenamefont {Kim},\ and\ \citenamefont
  {Sun}}]{yogurtcu2012mechanochemical}%
  \BibitemOpen
  \bibfield  {author} {\bibinfo {author} {\bibfnamefont {O.~N.}\ \bibnamefont
  {Yogurtcu}}, \bibinfo {author} {\bibfnamefont {J.~S.}\ \bibnamefont {Kim}}, \
  and\ \bibinfo {author} {\bibfnamefont {S.~X.}\ \bibnamefont {Sun}},\ }\href
  {\doibase 10.1016/j.bpj.2012.07.020} {\bibfield  {journal} {\bibinfo
  {journal} {Biophysical Journal}\ }\textbf {\bibinfo {volume} {103}},\
  \bibinfo {pages} {719} (\bibinfo {year} {2012})}\BibitemShut {NoStop}%
\bibitem [{\citenamefont {Nie}\ \emph {et~al.}(2014{\natexlab{a}})\citenamefont
  {Nie}, \citenamefont {Togashi}, \citenamefont {Sasaki}, \citenamefont
  {Takano}, \citenamefont {Sasai},\ and\ \citenamefont {Terada}}]{Nie2014}%
  \BibitemOpen
  \bibfield  {author} {\bibinfo {author} {\bibfnamefont {Q.-M.}\ \bibnamefont
  {Nie}}, \bibinfo {author} {\bibfnamefont {A.}~\bibnamefont {Togashi}},
  \bibinfo {author} {\bibfnamefont {T.~N.}\ \bibnamefont {Sasaki}}, \bibinfo
  {author} {\bibfnamefont {M.}~\bibnamefont {Takano}}, \bibinfo {author}
  {\bibfnamefont {M.}~\bibnamefont {Sasai}}, \ and\ \bibinfo {author}
  {\bibfnamefont {T.~P.}\ \bibnamefont {Terada}},\ }\href {\doibase
  10.1371/journal.pcbi.1003552} {\bibfield  {journal} {\bibinfo  {journal}
  {PLoS Computational Biology}\ }\textbf {\bibinfo {volume} {10}},\ \bibinfo
  {pages} {e1003552} (\bibinfo {year} {2014}{\natexlab{a}})}\BibitemShut
  {NoStop}%
\bibitem [{\citenamefont {Nie}\ \emph {et~al.}(2014{\natexlab{b}})\citenamefont
  {Nie}, \citenamefont {Sasai},\ and\ \citenamefont
  {Terada}}]{nie2014conformational}%
  \BibitemOpen
  \bibfield  {author} {\bibinfo {author} {\bibfnamefont {Q.-M.}\ \bibnamefont
  {Nie}}, \bibinfo {author} {\bibfnamefont {M.}~\bibnamefont {Sasai}}, \ and\
  \bibinfo {author} {\bibfnamefont {T.~P.}\ \bibnamefont {Terada}},\ }\href
  {\doibase 10.1039/c3cp54464h} {\bibfield  {journal} {\bibinfo  {journal}
  {Physical Chemistry Chemical Physics}\ }\textbf {\bibinfo {volume} {16}},\
  \bibinfo {pages} {6441} (\bibinfo {year} {2014}{\natexlab{b}})}\BibitemShut
  {NoStop}%
\bibitem [{\citenamefont {Barterls}\ \emph {et~al.}(1993)\citenamefont
  {Barterls}, \citenamefont {Cooke}, \citenamefont {Elliott},\ and\
  \citenamefont {Hughes}}]{barterls1993myosin}%
  \BibitemOpen
  \bibfield  {author} {\bibinfo {author} {\bibfnamefont {E.~M.}\ \bibnamefont
  {Barterls}}, \bibinfo {author} {\bibfnamefont {P.~H.}\ \bibnamefont {Cooke}},
  \bibinfo {author} {\bibfnamefont {G.~F.}\ \bibnamefont {Elliott}}, \ and\
  \bibinfo {author} {\bibfnamefont {R.~A.}\ \bibnamefont {Hughes}},\ }\href
  {\doibase 10.1016/0304-4165(93)90079-N} {\bibfield  {journal} {\bibinfo
  {journal} {Biochimica et Biophysica Acta (BBA) - General Subjects}\ }\textbf
  {\bibinfo {volume} {1157}},\ \bibinfo {pages} {63} (\bibinfo {year}
  {1993})}\BibitemShut {NoStop}%
\bibitem [{\citenamefont {Adelstein}\ and\ \citenamefont
  {Eisenberg}(1980)}]{adelstein1980regulation}%
  \BibitemOpen
  \bibfield  {author} {\bibinfo {author} {\bibfnamefont {R.~S.}\ \bibnamefont
  {Adelstein}}\ and\ \bibinfo {author} {\bibfnamefont {E.}~\bibnamefont
  {Eisenberg}},\ }\href {\doibase 10.1146/annurev.bi.49.070180.004421}
  {\bibfield  {journal} {\bibinfo  {journal} {Annual Review of Biochemistry}\
  }\textbf {\bibinfo {volume} {49}},\ \bibinfo {pages} {921} (\bibinfo {year}
  {1980})}\BibitemShut {NoStop}%
\bibitem [{\citenamefont {Gajewski}\ \emph {et~al.}(1986)\citenamefont
  {Gajewski}, \citenamefont {Steckler},\ and\ \citenamefont
  {Goldberg}}]{gajewski1986thermodynamics}%
  \BibitemOpen
  \bibfield  {author} {\bibinfo {author} {\bibfnamefont {E.}~\bibnamefont
  {Gajewski}}, \bibinfo {author} {\bibfnamefont {D.~K.}\ \bibnamefont
  {Steckler}}, \ and\ \bibinfo {author} {\bibfnamefont {R.~N.}\ \bibnamefont
  {Goldberg}},\ }\href {http://www.ncbi.nlm.nih.gov/pubmed/3528161} {\bibfield
  {journal} {\bibinfo  {journal} {The Journal of biological chemistry}\
  }\textbf {\bibinfo {volume} {261}},\ \bibinfo {pages} {12733} (\bibinfo
  {year} {1986})}\BibitemShut {NoStop}%
\bibitem [{\citenamefont {Hoffmann}(2016)}]{hoffmann2016molecular}%
  \BibitemOpen
  \bibfield  {author} {\bibinfo {author} {\bibfnamefont {P.~M.}\ \bibnamefont
  {Hoffmann}},\ }\href {\doibase 10.1088/0034-4885/79/3/032601} {\bibfield
  {journal} {\bibinfo  {journal} {Reports on Progress in Physics}\ }\textbf
  {\bibinfo {volume} {79}},\ \bibinfo {pages} {032601} (\bibinfo {year}
  {2016})}\BibitemShut {NoStop}%
\bibitem [{\citenamefont {{De Roeck}}\ and\ \citenamefont
  {Maes}(2007)}]{de2007symmetries}%
  \BibitemOpen
  \bibfield  {author} {\bibinfo {author} {\bibfnamefont {W.}~\bibnamefont {{De
  Roeck}}}\ and\ \bibinfo {author} {\bibfnamefont {C.}~\bibnamefont {Maes}},\
  }\href {\doibase 10.1103/PhysRevE.76.051117} {\bibfield  {journal} {\bibinfo
  {journal} {Physical Review E}\ }\textbf {\bibinfo {volume} {76}},\ \bibinfo
  {pages} {051117} (\bibinfo {year} {2007})},\ \Eprint
  {http://arxiv.org/abs/0610369} {arXiv:0610369 [cond-mat]} \BibitemShut
  {NoStop}%
\bibitem [{\citenamefont {Astumian}\ and\ \citenamefont
  {Bier}(1994)}]{astumian1994fluctuation}%
  \BibitemOpen
  \bibfield  {author} {\bibinfo {author} {\bibfnamefont {R.~D.}\ \bibnamefont
  {Astumian}}\ and\ \bibinfo {author} {\bibfnamefont {M.}~\bibnamefont
  {Bier}},\ }\href {\doibase 10.1103/PhysRevLett.72.1766} {\bibfield  {journal}
  {\bibinfo  {journal} {Physical Review Letters}\ }\textbf {\bibinfo {volume}
  {72}},\ \bibinfo {pages} {1766} (\bibinfo {year} {1994})}\BibitemShut
  {NoStop}%
\bibitem [{\citenamefont {Astumian}\ and\ \citenamefont
  {Bier}(1996)}]{astumian1996mechanochemical}%
  \BibitemOpen
  \bibfield  {author} {\bibinfo {author} {\bibfnamefont {R.~D.}\ \bibnamefont
  {Astumian}}\ and\ \bibinfo {author} {\bibfnamefont {M.}~\bibnamefont
  {Bier}},\ }\href {\doibase 10.1016/S0006-3495(96)79605-4} {\bibfield
  {journal} {\bibinfo  {journal} {Biophysical Journal}\ }\textbf {\bibinfo
  {volume} {70}},\ \bibinfo {pages} {637} (\bibinfo {year} {1996})}\BibitemShut
  {NoStop}%
\bibitem [{\citenamefont {J{\"{u}}licher}\ \emph {et~al.}(1997)\citenamefont
  {J{\"{u}}licher}, \citenamefont {Ajdari},\ and\ \citenamefont
  {Prost}}]{julicher1997modeling}%
  \BibitemOpen
  \bibfield  {author} {\bibinfo {author} {\bibfnamefont {F.}~\bibnamefont
  {J{\"{u}}licher}}, \bibinfo {author} {\bibfnamefont {A.}~\bibnamefont
  {Ajdari}}, \ and\ \bibinfo {author} {\bibfnamefont {J.}~\bibnamefont
  {Prost}},\ }\href {\doibase 10.1103/RevModPhys.69.1269} {\bibfield  {journal}
  {\bibinfo  {journal} {Reviews of Modern Physics}\ }\textbf {\bibinfo {volume}
  {69}},\ \bibinfo {pages} {1269} (\bibinfo {year} {1997})}\BibitemShut
  {NoStop}%
\bibitem [{\citenamefont {J{\"{u}}licher}\ and\ \citenamefont
  {Prost}(1997)}]{julicher1997spontaneous}%
  \BibitemOpen
  \bibfield  {author} {\bibinfo {author} {\bibfnamefont {F.}~\bibnamefont
  {J{\"{u}}licher}}\ and\ \bibinfo {author} {\bibfnamefont {J.}~\bibnamefont
  {Prost}},\ }\href {\doibase 10.1103/PhysRevLett.78.4510} {\bibfield
  {journal} {\bibinfo  {journal} {Physical Review Letters}\ }\textbf {\bibinfo
  {volume} {78}},\ \bibinfo {pages} {4510} (\bibinfo {year}
  {1997})}\BibitemShut {NoStop}%
\bibitem [{\citenamefont {Peskin}\ \emph {et~al.}(1994)\citenamefont {Peskin},
  \citenamefont {Ermentrout},\ and\ \citenamefont
  {Oster}}]{peskin1995correlation}%
  \BibitemOpen
  \bibfield  {author} {\bibinfo {author} {\bibfnamefont {C.~S.}\ \bibnamefont
  {Peskin}}, \bibinfo {author} {\bibfnamefont {G.~B.}\ \bibnamefont
  {Ermentrout}}, \ and\ \bibinfo {author} {\bibfnamefont {G.~F.}\ \bibnamefont
  {Oster}}\ }(\bibinfo  {publisher} {Springer New York},\ \bibinfo {address}
  {New York, NY},\ \bibinfo {year} {1994})\ pp.\ \bibinfo {pages}
  {479--489}\BibitemShut {NoStop}%
\bibitem [{\citenamefont {Huxley}(1969)}]{huxley1969mechanism}%
  \BibitemOpen
  \bibfield  {author} {\bibinfo {author} {\bibfnamefont {H.~E.}\ \bibnamefont
  {Huxley}},\ }\href {\doibase 10.1126/science.164.3886.1356} {\bibfield
  {journal} {\bibinfo  {journal} {Science}\ }\textbf {\bibinfo {volume}
  {164}},\ \bibinfo {pages} {1356} (\bibinfo {year} {1969})}\BibitemShut
  {NoStop}%
\bibitem [{\citenamefont {Huxley}\ and\ \citenamefont
  {Simmons}(1971)}]{huxley1971proposed}%
  \BibitemOpen
  \bibfield  {author} {\bibinfo {author} {\bibfnamefont {A.~F.}\ \bibnamefont
  {Huxley}}\ and\ \bibinfo {author} {\bibfnamefont {R.~M.}\ \bibnamefont
  {Simmons}},\ }\href {\doibase 10.1038/233533a0} {\bibfield  {journal}
  {\bibinfo  {journal} {Nature}\ }\textbf {\bibinfo {volume} {233}},\ \bibinfo
  {pages} {533} (\bibinfo {year} {1971})}\BibitemShut {NoStop}%
\bibitem [{\citenamefont {Sung}\ \emph {et~al.}(2017)\citenamefont {Sung},
  \citenamefont {Mortensen}, \citenamefont {Spudich},\ and\ \citenamefont
  {Flyvbjerg}}]{sung2017chapter}%
  \BibitemOpen
  \bibfield  {author} {\bibinfo {author} {\bibfnamefont {J.}~\bibnamefont
  {Sung}}, \bibinfo {author} {\bibfnamefont {K.}~\bibnamefont {Mortensen}},
  \bibinfo {author} {\bibfnamefont {J.}~\bibnamefont {Spudich}}, \ and\
  \bibinfo {author} {\bibfnamefont {H.}~\bibnamefont {Flyvbjerg}},\ }\href
  {\doibase 10.1016/bs.mie.2016.08.002} {\bibfield  {journal} {\bibinfo
  {journal} {Methods in enzymology}\ }\textbf {\bibinfo {volume} {582}},\
  \bibinfo {pages} {1} (\bibinfo {year} {2017})}\BibitemShut {NoStop}%
\bibitem [{\citenamefont {Brown}\ \emph {et~al.}(2009)\citenamefont {Brown},
  \citenamefont {Hategan}, \citenamefont {Safer}, \citenamefont {Goldman},\
  and\ \citenamefont {Discher}}]{brown2009cross-correlated}%
  \BibitemOpen
  \bibfield  {author} {\bibinfo {author} {\bibfnamefont {A.~E.}\ \bibnamefont
  {Brown}}, \bibinfo {author} {\bibfnamefont {A.}~\bibnamefont {Hategan}},
  \bibinfo {author} {\bibfnamefont {D.}~\bibnamefont {Safer}}, \bibinfo
  {author} {\bibfnamefont {Y.~E.}\ \bibnamefont {Goldman}}, \ and\ \bibinfo
  {author} {\bibfnamefont {D.~E.}\ \bibnamefont {Discher}},\ }\href {\doibase
  10.1016/j.bpj.2008.11.032} {\bibfield  {journal} {\bibinfo  {journal}
  {Biophysical Journal}\ }\textbf {\bibinfo {volume} {96}},\ \bibinfo {pages}
  {1952} (\bibinfo {year} {2009})}\BibitemShut {NoStop}%
\bibitem [{\citenamefont {Stam}\ \emph {et~al.}(2015)\citenamefont {Stam},
  \citenamefont {Alberts}, \citenamefont {Gardel},\ and\ \citenamefont
  {Munro}}]{stam2015isoforms}%
  \BibitemOpen
  \bibfield  {author} {\bibinfo {author} {\bibfnamefont {S.}~\bibnamefont
  {Stam}}, \bibinfo {author} {\bibfnamefont {J.}~\bibnamefont {Alberts}},
  \bibinfo {author} {\bibfnamefont {M.~L.}\ \bibnamefont {Gardel}}, \ and\
  \bibinfo {author} {\bibfnamefont {E.}~\bibnamefont {Munro}},\ }\href
  {\doibase 10.1016/j.bpj.2015.03.030} {\bibfield  {journal} {\bibinfo
  {journal} {Biophysical Journal}\ }\textbf {\bibinfo {volume} {108}},\
  \bibinfo {pages} {1997} (\bibinfo {year} {2015})}\BibitemShut {NoStop}%
\bibitem [{\citenamefont {Albert}\ \emph {et~al.}(2014)\citenamefont {Albert},
  \citenamefont {Erdmann},\ and\ \citenamefont {Schwarz}}]{Albert2014}%
  \BibitemOpen
  \bibfield  {author} {\bibinfo {author} {\bibfnamefont {P.~J.}\ \bibnamefont
  {Albert}}, \bibinfo {author} {\bibfnamefont {T.}~\bibnamefont {Erdmann}}, \
  and\ \bibinfo {author} {\bibfnamefont {U.~S.}\ \bibnamefont {Schwarz}},\
  }\href {\doibase 10.1088/1367-2630/16/9/093019} {\bibfield  {journal}
  {\bibinfo  {journal} {New Journal of Physics}\ }\textbf {\bibinfo {volume}
  {16}},\ \bibinfo {pages} {093019} (\bibinfo {year} {2014})},\ \Eprint
  {http://arxiv.org/abs/1404.1587} {arXiv:1404.1587} \BibitemShut {NoStop}%
\bibitem [{\citenamefont {van Kampen}(1981)}]{vanKampen1981stochastic}%
  \BibitemOpen
  \bibfield  {author} {\bibinfo {author} {\bibfnamefont {N.~G.}\ \bibnamefont
  {van Kampen}},\ }\href@noop {} {\emph {\bibinfo {title} {{Stochastic
  processes in chemistry and physics}}}}\ (\bibinfo  {publisher} {Elsevier -
  North Holland},\ \bibinfo {address} {Amsterdam},\ \bibinfo {year} {1981})\
  p.\ \bibinfo {pages} {464}\BibitemShut {NoStop}%
\bibitem [{\citenamefont {Bierbaum}\ and\ \citenamefont
  {Lipowsky}(2011)}]{Bierbaum2011}%
  \BibitemOpen
  \bibfield  {author} {\bibinfo {author} {\bibfnamefont {V.}~\bibnamefont
  {Bierbaum}}\ and\ \bibinfo {author} {\bibfnamefont {R.}~\bibnamefont
  {Lipowsky}},\ }\href {\doibase 10.1016/j.bpj.2011.02.012} {\bibfield
  {journal} {\bibinfo  {journal} {Biophysical Journal}\ }\textbf {\bibinfo
  {volume} {100}},\ \bibinfo {pages} {1747} (\bibinfo {year}
  {2011})}\BibitemShut {NoStop}%
\bibitem [{\citenamefont {Bierbaum}\ and\ \citenamefont
  {Lipowsky}(2013)}]{Bierbaum2013}%
  \BibitemOpen
  \bibfield  {author} {\bibinfo {author} {\bibfnamefont {V.}~\bibnamefont
  {Bierbaum}}\ and\ \bibinfo {author} {\bibfnamefont {R.}~\bibnamefont
  {Lipowsky}},\ }\href {\doibase 10.1371/journal.pone.0055366} {\bibfield
  {journal} {\bibinfo  {journal} {PLoS ONE}\ }\textbf {\bibinfo {volume} {8}},\
  \bibinfo {pages} {e55366} (\bibinfo {year} {2013})}\BibitemShut {NoStop}%
\bibitem [{\citenamefont {Pla{\c{c}}ais}\ \emph {et~al.}(2009)\citenamefont
  {Pla{\c{c}}ais}, \citenamefont {Balland}, \citenamefont {Gu{\'{e}}rin},
  \citenamefont {Joanny},\ and\ \citenamefont
  {Martin}}]{placcais2009spontaneous}%
  \BibitemOpen
  \bibfield  {author} {\bibinfo {author} {\bibfnamefont {P.-Y.}\ \bibnamefont
  {Pla{\c{c}}ais}}, \bibinfo {author} {\bibfnamefont {M.}~\bibnamefont
  {Balland}}, \bibinfo {author} {\bibfnamefont {T.}~\bibnamefont
  {Gu{\'{e}}rin}}, \bibinfo {author} {\bibfnamefont {J.-F.}\ \bibnamefont
  {Joanny}}, \ and\ \bibinfo {author} {\bibfnamefont {P.}~\bibnamefont
  {Martin}},\ }\href {\doibase 10.1103/PhysRevLett.103.158102} {\bibfield
  {journal} {\bibinfo  {journal} {Physical Review Letters}\ }\textbf {\bibinfo
  {volume} {103}},\ \bibinfo {pages} {158102} (\bibinfo {year}
  {2009})}\BibitemShut {NoStop}%
\bibitem [{\citenamefont {Saito}\ \emph {et~al.}(1994)\citenamefont {Saito},
  \citenamefont {Aoki}, \citenamefont {Aoki},\ and\ \citenamefont
  {Yanagida}}]{saito1994movement}%
  \BibitemOpen
  \bibfield  {author} {\bibinfo {author} {\bibfnamefont {K.}~\bibnamefont
  {Saito}}, \bibinfo {author} {\bibfnamefont {T.}~\bibnamefont {Aoki}},
  \bibinfo {author} {\bibfnamefont {T.}~\bibnamefont {Aoki}}, \ and\ \bibinfo
  {author} {\bibfnamefont {T.}~\bibnamefont {Yanagida}},\ }\href {\doibase
  10.1016/S0006-3495(94)80853-7} {\bibfield  {journal} {\bibinfo  {journal}
  {Biophysical Journal}\ }\textbf {\bibinfo {volume} {66}},\ \bibinfo {pages}
  {769} (\bibinfo {year} {1994})}\BibitemShut {NoStop}%
\bibitem [{\citenamefont {Kishino}\ and\ \citenamefont
  {Yanagida}(1988)}]{kishino1988force}%
  \BibitemOpen
  \bibfield  {author} {\bibinfo {author} {\bibfnamefont {A.}~\bibnamefont
  {Kishino}}\ and\ \bibinfo {author} {\bibfnamefont {T.}~\bibnamefont
  {Yanagida}},\ }\href {\doibase 10.1038/334074a0} {\bibfield  {journal}
  {\bibinfo  {journal} {Nature}\ }\textbf {\bibinfo {volume} {334}},\ \bibinfo
  {pages} {74} (\bibinfo {year} {1988})}\BibitemShut {NoStop}%
\bibitem [{\citenamefont {Finer}\ \emph {et~al.}(1994)\citenamefont {Finer},
  \citenamefont {Simmons},\ and\ \citenamefont {Spudich}}]{finer1994single}%
  \BibitemOpen
  \bibfield  {author} {\bibinfo {author} {\bibfnamefont {J.~T.}\ \bibnamefont
  {Finer}}, \bibinfo {author} {\bibfnamefont {R.~M.}\ \bibnamefont {Simmons}},
  \ and\ \bibinfo {author} {\bibfnamefont {J.~A.}\ \bibnamefont {Spudich}},\
  }\href {\doibase 10.1038/368113a0} {\bibfield  {journal} {\bibinfo  {journal}
  {Nature}\ }\textbf {\bibinfo {volume} {368}},\ \bibinfo {pages} {113}
  (\bibinfo {year} {1994})}\BibitemShut {NoStop}%
\bibitem [{\citenamefont {Broersma}(1960)}]{Broersma1960}%
  \BibitemOpen
  \bibfield  {author} {\bibinfo {author} {\bibfnamefont {S.}~\bibnamefont
  {Broersma}},\ }\href {\doibase 10.1063/1.1730995} {\bibfield  {journal}
  {\bibinfo  {journal} {The Journal of Chemical Physics}\ }\textbf {\bibinfo
  {volume} {32}},\ \bibinfo {pages} {1632} (\bibinfo {year}
  {1960})}\BibitemShut {NoStop}%
\bibitem [{\citenamefont {Broersma}(1981)}]{Broersma1981}%
  \BibitemOpen
  \bibfield  {author} {\bibinfo {author} {\bibfnamefont {S.}~\bibnamefont
  {Broersma}},\ }\href {\doibase 10.1063/1.441071} {\bibfield  {journal}
  {\bibinfo  {journal} {The Journal of Chemical Physics}\ }\textbf {\bibinfo
  {volume} {74}},\ \bibinfo {pages} {6989} (\bibinfo {year}
  {1981})}\BibitemShut {NoStop}%
\bibitem [{\citenamefont {Li}\ and\ \citenamefont
  {Tang}(2004)}]{li2004diffusion}%
  \BibitemOpen
  \bibfield  {author} {\bibinfo {author} {\bibfnamefont {G.}~\bibnamefont
  {Li}}\ and\ \bibinfo {author} {\bibfnamefont {J.~X.}\ \bibnamefont {Tang}},\
  }\href {\doibase 10.1103/PhysRevE.69.061921} {\bibfield  {journal} {\bibinfo
  {journal} {Physical Review E}\ }\textbf {\bibinfo {volume} {69}},\ \bibinfo
  {pages} {061921} (\bibinfo {year} {2004})}\BibitemShut {NoStop}%
\bibitem [{\citenamefont {Vilfan}\ and\ \citenamefont
  {Duke}(2003)}]{vilfan2003instabilities}%
  \BibitemOpen
  \bibfield  {author} {\bibinfo {author} {\bibfnamefont {A.}~\bibnamefont
  {Vilfan}}\ and\ \bibinfo {author} {\bibfnamefont {T.}~\bibnamefont {Duke}},\
  }\href {\doibase 10.1016/S0006-3495(03)74522-6} {\bibfield  {journal}
  {\bibinfo  {journal} {Biophysical Journal}\ }\textbf {\bibinfo {volume}
  {85}},\ \bibinfo {pages} {818} (\bibinfo {year} {2003})}\BibitemShut
  {NoStop}%
\bibitem [{\citenamefont {Blanchoin}\ \emph {et~al.}(2014)\citenamefont
  {Blanchoin}, \citenamefont {Boujemaa-Paterski}, \citenamefont {Sykes},\ and\
  \citenamefont {Plastino}}]{blanchoin2014actin}%
  \BibitemOpen
  \bibfield  {author} {\bibinfo {author} {\bibfnamefont {L.}~\bibnamefont
  {Blanchoin}}, \bibinfo {author} {\bibfnamefont {R.}~\bibnamefont
  {Boujemaa-Paterski}}, \bibinfo {author} {\bibfnamefont {C.}~\bibnamefont
  {Sykes}}, \ and\ \bibinfo {author} {\bibfnamefont {J.}~\bibnamefont
  {Plastino}},\ }\href {\doibase 10.1152/physrev.00018.2013} {\bibfield
  {journal} {\bibinfo  {journal} {Physiological Reviews}\ }\textbf {\bibinfo
  {volume} {94}},\ \bibinfo {pages} {235} (\bibinfo {year} {2014})}\BibitemShut
  {NoStop}%
\bibitem [{\citenamefont {Ennomani}\ \emph {et~al.}(2016)\citenamefont
  {Ennomani}, \citenamefont {Letort}, \citenamefont {Gu{\'{e}}rin},
  \citenamefont {Martiel}, \citenamefont {Cao}, \citenamefont
  {N{\'{e}}d{\'{e}}lec}, \citenamefont {{De La Cruz}}, \citenamefont
  {Th{\'{e}}ry},\ and\ \citenamefont {Blanchoin}}]{ennomani2016architecture}%
  \BibitemOpen
  \bibfield  {author} {\bibinfo {author} {\bibfnamefont {H.}~\bibnamefont
  {Ennomani}}, \bibinfo {author} {\bibfnamefont {G.}~\bibnamefont {Letort}},
  \bibinfo {author} {\bibfnamefont {C.}~\bibnamefont {Gu{\'{e}}rin}}, \bibinfo
  {author} {\bibfnamefont {J.-L.}\ \bibnamefont {Martiel}}, \bibinfo {author}
  {\bibfnamefont {W.}~\bibnamefont {Cao}}, \bibinfo {author} {\bibfnamefont
  {F.}~\bibnamefont {N{\'{e}}d{\'{e}}lec}}, \bibinfo {author} {\bibfnamefont
  {E.~M.}\ \bibnamefont {{De La Cruz}}}, \bibinfo {author} {\bibfnamefont
  {M.}~\bibnamefont {Th{\'{e}}ry}}, \ and\ \bibinfo {author} {\bibfnamefont
  {L.}~\bibnamefont {Blanchoin}},\ }\href {\doibase 10.1016/j.cub.2015.12.069}
  {\bibfield  {journal} {\bibinfo  {journal} {Current Biology}\ }\textbf
  {\bibinfo {volume} {26}},\ \bibinfo {pages} {616} (\bibinfo {year}
  {2016})}\BibitemShut {NoStop}%
\bibitem [{\citenamefont {Kloeden}\ and\ \citenamefont
  {Platen}(1989)}]{kloeden1989survey}%
  \BibitemOpen
  \bibfield  {author} {\bibinfo {author} {\bibfnamefont {P.~E.}\ \bibnamefont
  {Kloeden}}\ and\ \bibinfo {author} {\bibfnamefont {E.}~\bibnamefont
  {Platen}},\ }\href {\doibase 10.1007/BF01543857} {\bibfield  {journal}
  {\bibinfo  {journal} {Stochastic Hydrology and Hydraulics}\ }\textbf
  {\bibinfo {volume} {3}},\ \bibinfo {pages} {155} (\bibinfo {year}
  {1989})}\BibitemShut {NoStop}%
\bibitem [{\citenamefont {Maruyama}(1955)}]{maruyama1955continuous}%
  \BibitemOpen
  \bibfield  {author} {\bibinfo {author} {\bibfnamefont {G.}~\bibnamefont
  {Maruyama}},\ }\href {\doibase 10.1007/BF02846028} {\bibfield  {journal}
  {\bibinfo  {journal} {Rendiconti del Circolo Matematico di Palermo}\ }\textbf
  {\bibinfo {volume} {4}},\ \bibinfo {pages} {48} (\bibinfo {year}
  {1955})}\BibitemShut {NoStop}%
\bibitem [{\citenamefont {Rao}\ and\ \citenamefont
  {Arkin}(2003)}]{rao2003stochastic}%
  \BibitemOpen
  \bibfield  {author} {\bibinfo {author} {\bibfnamefont {C.~V.}\ \bibnamefont
  {Rao}}\ and\ \bibinfo {author} {\bibfnamefont {A.~P.}\ \bibnamefont
  {Arkin}},\ }\href {\doibase 10.1063/1.1545446} {\bibfield  {journal}
  {\bibinfo  {journal} {The Journal of chemical physics}\ }\textbf {\bibinfo
  {volume} {118}},\ \bibinfo {pages} {4999} (\bibinfo {year}
  {2003})}\BibitemShut {NoStop}%
\bibitem [{\citenamefont {Schwarz}\ \emph {et~al.}(2006)\citenamefont
  {Schwarz}, \citenamefont {Erdmann},\ and\ \citenamefont
  {Bischofs}}]{schwarz2006focal}%
  \BibitemOpen
  \bibfield  {author} {\bibinfo {author} {\bibfnamefont {U.~S.}\ \bibnamefont
  {Schwarz}}, \bibinfo {author} {\bibfnamefont {T.}~\bibnamefont {Erdmann}}, \
  and\ \bibinfo {author} {\bibfnamefont {I.~B.}\ \bibnamefont {Bischofs}},\
  }\href {\doibase 10.1016/j.biosystems.2005.05.019} {\bibfield  {journal}
  {\bibinfo  {journal} {Biosystems}\ }\textbf {\bibinfo {volume} {83}},\
  \bibinfo {pages} {225} (\bibinfo {year} {2006})}\BibitemShut {NoStop}%
\bibitem [{\citenamefont {Prager-Khoutorsky}\ \emph {et~al.}(2011)\citenamefont
  {Prager-Khoutorsky}, \citenamefont {Lichtenstein}, \citenamefont {Krishnan},
  \citenamefont {Rajendran}, \citenamefont {Mayo}, \citenamefont {Kam},
  \citenamefont {Geiger},\ and\ \citenamefont
  {Bershadsky}}]{prager2011fibroblast}%
  \BibitemOpen
  \bibfield  {author} {\bibinfo {author} {\bibfnamefont {M.}~\bibnamefont
  {Prager-Khoutorsky}}, \bibinfo {author} {\bibfnamefont {A.}~\bibnamefont
  {Lichtenstein}}, \bibinfo {author} {\bibfnamefont {R.}~\bibnamefont
  {Krishnan}}, \bibinfo {author} {\bibfnamefont {K.}~\bibnamefont {Rajendran}},
  \bibinfo {author} {\bibfnamefont {A.}~\bibnamefont {Mayo}}, \bibinfo {author}
  {\bibfnamefont {Z.}~\bibnamefont {Kam}}, \bibinfo {author} {\bibfnamefont
  {B.}~\bibnamefont {Geiger}}, \ and\ \bibinfo {author} {\bibfnamefont {A.~D.}\
  \bibnamefont {Bershadsky}},\ }\href {\doibase 10.1038/ncb2370} {\bibfield
  {journal} {\bibinfo  {journal} {Nature cell biology}\ }\textbf {\bibinfo
  {volume} {13}},\ \bibinfo {pages} {1457} (\bibinfo {year}
  {2011})}\BibitemShut {NoStop}%
\bibitem [{\citenamefont {Geiger}\ \emph {et~al.}(2009)\citenamefont {Geiger},
  \citenamefont {Spatz},\ and\ \citenamefont
  {Bershadsky}}]{geiger2009environmental}%
  \BibitemOpen
  \bibfield  {author} {\bibinfo {author} {\bibfnamefont {B.}~\bibnamefont
  {Geiger}}, \bibinfo {author} {\bibfnamefont {J.~P.}\ \bibnamefont {Spatz}}, \
  and\ \bibinfo {author} {\bibfnamefont {A.~D.}\ \bibnamefont {Bershadsky}},\
  }\href {\doibase 10.1038/nrm2593} {\bibfield  {journal} {\bibinfo  {journal}
  {Nature Reviews Molecular Cell Biology}\ }\textbf {\bibinfo {volume} {10}},\
  \bibinfo {pages} {21} (\bibinfo {year} {2009})}\BibitemShut {NoStop}%
\bibitem [{\citenamefont {{\'E}tienne}\ \emph {et~al.}(2015)\citenamefont
  {{\'E}tienne}, \citenamefont {Fouchard}, \citenamefont {Mitrossilis},
  \citenamefont {Bufi}, \citenamefont {Durand-Smet},\ and\ \citenamefont
  {Asnacios}}]{etienne2015cells}%
  \BibitemOpen
  \bibfield  {author} {\bibinfo {author} {\bibfnamefont {J.}~\bibnamefont
  {{\'E}tienne}}, \bibinfo {author} {\bibfnamefont {J.}~\bibnamefont
  {Fouchard}}, \bibinfo {author} {\bibfnamefont {D.}~\bibnamefont
  {Mitrossilis}}, \bibinfo {author} {\bibfnamefont {N.}~\bibnamefont {Bufi}},
  \bibinfo {author} {\bibfnamefont {P.}~\bibnamefont {Durand-Smet}}, \ and\
  \bibinfo {author} {\bibfnamefont {A.}~\bibnamefont {Asnacios}},\ }\href
  {\doibase 10.1073/pnas.1417113112} {\bibfield  {journal} {\bibinfo  {journal}
  {Proceedings of the National Academy of Sciences}\ }\textbf {\bibinfo
  {volume} {112}},\ \bibinfo {pages} {2740} (\bibinfo {year}
  {2015})}\BibitemShut {NoStop}%
\bibitem [{\citenamefont {Pe{\v{s}}ek}(2013)}]{Pesek2013}%
  \BibitemOpen
  \bibfield  {author} {\bibinfo {author} {\bibfnamefont {J.}~\bibnamefont
  {Pe{\v{s}}ek}},\ }\emph {\bibinfo {title} {{Heat processes in non-equilibrium
  stochastic systems}}},\ \href {http://arxiv.org/abs/1407.7434} {\bibinfo
  {type} {Phd thesis}},\ \bibinfo  {school} {Faculty of Mathematics and
  Physics, Charles University in Prague} (\bibinfo {year} {2013}),\ \Eprint
  {http://arxiv.org/abs/1407.7434} {arXiv:1407.7434} \BibitemShut {NoStop}%
\bibitem [{Note1()}]{Note1}%
  \BibitemOpen
  \bibinfo {note} {$\chi (\protect \text {condition}) = 1$ if condition is
  fulfilled otherwise $0$.}\BibitemShut {Stop}%
\bibitem [{\citenamefont {Schmiedl}\ and\ \citenamefont
  {Seifert}(2008)}]{Schmiedl2008}%
  \BibitemOpen
  \bibfield  {author} {\bibinfo {author} {\bibfnamefont {T.}~\bibnamefont
  {Schmiedl}}\ and\ \bibinfo {author} {\bibfnamefont {U.}~\bibnamefont
  {Seifert}},\ }\href {\doibase 10.1209/0295-5075/83/30005} {\bibfield
  {journal} {\bibinfo  {journal} {EPL (Europhysics Letters)}\ }\textbf
  {\bibinfo {volume} {83}},\ \bibinfo {pages} {30005} (\bibinfo {year}
  {2008})},\ \Eprint {http://arxiv.org/abs/0801.3743} {arXiv:0801.3743}
  \BibitemShut {NoStop}%
\bibitem [{\citenamefont {Boksenbojm}\ and\ \citenamefont
  {Wynants}(2009)}]{boksenbojm2009entropy}%
  \BibitemOpen
  \bibfield  {author} {\bibinfo {author} {\bibfnamefont {E.}~\bibnamefont
  {Boksenbojm}}\ and\ \bibinfo {author} {\bibfnamefont {B.}~\bibnamefont
  {Wynants}},\ }\href {\doibase 10.1088/1751-8113/42/44/445003} {\bibfield
  {journal} {\bibinfo  {journal} {Journal of Physics A: Mathematical and
  Theoretical}\ }\textbf {\bibinfo {volume} {42}},\ \bibinfo {pages} {445003}
  (\bibinfo {year} {2009})},\ \Eprint {http://arxiv.org/abs/0906.4222}
  {arXiv:0906.4222} \BibitemShut {NoStop}%
\bibitem [{\citenamefont {Parmeggiani}\ \emph {et~al.}(1999)\citenamefont
  {Parmeggiani}, \citenamefont {J{\"{u}}licher}, \citenamefont {Ajdari},\ and\
  \citenamefont {Prost}}]{parmeggiani1999energy}%
  \BibitemOpen
  \bibfield  {author} {\bibinfo {author} {\bibfnamefont {A.}~\bibnamefont
  {Parmeggiani}}, \bibinfo {author} {\bibfnamefont {F.}~\bibnamefont
  {J{\"{u}}licher}}, \bibinfo {author} {\bibfnamefont {A.}~\bibnamefont
  {Ajdari}}, \ and\ \bibinfo {author} {\bibfnamefont {J.}~\bibnamefont
  {Prost}},\ }\href {\doibase 10.1103/PhysRevE.60.2127} {\bibfield  {journal}
  {\bibinfo  {journal} {Physical Review E}\ }\textbf {\bibinfo {volume} {60}},\
  \bibinfo {pages} {2127} (\bibinfo {year} {1999})}\BibitemShut {NoStop}%
\bibitem [{\citenamefont {Parrondo}\ \emph {et~al.}(1998)\citenamefont
  {Parrondo}, \citenamefont {Blanco}, \citenamefont {Cao},\ and\ \citenamefont
  {Brito}}]{parrondo1998efficiency}%
  \BibitemOpen
  \bibfield  {author} {\bibinfo {author} {\bibfnamefont {J.~M.~R.}\
  \bibnamefont {Parrondo}}, \bibinfo {author} {\bibfnamefont {J.~M.}\
  \bibnamefont {Blanco}}, \bibinfo {author} {\bibfnamefont {F.~J.}\
  \bibnamefont {Cao}}, \ and\ \bibinfo {author} {\bibfnamefont
  {R.}~\bibnamefont {Brito}},\ }\href {\doibase 10.1209/epl/i1998-00348-5}
  {\bibfield  {journal} {\bibinfo  {journal} {Europhysics Letters (EPL)}\
  }\textbf {\bibinfo {volume} {43}},\ \bibinfo {pages} {248} (\bibinfo {year}
  {1998})},\ \Eprint {http://arxiv.org/abs/43303} {arXiv:43303} \BibitemShut
  {NoStop}%
\bibitem [{\citenamefont {Zhang}(2009)}]{zhang2009efficiency}%
  \BibitemOpen
  \bibfield  {author} {\bibinfo {author} {\bibfnamefont {Y.}~\bibnamefont
  {Zhang}},\ }\href {\doibase 10.1007/s10955-009-9695-3} {\bibfield  {journal}
  {\bibinfo  {journal} {Journal of Statistical Physics}\ }\textbf {\bibinfo
  {volume} {134}},\ \bibinfo {pages} {669} (\bibinfo {year}
  {2009})}\BibitemShut {NoStop}%
\bibitem [{\citenamefont {Maskow}\ and\ \citenamefont
  {Paufler}(2015)}]{Maskow2015}%
  \BibitemOpen
  \bibfield  {author} {\bibinfo {author} {\bibfnamefont {T.}~\bibnamefont
  {Maskow}}\ and\ \bibinfo {author} {\bibfnamefont {S.}~\bibnamefont
  {Paufler}},\ }\href {\doibase 10.1016/j.ymeth.2014.10.035} {\bibfield
  {journal} {\bibinfo  {journal} {Methods}\ }\textbf {\bibinfo {volume} {76}},\
  \bibinfo {pages} {3} (\bibinfo {year} {2015})}\BibitemShut {NoStop}%
\bibitem [{\citenamefont {Maes}\ and\ \citenamefont
  {Neto{\v{c}}n{\'{y}}}(2010)}]{Maes2010}%
  \BibitemOpen
  \bibfield  {author} {\bibinfo {author} {\bibfnamefont {C.}~\bibnamefont
  {Maes}}\ and\ \bibinfo {author} {\bibfnamefont {K.}~\bibnamefont
  {Neto{\v{c}}n{\'{y}}}},\ }\href {\doibase 10.1063/1.3274819} {\bibfield
  {journal} {\bibinfo  {journal} {Journal of Mathematical Physics}\ }\textbf
  {\bibinfo {volume} {51}},\ \bibinfo {pages} {015219} (\bibinfo {year}
  {2010})}\BibitemShut {NoStop}%
\end{thebibliography}%

\end{document}